\documentclass[final,5p,times,twocolumn,authoryear]{elsarticle}

\graphicspath{{assets/}}
\DeclareGraphicsExtensions{.pdf,.jpeg,.png}

\usepackage{mathrsfs}
\usepackage{amsmath}
\usepackage{amssymb}
\usepackage{amsthm}
\usepackage{bbm}
\usepackage{array}
\usepackage{xcolor}

\usepackage{algorithm}
\usepackage{algorithmic}

\usepackage[caption=false,font=footnotesize]{subfig}

\usepackage{url}
\usepackage[hidelinks]{hyperref}

\theoremstyle{plain}
\newtheorem{theorem}{Theorem}

\newtheorem{proposition}{Proposition}
\newtheorem{corollary}{Corollary}

\theoremstyle{definition}
\newtheorem{assumption}{Assumption}
\newtheorem{definition}{Definition}

\theoremstyle{remark}
\newtheorem{remark}{Remark}

\journal{Automatica}

\begin{document}

\begin{frontmatter}

\title{Distributed State Estimation over Wireless MIMO Fading Channels with Imperfect CSI}

\author[ZJU]{Xinming Yu}
\ead{yuxinming@zju.edu.cn}

\author[ZJU]{Chan Wang}
\ead{0617464@zju.edu.cn}

\author[ZJU]{Ming-min Zhao}
\ead{zmmblack@zju.edu.cn}

\author[ZJU]{An Liu}
\ead{anliu@zju.edu.cn}

\author[ZJU]{Songfu Cai\corref{cor1}}
\cortext[cor1]{Corresponding author}
\ead{sfcai@zju.edu.cn}

\affiliation[ZJU]{%
    organization={College of Information Science and Electronic Engineering, Zhejiang University},
    city={Hangzhou},
    country={China}
}

\begin{abstract}
We consider distributed state estimation of a linear plant under stabilizing feedback over wireless multiple-input multiple-output (MIMO) fading channels, where channel-estimation noise corrupts the channel state information (CSI) used for information fusion. We propose a distributed estimator that combines local prediction, communication over wireless MIMO channels, and a posterior update based on local information and communicated data. Each sensor uses noisy pilots to estimate its aggregated wireless MIMO fading channel, and a pilot-noise bias correction recovers the required neighboring-data differences in conditional mean. We first analyze the mean estimation error and show that a Chebyshev design of the communication weights achieves a prescribed decay factor $d\in(0,1)$ satisfying the sensing conditions, with finitely many communication updates per plant sample at every finite signal-to-noise ratio (SNR) gain. We then characterize mean-square stability and the effects of SNR gain and receive-array size on the state estimation mean-square error (MSE), including its transient and steady-state value. In the ideal-communication limit, we show that the square root of the absolute MSE deviation from its steady-state value decays as $o(d^k)$. Numerical simulations verify the theoretical results and show substantial performance gain compared to various state-of-the-art baselines.

\end{abstract}


\end{frontmatter}


\section{Introduction}
Distributed state estimation allows sensors with partial observations to reconstruct the state of a dynamical system by exchanging information. The resulting estimates support monitoring, decision making, and feedback control~\citep{rego2019survey,jia2021resource,zheng2024heterogeneous} in applications such as industrial and environmental monitoring~\citep{jia2021resource}, smart grids~\citep{cheng2024multisource}, target tracking~\citep{zhang2022twotarget}, and cooperative autonomous systems~\citep{bai2024cooperative}. In wireless networked systems, a typical distributed state-estimation architecture consists of spatially distributed wireless sensors that collect partial and noisy measurements and exchange locally acquired information to obtain a global estimate of the underlying dynamical-system state.
Wireless fading, receiver noise, and errors in channel state information (CSI) affect this exchange and can prevent the sensor network from attaining the required estimation accuracy and convergence rate.

The relation between local sensing and information exchange is well understood for reliable communication channels. The centralized Kalman filter gives the minimum mean-square error (MMSE) estimate for linear Gaussian systems~\citep{kalman1960}. Distributed implementations replace centralized information processing by consensus, diffusion, or consensus-plus-innovations updates~\citep{olfatisaber2007dkf,cattivelli2010diffusion,das2017consensus,battilotti2021stability,battilotti2025reduced}. For example, the consensus-based distributed Kalman filter of \citet{olfatisaber2007dkf} recovers the centralized estimate when exact consensus is attained on the required measurement and covariance information. The diffusion strategy of \citet{cattivelli2010diffusion} provides unbiasedness and steady-state error-covariance guarantees. Under global detectability and sufficient network capacity, \citet{das2017consensus} established mean-square convergence for a consensus-plus-innovations filter. Other approaches use communication-efficient consensus~\citep{battilotti2021stability,battilotti2025reduced}, graph-theoretic detectability conditions~\citep{park2017omniscience,mitra2018distributed}, or optimization and Bayesian consensus~\citep{yang2025network,battistelli2014kl}. These results explain how measurements, plant dynamics, and graph connectivity jointly determine state estimation performance. Their communication models, however, generally assume error-free information exchange or describe link changes without the signal distortion introduced by a physical wireless channel.

Existing networked state estimation studies have incorporated several forms of unreliable communication. For intermittent observations, \citet{sinopoli2004intermittent} characterized estimation stability in terms of packet-dropping rates through an iterative linear matrix inequality (LMI) feasibility problem. In a multisensor distributed setting, \citet{zhou2019packetdrops} derived a modified algebraic Riccati equation and an LMI-computable admissible dropout-rate margin. Buffered transmissions under intermittent access and deception attacks were considered by \citet{huang2025buffer}. For analog-erasure channels, \citet{mitra2018secureerasure} related mean-square stability to plant instability, graph robustness, and the effective erasure probability. When link failures are not directly observed, the filter of \citet{battilotti2018unknownlinks} includes local failure detection and evaluates estimation performance numerically. Beyond binary delivery, \citet{xu2018roundrobin} obtained periodic LMI stability conditions for scalar-fading channels under Round-Robin scheduling. For temporally correlated scalar fading, \citet{li2021timecorrelated} used an augmented plant--channel model to minimize a finite-horizon covariance upper bound. In these models~\citep{zhou2019packetdrops,huang2025buffer,mitra2018secureerasure,battilotti2018unknownlinks,xu2018roundrobin,li2021timecorrelated}, data communication over each link is represented by an on--off variable or a scalar gain. Such a representation does not describe the mixing of state coordinates by a full wireless multiple-input multiple-output (MIMO) fading channel matrix.

A full MIMO fading channel mixes the transmitted coordinates, so different state directions can experience different attenuation and information loss. Most estimation results for these channels concern a remote sensor and a central estimator. For a static MIMO additive white Gaussian noise channel, \citet{jahangiri2025finitemse} studied finite mean-square error (MSE) conditions for linear encoding and decoding of vector Gauss--Markov sources. Parallel-channel models retain several transmission dimensions while excluding instantaneous mixing between them. Within this class, \citet{jin2024parallel} established necessary and sufficient power conditions for mean-square-stable remote estimation, and \citet{liu2023markov} considered multiple linear processes with correlated finite-state Markov fading across time and channels. For wireless MIMO fading channels, \citet{parseh2017distortion} combined orthogonal space--time coding and Kalman filtering to analyze distortion outage and antenna diversity. Allowing the effective measurement matrix to lose rank at individual time instants, \citet{cai2020framework} decomposed the Riccati recursion into observable and unobservable components and derived sufficient communication-resource conditions for stability.

Distributed state estimation further requires that information pass between sensors without a fusion center. Existing distributed designs have considered diagonal Rayleigh fading~\citep{han2023rayleigh}, diagonal stochastic fading in gossip and proportional--integral observers~\citep{wang2026gossip}, and diagonal fading on direct and relay links~\citep{han2024relay}. Diagonal channels model attenuation of individual coordinates but exclude the coordinate mixing caused by a full fading matrix. Moreover, these channel-based estimation studies~\citep{jahangiri2025finitemse,parseh2017distortion,cai2020framework,han2023rayleigh,wang2026gossip,han2024relay,liu2023markov,jin2024parallel} do not account for the CSI uncertainty introduced by channel-estimation noise. Consequently, those guarantees do not directly characterize a sensor network communicating over full MIMO fading matrices when each receiver has only a noisy estimate of the sum of its incoming channels.

We consider distributed state estimation over wireless MIMO fading channels with imperfect CSI induced by channel-estimation noises. We aim to recover the plant state at each sensor with a prescribed convergence rate. We establish prescribed decay of the mean state estimation error $\bar{\mathbf e}(k)\triangleq\mathbb E[\mathbf e(k)]$, where $\mathbf e(k)$ stacks the local state estimation errors. We also give conditions under which $|\mathcal M(k)-\mathcal M_\star|^{1/2}=o(d^k)$ for a prescribed $d\in(0,1)$. Here $\mathcal M(k)$ is the state estimation MSE and $\mathcal M_\star$ is its steady-state value. The main contributions are as follows.

\begin{itemize}
\item Distributed estimator using imperfect aggregate CSI.
We propose a distributed estimator that combines local prediction, analog over-the-air (OTA) aggregation of neighboring data, and a posterior update based on local information and communicated data. The fusion update acts only on coordinates requiring information from neighboring sensors. Their estimates propagate over multiple hops during $T$ communication cycles. Node $i$ forms the fusion signal using the received data and the noisy estimate $\hat{\mathbf H}_i^\tau(k)$ of its incoming channel sum, without requiring perfect CSI.

\item Prescribed mean-error convergence.
We compensate for the mean bias caused by pilot noise so that the conditional-mean fusion signal equals the sum of neighboring data differences. For any prescribed $d\in(0,1)$ satisfying the sensing conditions, a Chebyshev design of the fusion weights and communication count $T$ guarantees $\|\bar{\mathbf e}(k)\|=o(d^k)$ for every finite signal-to-noise ratio (SNR) gain $\rho$.

\item Prescribed MSE convergence and impacts of communication resources.
We explicitly characterize how the SNR gain $\rho$ and receive-array size $N_r$ affect the transient convergence bounds and steady-state MSE $\mathcal M_\star$. In the joint ideal-communication limit $\rho,N_r\to\infty$, the MSE approaches $\mathcal M_\star$ at the prescribed rate \mbox{$|\mathcal M(k)-\mathcal M_\star|^{1/2}=o(d^k)$}. We also identify a mean-square stability condition under which $\mathcal M_\star\to0$ as the process-noise covariance $\mathbf W_w\to\mathbf0$ and $\rho\to\infty$, even with fixed $N_r$.
\end{itemize}

\subsection{Notation}
For a square matrix $\mathbf A$, $\mathbf A^\top$, $r(\mathbf A)$, and $\|\mathbf A\|_{\mathrm F}$ denote its transpose, spectral radius, and Frobenius norm, respectively. The Euclidean norm of a vector is $\|\cdot\|$. The symbols $\mathbb E[\cdot]$, $\operatorname{tr}(\cdot)$, $\otimes$, $\operatorname{col}(\cdot)$, and $\operatorname{blkdiag}(\cdot)$ denote expectation, trace, Kronecker product, vertical block stacking, and block-diagonal stacking. The vectorization $\operatorname{vec}(\mathbf A)$ stacks the columns of $\mathbf A$. The spectrum of a square matrix is $\sigma(\mathbf A)$, its spectral norm is $\|\mathbf A\|_2$, and $z^*$ denotes the complex conjugate of a scalar $z$. Empty sums are defined as zero.

\section{System Model}\label{sec:system_model}
We first specify the plant state $\mathbf x(k)$ and the communication observations, then define the state estimation MSE $\mathcal M(k)$ and its required convergence factor $d$. These definitions determine which state coordinates need information from neighboring sensors.

\subsection{Dynamic Plant Model}\label{sec:dynamic_plant}
The plant state $\mathbf x(t)$ evolves under the control input $\mathbf u(t)$ and process disturbance $\mathbf G_{\mathrm c}\,\mathrm d\mathbf W(t)$ according to
\begin{equation}
    \mathrm d\mathbf{x}(t)
    =\mathbf{A}_{\mathrm c}\mathbf{x}(t)\,\mathrm dt
    +\mathbf{B}_{\mathrm c}\mathbf{u}(t)\,\mathrm dt
    +\mathbf{G}_{\mathrm c}\,\mathrm d\mathbf{W}(t),
    \label{eq:plant_ct}
\end{equation}
where $\mathbf{x}(t)\in\mathbb{R}^{N}$ and $\mathbf{u}(t)\in\mathbb{R}^{Q}$ are the plant state and control input, respectively, and $\mathbf{x}(0)=\mathbf{x}_0$. The matrices $\mathbf{A}_{\mathrm c}\in\mathbb{R}^{N\times N}$, $\mathbf{B}_{\mathrm c}\in\mathbb{R}^{N\times Q}$, and $\mathbf{G}_{\mathrm c}\in\mathbb{R}^{N\times N}$ are the continuous-time state, input, and disturbance matrices, respectively. Moreover, $\mathbf{W}(t)$ is an $N$-dimensional standard Wiener process independent of $\mathbf{x}_0$, which is assumed to have a finite second moment.

Sensors sample the plant once every $\tau_{\mathrm{ctrl}}$ units of time, and the input $\mathbf u(t)$ is held constant between samples. The sampled state then satisfies
\begin{equation}
    \mathbf{x}(k+1)
    =\mathbf{A}\mathbf{x}(k)+\mathbf{B}\mathbf{u}(k)+\mathbf{w}(k),
    \quad k\in\mathbb{Z}_{\geq0},
    \label{eq:plant_open_loop}
\end{equation}
where the discrete-time state and input matrices are
\begin{equation}
\mathbf{A}\triangleq\exp\!\left(\mathbf{A}_{\mathrm c}\tau_{\mathrm{ctrl}}\right),
\quad
\mathbf{B}\triangleq\int_{0}^{\tau_{\mathrm{ctrl}}}\exp\!\left(\mathbf{A}_{\mathrm c}s\right)\mathbf{B}_{\mathrm c}\,\mathrm{d}s.
\label{eq:zoh_discretization}
\end{equation}
The accumulated process disturbances $\mathbf w(k)$ are independent and identically distributed (i.i.d.) zero-mean Gaussian vectors. Their covariance, which quantifies the process-noise magnitude in the subsequent MSE analysis, is
\begin{equation}
\mathbf W_w\triangleq
\int_0^{\tau_{\mathrm{ctrl}}}
\exp(\mathbf A_{\mathrm c}s)\mathbf G_{\mathrm c}\mathbf G_{\mathrm c}^{\top}
\exp(\mathbf A_{\mathrm c}^{\top}s)\,\mathrm ds\succeq\mathbf0.
\label{eq:discrete_process_covariance}
\end{equation}
The plant is operated with a given stabilizing state feedback $\mathbf u(k)=-\mathbf K\mathbf x(k)$, where $\mathbf K\in\mathbb R^{Q\times N}$ is constant. Thus, the state to be estimated evolves according to
\begin{equation}
    \mathbf{x}(k+1)=\hat{\mathbf{A}}\mathbf{x}(k)+\mathbf{w}(k),
    \label{eq:plant_closed_loop}
\end{equation}
where $\hat{\mathbf{A}}\triangleq\mathbf{A}-\mathbf{B}\mathbf{K}\in\mathbb{R}^{N\times N}$ and $r(\hat{\mathbf{A}})<1$, with $r(\cdot)$ denoting the spectral radius.

\subsection{Wireless Communication Model}
\label{sec:WC_Network}
The plant is monitored by $M$ sensing nodes indexed by
$\mathcal V=\{1,\ldots,M\}$. The communication links form a connected
undirected graph $\mathcal G=(\mathcal V,\mathcal E)$. The edge set
$\mathcal E\subseteq\mathcal V\times\mathcal V$ specifies
the bidirectional links between nodes. The one-hop neighbor set of node $i$ is $\mathcal N_i\triangleq\{j\in\mathcal V:(i,j)\in\mathcal E\}$.
The communication topology is characterized by the graph Laplacian
$\mathbf L\in\mathbb R^{M\times M}$, where
$[\mathbf L]_{ii}=|\mathcal N_i|$, $[\mathbf L]_{ij}=-1$ for
$j\in\mathcal N_i$, and $[\mathbf L]_{ij}=0$ otherwise.

Information may need to pass through several neighbors before reaching a sensor. We allow $T$ communication updates between successive plant samples, with the timing specified as follows.

\begin{assumption}[Communication Cycles]
\label{as:time_scales}
Each control cycle of duration $\tau_{\mathrm{ctrl}}$ contains $T=\tau_{\mathrm{ctrl}}/\tau_{\mathrm{comm}}\in\mathbb Z^{+}$ communication cycles, each of duration $\tau_{\mathrm{comm}}$. These communication cycles
 are indexed by $\tau\in\{0,\ldots,T-1\}$. The sampled state $\mathbf{x}(k)$ is treated as fixed during the $T$ communication cycles of control cycle $k$.
\end{assumption}

Each node has $N_t=N$ transmit antennas and $N_r$ receive antennas. During communication cycle $\tau$ of control cycle $k$, node $i$ broadcasts its data vector $\mathbf z_i^\tau(k)\in\mathbb R^N$ to the neighbors in $\mathcal N_i$ by analog OTA aggregation~\citep{cai2018modulation}. The neighboring transmissions share the spectrum and aggregate at the receiver, so node $i$ observes
\begin{equation}
\mathbf r_i^\tau(k)
=\sqrt{\rho}\sum_{j\in\mathcal N_i}
\mathbf H_{ij}^\tau(k)\mathbf z_j^\tau(k)
+\boldsymbol\eta_i^\tau(k),
\label{eq:received_data}
\end{equation}
where $\mathbf r_i^\tau(k)\in\mathbb R^{N_r}$ is the received superimposed data, $\rho>0$ is the transmit SNR gain, $\mathbf H_{ij}^{\tau}(k)\in\mathbb R^{N_r\times N}$ is the wireless MIMO fading channel from  the $j$th node to the $i$th node, and $\boldsymbol\eta_i^\tau(k)\sim\mathcal N(\mathbf0,\mathbf I_{N_r})$ is the i.i.d. Gaussian channel noise.

To process the superimposed data $\mathbf r_i^\tau(k)$, node $i$ needs information about the wireless MIMO fading channels $\mathbf H_{ij}^\tau(k)$. Each communication slot therefore carries both data and a common known nonsingular pilot matrix $\mathbf X_p\in\mathbb R^{N\times N}$. The pilot and data use the same channel realization. Their aggregate pilot observation and the resulting CSI estimate are
\begin{equation}
\mathbf R_i^\tau(k)
=\sqrt{\rho}\sum_{j\in\mathcal N_i}
\mathbf H_{ij}^\tau(k)\mathbf X_p
+\mathbf N_{p,i}^\tau(k),
\label{eq:aggregated_pilot}
\end{equation}
\begin{equation}
\hat{\mathbf H}_i^\tau(k)
\triangleq
\frac{\mathbf R_i^\tau(k)\mathbf X_p^{-1}}{\sqrt\rho}
=\sum_{j\in\mathcal N_i}\mathbf H_{ij}^\tau(k)
+\frac{\breve{\mathbf N}_{p,i}^\tau(k)}{\sqrt\rho},
\label{eq:aggregated_csi}
\end{equation}
where
$\breve{\mathbf N}_{p,i}^\tau(k)
\triangleq\mathbf N_{p,i}^\tau(k)\mathbf X_p^{-1}$.
Here, $\mathbf N_{p,i}^\tau(k)\in\mathbb R^{N_r\times N}$ has independent $\mathcal N(0,1)$ entries. Since all neighbors transmit the same pilot, $\hat{\mathbf H}_i^\tau(k)$ estimates the channel sum $\sum_{j\in\mathcal N_i}\mathbf H_{ij}^\tau(k)$, rather than each individual link. The additive term $\breve{\mathbf N}_{p,i}^\tau(k)/\sqrt\rho$ is the aggregate CSI estimation error. Consequently, node $i$ processes its received data $\mathbf r_i^\tau(k)$ using noisy aggregate CSI $\hat{\mathbf H}_i^\tau(k)$.

We model the wireless MIMO channel matrix $\mathbf H_{ij}^\tau(k)$ in the data observation~\eqref{eq:received_data} and pilot observation~\eqref{eq:aggregated_pilot} as follows.
\begin{assumption}[MIMO Wireless Fading Channel Model]
\label{as:Wireless_Fading_Channels}
Each MIMO channel matrix $\mathbf H_{ij}^{\tau}(k)$ remains constant within its communication cycle and has i.i.d. $\mathcal N(0,1)$ entries. The matrices are independent for distinct index tuples $(i,j,\tau,k)$, including distinct directed communication pairs.
\end{assumption}

We also specify the independence of the plant and communication disturbances.
\begin{assumption}[Mutual Independence]
\label{as:statistical_independence}
The plant initial state $\mathbf x(0)$, the process-disturbance
sequence $\{\mathbf w(k)\}_{k\geq0}$, the MIMO channel
realizations, and the additive receiver noises associated with all
data and pilot transmissions are mutually independent. The receiver-noise realizations
are also independent across sensing nodes, control cycles, and communication
cycles, and between data and pilot transmissions.
\end{assumption}

\subsection{Local Information and State Estimation}
\label{sec:local_info_and_state_esti}
Node $i$ observes the plant through the local measurement
\begin{equation}
    \mathbf{y}_i(k)=\mathbf{C}_i\mathbf{x}(k),
    \label{eq:observation}
\end{equation}
where $\mathbf C_i\in\mathbb R^{c_i\times N}$ is the observation matrix. Local sensing need not determine the complete state $\mathbf x(k)$. The missing information must be obtained from data transmitted by neighboring nodes.

At the beginning of control cycle $k$, node $i$ has the current measurement $\mathbf y_i(k)$, its measurement history, and the received data and CSI estimates from completed control cycles. Represent this information by
\begin{equation}
\begin{aligned}
\mathscr{I}_i(k) = \sigma \big(
&\mathscr{I}_i^{\mathrm{init}},
\{\mathbf{y}_i(t): 0 \leq t \leq k\},\\
&\{\mathbf r_i^\tau(t), \hat{\mathbf{H}}_i^\tau(t):
0 \leq t < k,\ 0 \leq \tau < T\}
\big).
\end{aligned}
\end{equation}
The sigma-algebra $\mathscr I_i^{\mathrm{init}}$ contains the initial estimates $\hat{\mathbf x}_i(-1)$ and $\hat{\mathbf x}_i(0)$. Throughout the analysis, these estimates and the plant initial state $\mathbf x_0$ have finite second moments and are jointly independent of all subsequent channels, process disturbances, and receiver noises. Their joint distribution is held fixed when the SNR gain $\rho$ or receive-array size $N_r$ varies. Algorithm~\ref{alg:tracking_protocol} sets $\hat{\mathbf x}_i(-1)=\hat{\mathbf x}_i(0)=\mathbf0_N$ at every node.

All estimator updates are causal, so the transmitted data $\mathbf z_i^\tau(k)$ can depend only on observations available before communication cycle $\tau$. At time $k$, node $i$ computes its state estimate $\hat{\mathbf x}_i(k)\in\mathbb R^N$ from $\mathscr I_i(k)$, i.e., $\sigma(\hat{\mathbf x}_i(k))\subseteq\mathscr I_i(k)$.
\begin{remark}[Imperfect CSI]
Existing channel-based state estimation methods commonly assume that the channel information used by the estimator is known exactly. For example, \citet{DEY2009KFfaded} assume exact fading gains at the fusion center, and \citet{LI2018TransmitPower} use instantaneous channel power gains without estimation error. In practice, CSI is usually acquired from noisy pilots~\citep{HASSIBI2003Training,TONG2004Pilot}. We therefore use the imperfect channel-sum estimate $\hat{\mathbf H}_i^\tau(k)$ in~\eqref{eq:aggregated_csi} and explicitly include its estimation error $\breve{\mathbf N}_{p,i}^\tau(k)/\sqrt\rho$ in the state estimation analysis.
\end{remark}

Define the local error $\mathbf e_i(k)$ and stacked network error $\mathbf e(k)$ by
\begin{equation}
\mathbf{e}_i(k) \triangleq \hat{\mathbf{x}}_i(k)-\mathbf{x}(k),\quad
\mathbf{e}(k) \triangleq
\operatorname{col}_{i\in\mathcal{V}}\mathbf{e}_i(k).
\end{equation}
The network state estimation MSE is
\begin{equation}
\mathcal{M}(k) \triangleq
\mathbb{E}\left[\|\mathbf{e}(k)\|^2\right]
=
\sum_{i\in\mathcal{V}}
\mathbb{E}\left[\|\mathbf{e}_i(k)\|^2\right].
\end{equation}
The process disturbance $\mathbf w(k)$ can sustain a nonzero state estimation MSE. We therefore measure how rapidly $\mathcal M(k)$ approaches its stationary value $\mathcal M_\star$, called the error platform. The target factor $d$ is imposed on the root of the absolute difference $|\mathcal M(k)-\mathcal M_\star|$ as follows.

\begin{definition}[Prescribed Root-MSE Convergence]
\label{def:converge_speed}
For a prescribed factor $d\in(0,1)$, the network estimator attains the
prescribed platform-convergence rate if a finite limit $\mathcal{M}_{\star} \triangleq
\lim_{k\to\infty}\mathcal{M}(k)$ exists and
\begin{equation}
\limsup_{k\to\infty}
\left|\mathcal{M}(k)-\mathcal{M}_{\star}\right|^{1/(2k)}
< d.
\label{eq:prescribed_d}
\end{equation}
\end{definition}
In particular, \eqref{eq:prescribed_d} implies
$\left|\mathcal{M}(k)-\mathcal{M}_{\star}\right|^{1/2}
= o(d^k)$
and
$\mathcal{M}(k)-\mathcal{M}_{\star}
= o(d^{2k})$.
The factor $d$ therefore specifies the speed of convergence to the platform
$\mathcal M_\star$, whose value determines the long-term estimation accuracy.
Persistent process disturbances can keep $\mathcal M_\star$ positive even
when the prescribed rate is attained. A smaller $d$ requires the transient
deviation $|\mathcal M(k)-\mathcal M_\star|$ to vanish more quickly.

The rate in Definition~\ref{def:converge_speed} depends on the state directions that each sensor can correct. To identify their natural decay rates, write the closed-loop matrix $\hat{\mathbf A}$ in Jordan form,
\begin{equation}
\hat{\mathbf{A}}
=\mathbf{V}\mathbf{J}\mathbf{V}^{-1},
\qquad
\mathbf{J}
=\operatorname{blkdiag}
\left(\mathbf{J}_1,\ldots,\mathbf{J}_{n_{\mathrm J}}\right),
\label{eq:jordan_decomposition}
\end{equation}
where $\mathbf{V}\in\mathbb{C}^{N\times N}$ is nonsingular and $n_{\mathrm J}$ denotes the total number of Jordan blocks. For block $l\in\{1,\ldots,n_{\mathrm J}\}$, we have
$\mathbf{J}_l=\lambda_l\mathbf{I}_{m_l}+\mathbf{N}_l$, with eigenvalue $\lambda_l$, size $m_l$, and nilpotent
part $\mathbf{N}_l^{m_l}=\mathbf{0}$, where
$\sum_{l=1}^{n_{\mathrm J}}m_l=N$. Partition $\mathbf V=
[\,\mathbf V_1\ \cdots\ \mathbf V_{n_{\mathrm J}}\,],$ and
$\mathbf V^{-1}=\operatorname{col}_l\mathbf W_{l}^{\top}$. Under the change of coordinates $\mathbf{x}^{\mathrm J}(k)\triangleq\mathbf{V}^{-1}\mathbf{x}(k)$, the sub-state associated with the $l$th Jordan block is $\mathbf{x}_l^{\mathrm J}(k)\triangleq\mathbf{W}_l^{\top}\mathbf{x}(k)\in\mathbb C^{m_l}$.
The basis matrices satisfy $\hat{\mathbf A}\mathbf V_l=\mathbf V_l\mathbf J_l$ and $\mathbf W_l^{\top}\hat{\mathbf A}=\mathbf J_l\mathbf W_l^{\top}$.
We refer to the dynamics associated with an eigenvalue $\lambda_l$ as a mode.

\begin{proposition}[Natural Modal Root-MSE Rate]
\label{prop:natural_rootMSE_rate}
For the sub-state $\mathbf{x}_l^{\mathrm J}(k)$ associated with $\lambda_l$, the limit
$\lim_{t\to\infty}
\mathbb{E}\left[\|\mathbf{x}_l^{\mathrm J}(t)\|^2\right]$
exists and is finite. Moreover,
\begin{equation}
\limsup_{k\to\infty}
\left|
\mathbb{E}\left[\|\mathbf{x}_l^{\mathrm J}(k)\|^2\right]-\lim_{t\to\infty}
\mathbb{E}\left[\|\mathbf{x}_l^{\mathrm J}(t)\|^2\right]
\right|^{1/(2k)}
\leq |\lambda_l|.
\end{equation}
Thus, for $|\lambda_l|<d$, the zero estimate $\hat{\mathbf x}_l^{\mathrm J}(k)=\mathbf0$ attains the rate in Definition~\ref{def:converge_speed} for this sub-state.
\end{proposition}

Proposition~\ref{prop:natural_rootMSE_rate} shows that modes with $|\lambda_l|<d$ meet the target through natural decay. Modes with $|\lambda_l|\geq d$ require correction to guarantee the prescribed rate. We call them performance-limiting modes and collect their block indices in
\begin{equation}
    \mathcal{I}_{\mathrm{pl}}
    \triangleq
    \left\{
    l\in\{1,\ldots,n_{\mathrm J}\}
    \,\middle|\,
    |\lambda_l|\geq d
    \right\}.
    \label{eq:critical_modes}
\end{equation}
\subsection{Distributed Sensing Model}\label{sec:sensing_network}
A performance-limiting sub-state $\mathbf x_l^{\mathrm J}(k)$ must be corrected using measurements available somewhere in the network. We first require sufficient information in the stacked measurement matrix $\mathbf C\triangleq\operatorname{col}_{i\in\mathcal V}\mathbf C_i\in\mathbb R^{(\sum_{i\in\mathcal V}c_i)\times N}$.
\begin{assumption}[Global Observability]
\label{as:global_modal_source_observability}
The plant is globally observable by the entire sensing network. Specifically, the pair $(\hat{\mathbf{A}},\mathbf{C})$ is observable.
\end{assumption}
Global observability of $(\hat{\mathbf A},\mathbf C)$ does not mean that node $i$ can reconstruct every sub-state from $\mathbf y_i(k)$. Substituting $\mathbf x(k)=\mathbf V\mathbf x^{\mathrm J}(k)$ into~\eqref{eq:observation} gives $\mathbf y_i(k)=\sum_l\mathbf C_i\mathbf V_l\mathbf x_l^{\mathrm J}(k)$. Thus, $\mathbf C_i\mathbf V_l$ specifies how block $l$ appears in the local measurement. The measurements available at one node may reveal only a portion of a Jordan chain. Let $\mathbf v_{l,r}$ be the $r$th column of $\mathbf V_l$. For $l\in\mathcal I_{\mathrm{pl}}$, define the first visible coordinate by
\begin{equation}
r_{i,l}\triangleq\min\{r:\mathbf C_i\mathbf v_{l,r}\neq\mathbf0\},
\label{eq:first_visible_jordan_coordinate}
\end{equation}
where $r_{i,l}=m_l+1$ if $\mathbf C_i\mathbf V_l=\mathbf0$. In the standard upper-triangular Jordan convention, coordinates $1,\ldots,r_{i,l}-1$ do not affect the local output, whereas the tail $r_{i,l},\ldots,m_l$ is observable when that chain is considered alone. The tail dynamics do not depend on the preceding coordinates.

To use common communication coordinates, partition each performance-limiting chain at all node-wise visibility boundaries. Write the distinct elements of
$\{1,m_l+1\}\cup\{r_{i,l}:i\in\mathcal V\}$ as
$1=a_{l,1}<\cdots<a_{l,n_l^{\mathrm{mb}}+1}=m_l+1$.
Segment $(l,s)$ contains coordinates $a_{l,s},\ldots,a_{l,s+1}-1$ and has size $m_{l,s}=a_{l,s+1}-a_{l,s}$. For $l\notin\mathcal I_{\mathrm{pl}}$, take a single segment, with $n_l^{\mathrm{mb}}=1$, $a_{l,1}=1$, and $a_{l,2}=m_l+1$. Let $\mathbf V_{l,s}$ contain the corresponding columns of $\mathbf V_l$, and let $\mathbf W_{l,s}^{\top}$ contain the matching rows of $\mathbf W_l^\top$. The diagonal Jordan subblock is denoted by $\mathbf J_{l,s}$.
The segment state $\mathbf x_{l,s}^{\mathrm J}(k)\triangleq\mathbf W_{l,s}^\top\mathbf x(k)$ satisfies
\begin{equation}
\begin{aligned}
\mathbf x_{l,s}^{\mathrm J}(k+1)
={}&\mathbf J_{l,s}\mathbf x_{l,s}^{\mathrm J}(k)\\
&+\sum_{t>s}\mathbf J_{l,s,t}\mathbf x_{l,t}^{\mathrm J}(k)
+\mathbf W_{l,s}^\top\mathbf w(k),
\end{aligned}
\label{eq:jordan_segment_dynamics}
\end{equation}
where $\mathbf J_{l,s,t}$ is the corresponding off-diagonal subblock of $\mathbf J_l$. These chain connections act only from later segments to earlier ones. We refer to the common segments as Jordan mini-blocks.

Let $\mathcal B\triangleq\{(l,s):1\le l\le n_{\mathrm J},\,1\le s\le n_l^{\mathrm{mb}}\}$ be their index set. Assign the network-assisted and locally retained segments at node $i$ by
\begin{equation}
\begin{aligned}
\mathcal U_i
&\triangleq\{(l,s)\in\mathcal B:
l\in\mathcal I_{\mathrm{pl}},\ a_{l,s}<r_{i,l}\},\\
\mathcal O_i&\triangleq\mathcal B\setminus\mathcal U_i.
\end{aligned}
\label{eq:local_unobservable_modes}
\end{equation}
Thus $\mathcal U_i$ contains the locally unobservable prefixes that require neighboring information. The set $\mathcal O_i$ contains the locally retained tails and all naturally fast blocks with $|\lambda_l|<d$. The latter need not be locally observable.
To construct updates for these two groups, collect their coordinates as
\begin{equation}
\begin{aligned}
\mathbf x_{i,\mathcal U}^{\mathrm J}(k)
&\triangleq\operatorname{col}_{(l,s)\in\mathcal U_i}
\mathbf x_{l,s}^{\mathrm J}(k),\\
\mathbf x_{i,\mathcal O}^{\mathrm J}(k)
&\triangleq\operatorname{col}_{(l,s)\in\mathcal O_i}
\mathbf x_{l,s}^{\mathrm J}(k).
\end{aligned}
\label{eq:nodewise_jordan_coordinates}
\end{equation}
The corresponding rows of the dual basis are
$(\mathbf W_i^{\mathcal U})^\top\triangleq\operatorname{col}\{\mathbf W_{l,s}^\top:(l,s)\in\mathcal U_i\}$ and
$(\mathbf W_i^{\mathcal O})^\top\triangleq\operatorname{col}\{\mathbf W_{l,s}^\top:(l,s)\in\mathcal O_i\}$.
Consequently, $\mathbf x_{i,\mathcal U}^{\mathrm J}(k)=(\mathbf W_i^{\mathcal U})^\top\mathbf x(k)$ and $\mathbf x_{i,\mathcal O}^{\mathrm J}(k)=(\mathbf W_i^{\mathcal O})^\top\mathbf x(k)$.

Collect the associated basis columns as $\mathbf V_i^{\mathcal U}\triangleq[\mathbf V_{l,s}:(l,s)\in\mathcal U_i]$ and $\mathbf V_i^{\mathcal O}\triangleq[\mathbf V_{l,s}:(l,s)\in\mathcal O_i]$. Use the same chain order in each collection, with the corresponding zero-dimensional matrices for empty collections.
The dual-basis identities give
\begin{equation}
\mathbf x(k)=\mathbf V_i^{\mathcal U}\mathbf x_{i,\mathcal U}^{\mathrm J}(k)
+\mathbf V_i^{\mathcal O}\mathbf x_{i,\mathcal O}^{\mathrm J}(k).
\label{eq:nodewise_state_decomposition}
\end{equation}
By the visibility boundary and the direction of the Jordan-chain connections,
\begin{equation}
\mathbf C_i\mathbf V_i^{\mathcal U}=\mathbf0,
\qquad
(\mathbf W_i^{\mathcal O})^\top\hat{\mathbf A}
\mathbf V_i^{\mathcal U}=\mathbf0.
\label{eq:nodewise_separation_identities}
\end{equation}
Consequently, the original local measurement is
\begin{equation}
\mathbf y_i(k)=
\mathbf C_i\mathbf V_i^{\mathcal O}
\mathbf x_{i,\mathcal O}^{\mathrm J}(k).
\label{eq:nodewise_jordan_measurement_decomposition}
\end{equation}
The state matrix governing $\mathbf x_{i,\mathcal O}^{\mathrm J}(k)$ is $\mathbf{J}_i^{\mathcal{O}}\triangleq(\mathbf W_i^{\mathcal O})^\top\hat{\mathbf A}\mathbf V_i^{\mathcal O}$,
whose dimension is $n_i^{\mathcal{O}}\triangleq\sum_{(l,s)\in\mathcal{O}_i}m_{l,s}$.
It includes the connections between retained segments of the same chain. Equation~\eqref{eq:nodewise_separation_identities} makes this branch independent of the network-assisted coordinates, although it can drive those coordinates.

For repeated eigenvalues, the retained tails must also be distinguishable jointly from the local measurements.
\begin{assumption}[Local Distinguishability of Retained Jordan Chains]
\label{as:network_decoupled_local_detectability}
For every node $i\in\mathcal V$ and every eigenvalue $\lambda$ satisfying $|\lambda|\ge d$, the vectors
\[
\{\mathbf C_i\mathbf v_{l,r_{i,l}}:
\lambda_l=\lambda,\ r_{i,l}\le m_l\}
\]
are linearly independent.
\end{assumption}
The listed vectors are the output images of the eigenvectors of the retained tails at $\lambda$. By the Popov--Belevitch--Hautus (PBH) criterion~\citep{kailath1980linear}, the assumption is equivalent to $d$-detectability of $(\mathbf J_i^{\mathcal O},\mathbf C_i\mathbf V_i^{\mathcal O})$, namely
\begin{equation}
\operatorname{rank}
\begin{bmatrix}
\lambda\mathbf I_{n_i^{\mathcal O}}-\mathbf J_i^{\mathcal O}\\
\mathbf C_i\mathbf V_i^{\mathcal O}
\end{bmatrix}
=n_i^{\mathcal O}
\label{eq:effective_anchored_detectability}
\end{equation}
for $\lambda\in\sigma(\mathbf J_i^{\mathcal O})$ with $|\lambda|\ge d$. Modes with $|\lambda|<d$ may remain undetectable. For a real plant, choose real bases for real chains and conjugate-paired bases for conjugate chains. The latter then have identical visibility boundaries and matching partitions.

Under this assumption, a node with $(l,s)\in\mathcal O_i$ and $l\in\mathcal I_{\mathrm{pl}}$ can estimate the performance-limiting sub-state $\mathbf x_{l,s}^{\mathrm J}(k)$ locally at the prescribed rate and supply that estimate to other nodes. We call it a modal source for mini-block $(l,s)$. A node with $(l,s)\in\mathcal U_i$ is a non-source. The corresponding node sets are
\begin{equation}
\begin{aligned}
\mathcal V_{l,s}^{\mathcal U}
&\triangleq\{i\in\mathcal V:(l,s)\in\mathcal U_i\},\\
\mathcal V_{l,s}^{\mathcal O}
&\triangleq\mathcal V\setminus\mathcal V_{l,s}^{\mathcal U}.
\end{aligned}
\label{eq:pcsi_modal_node_sets}
\end{equation}
For every $l\in\mathcal I_{\mathrm{pl}}$ and $1\le s\le n_l^{\mathrm{mb}}$, the estimates supplied by $\mathcal V_{l,s}^{\mathcal O}$ provide the measurement information needed by the nodes in $\mathcal V_{l,s}^{\mathcal U}$. Global observability ensures that a source exists.
\begin{proposition}[Existence of Modal Sources]
\label{prop:existence_of_modal_sources}
Under Assumption~\ref{as:global_modal_source_observability}, every performance-limiting Jordan mini-block has at least one modal source. That is,
\begin{equation}
\mathcal V_{l,s}^{\mathcal O}\neq\varnothing,
\qquad l\in\mathcal I_{\mathrm{pl}},\ 1\le s\le n_l^{\mathrm{mb}}.
\label{eq:modal_source_condition}
\end{equation}
\end{proposition}
\begin{proof}
Please see~\ref{app:proof_existence_of_modal_sources}.
\end{proof}
A source in $\mathcal V_{l,s}^{\mathcal O}$ need not be a direct neighbor of every node in $\mathcal V_{l,s}^{\mathcal U}$. Connectivity of $\mathcal G$ provides a path, and successive communication cycles relay its estimate of $\mathbf x_{l,s}^{\mathrm J}(k)$ along that path. Section~\ref{sec:distributed_estimator_design} specifies how each sensor combines these received estimates with its local correction.

\section{Distributed State Estimation with Imperfect CSI}
\label{sec:distributed_estimator_design}
The Jordan-chain partition in Section~\ref{sec:sensing_network}
assigns the coordinates in $\mathcal O_i$ to local correction
and those in $\mathcal U_i$ to communication.
At control cycle $k$, node $i$ first uses $\mathbf y_i(k)$ to
predict $\mathbf x(k+1)$. The nodes exchange these predictions
through $T$ communication cycles and obtain the final data
$\mathbf z_i^T(k)$. At the beginning of control cycle $k+1$,
node $i$ combines $\mathbf z_i^T(k)$ with its stored estimates
to form $\hat{\mathbf x}_i(k+1)$. We specify these three
updates in their execution order.

\subsection{Local Prediction}
\label{subsec:local_prediction}
Using the current measurement $\mathbf y_i(k)$ and estimate
$\hat{\mathbf x}_i(k)$, node $i$ computes the prior estimate
\begin{equation}
\hat{\mathbf x}_i(k+1\mid k)
=\hat{\mathbf A}\hat{\mathbf x}_i(k)
+\mathbf K_i\big(\mathbf y_i(k)-\mathbf C_i\hat{\mathbf x}_i(k)\big).
\label{eq:local_prior_estimate}
\end{equation}
For $n_i^{\mathcal O}>0$, choose
$\mathbf K_i=\mathbf V_i^{\mathcal O}\mathbf G_i\in\mathbb R^{N\times c_i}$, where
$\mathbf G_i\in\mathbb C^{n_i^{\mathcal O}\times c_i}$ satisfies
$r(\mathbf J_i^{\mathcal O}-\mathbf G_i\mathbf C_i\mathbf V_i^{\mathcal O})<d$.
Assumption~\ref{as:network_decoupled_local_detectability} ensures the existence of this gain.
The gains are paired on conjugate Jordan blocks so that $\mathbf K_i$ is real.
For $n_i^{\mathcal O}=0$, set $\mathbf K_i=\mathbf0_{N\times c_i}$.
The gain corrects only the locally retained coordinates. Their dynamics and measurements are independent of the network-assisted prefixes by
\eqref{eq:nodewise_separation_identities}.
The coordinates in $\mathcal U_i$ are updated using the
neighboring predictions in the next step.

\subsection{Information Fusion and Communication}
\label{subsec:fast_time_fusion}
Node $i$ initializes its data with the prior estimate
$\hat{\mathbf x}_i(k+1\mid k)$ and stores a second copy,
\begin{equation}
\mathbf z_i^{-1}(k)=\mathbf z_i^0(k)
\triangleq\hat{\mathbf x}_i(k+1\mid k).
\label{eq:initial_information_state}
\end{equation}
The vector $\mathbf z_i^{-1}(k)$ supplies the memory needed
by the communication update and does not require an extra
transmission. During communication cycle $\tau$, node $i$
transmits $\mathbf z_i^\tau(k)$ and the pilot matrix
$\mathbf X_p$. It uses the received data $\mathbf r_i^\tau(k)$
in~\eqref{eq:received_data} and the aggregate CSI estimate
$\hat{\mathbf H}_i^\tau(k)$ in~\eqref{eq:aggregated_csi}
to form the fusion signal
\begin{equation}
\begin{aligned}
\mathbf s_i^\tau(k)\triangleq\frac1{N_r}
\Bigg\{&
(\hat{\mathbf H}_i^\tau(k))^\top
\frac{\mathbf r_i^\tau(k)}{\sqrt\rho}
-\Big[(\hat{\mathbf H}_i^\tau(k))^\top
\hat{\mathbf H}_i^\tau(k)\\
&\phantom{-\Big[}
-\frac{N_r}{\rho}\mathbf X_p^{-\top}\mathbf X_p^{-1}\Big]
\mathbf z_i^\tau(k)\Bigg\}.
\end{aligned}
\label{eq:regularized_ota_term}
\end{equation}
The subtraction of
$N_r\mathbf X_p^{-\top}\mathbf X_p^{-1}/\rho$ from
$(\hat{\mathbf H}_i^\tau(k))^\top\hat{\mathbf H}_i^\tau(k)$
in~\eqref{eq:regularized_ota_term} is the
pilot-noise bias correction. Indeed, the pilot error
$\breve{\mathbf N}_{p,i}^\tau(k)/\sqrt\rho$
in~\eqref{eq:aggregated_csi} adds precisely this deterministic
matrix to the expected channel product.

Under Assumptions~\ref{as:Wireless_Fading_Channels}
and~\ref{as:statistical_independence}, the current channels
$\{\mathbf H_{ij}^\tau(k)\}_{j\in\mathcal N_i}$ and noises
$\boldsymbol\eta_i^\tau(k),\mathbf N_{p,i}^\tau(k)$
are independent of the data formed before transmission.
Conditioning on
$\{\mathbf z_j^\tau(k)\}_{j\in\mathcal N_i\cup\{i\}}$,
the received-data term
$(\hat{\mathbf H}_i^\tau(k))^\top\mathbf r_i^\tau(k)/(N_r\sqrt\rho)$
has mean $\sum_{j\in\mathcal N_i}\mathbf z_j^\tau(k)$.
After the bias correction, the subtracted self-data term has
mean $|\mathcal N_i|\mathbf z_i^\tau(k)$.
Their difference therefore satisfies
\begin{equation}
\begin{aligned}
\bar{\mathbf s}_i^\tau(k)
\triangleq{}&\mathbb E\!\left[\mathbf s_i^\tau(k)
\,\middle|\,
\{\mathbf z_j^\tau(k)\}_{j\in\mathcal N_i\cup\{i\}}\right]\\
={}&\sum_{j\in\mathcal N_i}
\big(\mathbf z_j^\tau(k)-\mathbf z_i^\tau(k)\big).
\end{aligned}
\label{eq:unbiased_ota_fusion}
\end{equation}
Thus $\bar{\mathbf s}_i^\tau(k)$ equals the sum of the
data differences required by graph-based information fusion.
The actual signal $\mathbf s_i^\tau(k)$ fluctuates around
this value because the channels and CSI are random.

Only the coordinates indexed by $\mathcal U_i$ need these
neighboring data. Using the segment bases from
Section~\ref{sec:sensing_network}, for $l\in\mathcal I_{\mathrm{pl}}$ and $1\le s\le n_l^{\mathrm{mb}}$, define the corresponding
coordinate projectors
\begin{equation}
\mathbf P_{l,s}^{\mathcal U_i}\triangleq
\begin{cases}
\mathbf V_{l,s}\mathbf W_{l,s}^\top,&(l,s)\in\mathcal U_i,\\
\mathbf0_{N\times N},&(l,s)\in\mathcal O_i,
\end{cases}
\label{eq:unobservable_projection}
\end{equation}
\begin{equation}
\mathbf P^{\mathcal O_i}\triangleq
\mathbf I_N-\sum_{(l,s)\in\mathcal U_i}\mathbf P_{l,s}^{\mathcal U_i}.
\label{eq:locally_anchored_projection}
\end{equation}
The term $\mathbf P^{\mathcal O_i}\mathbf z_i^0(k)$ retains
the locally corrected prediction. For each $(l,s)\in\mathcal U_i$,
$\mathbf P_{l,s}^{\mathcal U_i}\mathbf s_i^\tau(k)$ selects the
neighboring-data correction for segment $(l,s)$.
Combining this correction with the current and preceding
data gives the update
\begin{equation}
\begin{aligned}
\mathbf z_i^{\tau+1}(k)
={}&\mathbf P^{\mathcal O_i}\mathbf z_i^0(k)\\
&+\sum_{(l,s)\in\mathcal U_i}\mathbf P_{l,s}^{\mathcal U_i}
\Big[\omega_{l,s}^\tau
\big(\mathbf z_i^\tau(k)+\gamma_{l,s}\mathbf s_i^\tau(k)\big)\\
&\phantom{+\sum_{(l,s)\in\mathcal U_i}\mathbf P_{l,s}^{\mathcal U_i}\Big[}
+(1-\omega_{l,s}^\tau)\mathbf z_i^{\tau-1}(k)\Big].
\end{aligned}
\label{eq:update_z}
\end{equation}
Here $\gamma_{l,s}$ and $\omega_{l,s}^\tau$ are prescribed scalar
weights, independent of $k$ and all wireless MIMO channel and noise
realizations. The weight $\omega_{l,s}^\tau$ may vary over the communication
cycles $\tau$.
If $\mathcal V_{l,s}^{\mathcal U}=\varnothing$, take
$\gamma_{l,s}=0$ and $\omega_{l,s}^\tau=1$; all associated
projectors are zero.
In~\eqref{eq:update_z}, the term
$\omega_{l,s}^\tau\gamma_{l,s}\mathbf P_{l,s}^{\mathcal U_i}\mathbf s_i^\tau(k)$
incorporates the newly received data, while
$(1-\omega_{l,s}^\tau)\mathbf P_{l,s}^{\mathcal U_i}\mathbf z_i^{\tau-1}(k)$
uses the stored data. This two-vector update allows the weights
to control how quickly source estimates reach non-source
nodes. Section~\ref{subsec:mean_chebyshev_design} selects
$\gamma_{l,s},\omega_{l,s}^\tau$ and $T$ for the required rate.
For $\tau+1<T$, node $i$ transmits
$\mathbf z_i^{\tau+1}(k)$ in the next communication cycle.
The final vector $\mathbf z_i^T(k)$ enters the posterior update.

\subsection{Posterior Estimation}
\label{subsec:local_posterior_estimation}
The posterior estimate $\hat{\mathbf x}_i(k+1)$ uses
$\mathbf z_i^T(k)$ to correct the prior estimate
$\hat{\mathbf x}_i(k+1\mid k)$.
It also uses the stored estimates $\hat{\mathbf x}_i(k)$ and
$\hat{\mathbf x}_i(k-1)$ to shape the error evolution across
control cycles.
Choose $d_-\in(0,d)$, set
$\beta_l=d_-^2/\lambda_l$ for $l\in\mathcal I_{\mathrm{pl}}$,
and form the weighting matrices
\begin{equation}
\begin{aligned}
\mathbf M_{0,i}
&\triangleq-\sum_{(l,s)\in\mathcal U_i}
\beta_l\mathbf V_{l,s}\mathbf J_{l,s}^{-1}\mathbf W_{l,s}^\top,\\
\mathbf M_{\beta,i}
&\triangleq\sum_{(l,s)\in\mathcal U_i}
\beta_l\mathbf V_{l,s}\mathbf W_{l,s}^\top.
\end{aligned}
\label{eq:affine_synthesis_matrices}
\end{equation}
The inverse $\mathbf J_{l,s}^{-1}$ exists because
$|\lambda_l|\geq d>0$ for $(l,s)\in\mathcal U_i$.
At the beginning of control cycle $k+1$, node $i$ computes
\begin{equation}
\begin{aligned}
\hat{\mathbf x}_i(k+1)
={}&\hat{\mathbf x}_i(k+1\mid k)\\
&+(\mathbf I_N-\mathbf M_{0,i})
\big[\mathbf z_i^T(k)-\hat{\mathbf x}_i(k+1\mid k)\big]\\
&+\mathbf M_{\beta,i}
\big[\hat{\mathbf x}_i(k)-\hat{\mathbf A}\hat{\mathbf x}_i(k-1)\big].
\end{aligned}
\label{eq:local_posterior_estimate}
\end{equation}
The first correction in~\eqref{eq:local_posterior_estimate}
is the change $\mathbf z_i^T(k)-\hat{\mathbf x}_i(k+1\mid k)$
produced by communication. It acts only on $\mathcal U_i$, i.e.,
\eqref{eq:locally_anchored_projection} and~\eqref{eq:update_z}
give
$\mathbf P^{\mathcal O_i}
[\mathbf z_i^T(k)-\hat{\mathbf x}_i(k+1\mid k)]=\mathbf0_N$.
Moreover,
$\mathbf P^{\mathcal O_i}\mathbf M_{0,i}
=\mathbf P^{\mathcal O_i}\mathbf M_{\beta,i}=\mathbf0_{N\times N}$.
Hence the posterior keeps the local prediction unchanged on
$\mathcal O_i$, i.e.,
$\mathbf P^{\mathcal O_i}\hat{\mathbf x}_i(k+1)
=\mathbf P^{\mathcal O_i}\hat{\mathbf x}_i(k+1\mid k)$.

On $\mathcal U_i$, the history correction
$\mathbf M_{\beta,i}
[\hat{\mathbf x}_i(k)-\hat{\mathbf A}\hat{\mathbf x}_i(k-1)]$
introduces dependence on the preceding posterior estimate.
The resulting posterior error depends on both
$\mathbf e_i(k)$ and $\mathbf e_i(k-1)$.
Section~\ref{sec:mean_convergence_analysis} uses this dependence
to choose the communication weights for decay faster than
$d^k$. Algorithm~\ref{alg:tracking_protocol} summarizes
prediction, communication, and posterior estimation.
\begin{algorithm}[t]
\caption{Distributed State Estimation for Prescribed Mean-Error and MSE Convergence}
\label{alg:tracking_protocol}
\begin{algorithmic}[1]
\REQUIRE Node $i\in\mathcal V$ with matrices
$\hat{\mathbf A}$, $\mathbf C_i$, $\mathbf K_i$, $\mathbf X_p$,
parameters $T$, $\rho$, $N_r$, weights
$\{\gamma_{l,s}\}_{(l,s)\in\mathcal U_i}$,
$\{\omega_{l,s}^\tau\}_{(l,s)\in\mathcal U_i,\,0\le\tau<T}$,
projectors $\{\mathbf P_{l,s}^{\mathcal U_i}\}_{(l,s)\in\mathcal U_i}$,
$\mathbf P^{\mathcal O_i}$, and matrices
$\mathbf M_{0,i}$, $\mathbf M_{\beta,i}$.
\STATE Set $\hat{\mathbf x}_i(-1)=\hat{\mathbf x}_i(0)=\mathbf0_N$.
\FOR{each control cycle $k=0,1,2,\ldots$}
    \STATE Obtain $\mathbf y_i(k)$ and compute
    $\hat{\mathbf x}_i(k+1\mid k)$ using
    \eqref{eq:local_prior_estimate}.
    \STATE Initialize $\mathbf z_i^{-1}(k)$ and $\mathbf z_i^0(k)$
    using \eqref{eq:initial_information_state}.
    \FOR{each communication cycle $\tau\in\{0,\ldots,T-1\}$}
        \STATE Transmit $\mathbf z_i^\tau(k)$ and $\mathbf X_p$, and
        receive $\mathbf r_i^\tau(k)$ and $\mathbf R_i^\tau(k)$.
        \STATE Compute $\hat{\mathbf H}_i^\tau(k)$ using
        \eqref{eq:aggregated_csi}.
        \STATE Form $\mathbf s_i^\tau(k)$ using
        \eqref{eq:regularized_ota_term}.
        \STATE Compute $\mathbf z_i^{\tau+1}(k)$ using
        \eqref{eq:update_z}.
    \ENDFOR
    \STATE At the beginning of control cycle $k+1$, compute
    $\hat{\mathbf x}_i(k+1)$ using \eqref{eq:local_posterior_estimate}.
\ENDFOR
\end{algorithmic}
\end{algorithm}
\section{Mean-Error Convergence Analysis}
\label{sec:mean_convergence_analysis}
The estimator uses local measurements to correct the
$\mathcal O_i$ coordinates and neighboring data to correct
the $\mathcal U_i$ coordinates. We first express the mean
posterior error $\bar{\mathbf e}(k)=\mathbb E[\mathbf e(k)]$
in terms of these two operations. The local gains determine
the decay of the $\mathcal O_i$ errors. The communication
weights $\gamma_{l,s},\omega_{l,s}^\tau$ and the cycle count $T$
then determine the decay of the $\mathcal U_i$ errors.

\subsection{Global Mean-Error Dynamics}
\label{subsec:mean_error_dynamics}
Define the prior error
$\mathbf e_i(k+1\mid k)=\hat{\mathbf x}_i(k+1\mid k)-\mathbf x(k+1)$
and stack it as
$\mathbf e(k+1\mid k)=\operatorname{col}_{i=1}^{M}\mathbf e_i(k+1\mid k)$.
The posterior error $\mathbf e(k+1)$ was defined in
Section~\ref{sec:local_info_and_state_esti}.
Set $\tilde{\mathbf A}=\mathbf I_M\otimes\hat{\mathbf A}$,
$\tilde{\mathbf K}=\operatorname{blkdiag}_{i=1}^{M}(\mathbf K_i)$,
and $\tilde{\mathbf C}=\operatorname{blkdiag}_{i=1}^{M}(\mathbf C_i)$.
Subtracting the plant dynamics \eqref{eq:plant_closed_loop}
from the predictions \eqref{eq:local_prior_estimate} gives
\begin{equation}
\mathbf e(k+1\mid k)
=(\tilde{\mathbf A}-\tilde{\mathbf K}\tilde{\mathbf C})\mathbf e(k)
-\mathbf1_M\otimes\mathbf w(k).
\label{eq:pcsi_stacked_prior_error}
\end{equation}
The vector $\mathbf1_M\otimes\mathbf w(k)$ contains the same
plant disturbance $\mathbf w(k)$ at every node. Its entries across
nodes are therefore correlated.

The communicated data $\mathbf z_i^\tau(k)$ also estimate
$\mathbf x(k+1)$. Define their errors by
$\boldsymbol\epsilon_i^\tau(k)=\mathbf z_i^\tau(k)-\mathbf x(k+1)$
and $\boldsymbol\epsilon^\tau(k)=\operatorname{col}_{i=1}^{M}
\boldsymbol\epsilon_i^\tau(k)$, for $-1\leq\tau\leq T$.
Initialization~\eqref{eq:initial_information_state} gives
$\boldsymbol\epsilon^{-1}(k)=\boldsymbol\epsilon^0(k)=\mathbf e(k+1\mid k)$.
Thus the communication correction in the posterior update is
$\boldsymbol\epsilon^T(k)-\mathbf e(k+1\mid k)$.
With $\tilde{\mathbf M}_0=\operatorname{blkdiag}_{i=1}^{M}(\mathbf M_{0,i})$
and $\tilde{\mathbf M}_\beta=\operatorname{blkdiag}_{i=1}^{M}(\mathbf M_{\beta,i})$,
subtracting the plant state from~\eqref{eq:local_posterior_estimate}
gives, for $k\geq1$,
\begin{equation}
\begin{aligned}
\mathbf e(k+1)
={}&\tilde{\mathbf M}_0\mathbf e(k+1\mid k)
+(\mathbf I_{MN}-\tilde{\mathbf M}_0)\boldsymbol\epsilon^T(k)\\
&+\tilde{\mathbf M}_\beta
\big[\mathbf e(k)-\tilde{\mathbf A}\mathbf e(k-1)
+\mathbf1_M\otimes\mathbf w(k-1)\big].
\end{aligned}
\label{eq:pcsi_stacked_posterior_error}
\end{equation}
To close the mean of~\eqref{eq:pcsi_stacked_posterior_error},
we need the relation between the terminal mean data error
$\mathbb E[\boldsymbol\epsilon^T(k)]$ and the initial mean
prior error $\mathbb E[\mathbf e(k+1\mid k)]$.
We obtain it from the communication updates.
Since subtracting the common state leaves pairwise data
differences unchanged,
$\mathbf z_j^\tau(k)-\mathbf z_i^\tau(k)
=\boldsymbol\epsilon_j^\tau(k)-\boldsymbol\epsilon_i^\tau(k)$,
the conditional fusion mean~\eqref{eq:unbiased_ota_fusion}
becomes
\begin{equation}
\bar{\mathbf s}_i^\tau(k)
=\sum_{j\in\mathcal N_i}
\big[\boldsymbol\epsilon_j^\tau(k)
-\boldsymbol\epsilon_i^\tau(k)\big].
\label{eq:pcsi_fusion_error_form}
\end{equation}
Set $\mathbf s^\tau(k)=\operatorname{col}_{i=1}^{M}\mathbf s_i^\tau(k)$,
$\bar{\boldsymbol\epsilon}^\tau(k)=\mathbb E[\boldsymbol\epsilon^\tau(k)]$,
and $\bar{\mathbf L}=\mathbf L\otimes\mathbf I_N$.
Taking expectations in~\eqref{eq:pcsi_fusion_error_form} gives
$\mathbb E[\mathbf s^\tau(k)]=-\bar{\mathbf L}\bar{\boldsymbol\epsilon}^\tau(k)$.
Thus the mean fusion signal applies the graph Laplacian
$\bar{\mathbf L}$ to the mean data error.

To obtain the communication dynamics, subtract
$\mathbf x(k+1)$ from~\eqref{eq:update_z}.
The identity
$\mathbf P^{\mathcal O_i}+\sum_{(l,s)\in\mathcal U_i}\mathbf P_{l,s}^{\mathcal U_i}=\mathbf I_N$
and the weights $\omega_{l,s}^\tau+(1-\omega_{l,s}^\tau)=1$
cancel the common state.
Define the stacked projectors
$\tilde{\mathbf P}^{\mathcal O}=\operatorname{blkdiag}_{i=1}^{M}(\mathbf P^{\mathcal O_i})$,
$\tilde{\mathbf P}_{l,s}^{\mathcal U}=\operatorname{blkdiag}_{i=1}^{M}(\mathbf P_{l,s}^{\mathcal U_i})$,
and weights $\mathbf\Gamma_{l,s}=\gamma_{l,s}\mathbf I_{MN}$.
The resulting data-error update is
\begin{equation}
\begin{aligned}
\boldsymbol\epsilon^{\tau+1}(k)
={}&\tilde{\mathbf P}^{\mathcal O}\boldsymbol\epsilon^0(k)
+\sum_{\substack{l\in\mathcal I_{\mathrm{pl}}\\1\le s\le n_l^{\mathrm{mb}}}}
\tilde{\mathbf P}_{l,s}^{\mathcal U}
\Big[\omega_{l,s}^\tau
\big(\boldsymbol\epsilon^\tau(k)+\mathbf\Gamma_{l,s}\mathbf s^\tau(k)\big)\\
&\phantom{+\sum_{\substack{l\in\mathcal I_{\mathrm{pl}}\\1\le s\le n_l^{\mathrm{mb}}}}\tilde{\mathbf P}_{l,s}^{\mathcal U}\Big[}
+(1-\omega_{l,s}^\tau)\boldsymbol\epsilon^{\tau-1}(k)\Big].
\end{aligned}
\label{eq:pcsi_stacked_internal_error}
\end{equation}
In~\eqref{eq:pcsi_stacked_internal_error}, the channels and noises
of communication cycle $\tau$ act through $\mathbf s^\tau(k)$.
Its mean is $-\bar{\mathbf L}\bar{\boldsymbol\epsilon}^\tau(k)$,
where $\bar{\boldsymbol\epsilon}^\tau(k)=\mathbb E[\boldsymbol\epsilon^\tau(k)]$.
For fixed projectors and weights, the mean communication update
is therefore a deterministic linear map of the initial prior error
$\bar{\mathbf e}(k+1\mid k)=\mathbb E[\mathbf e(k+1\mid k)]$.
Write $\bar{\boldsymbol\epsilon}^\tau(k)
=\mathbf\Psi_\tau\bar{\mathbf e}(k+1\mid k)$.
The matrix $\mathbf\Psi_\tau\in\mathbb R^{MN\times MN}$ is computed by
\begin{equation}
\begin{aligned}
\mathbf\Psi_{-1}&=\mathbf\Psi_0=\mathbf I_{MN},\\
\mathbf\Psi_{\tau+1}
&=\tilde{\mathbf P}^{\mathcal O}
+\sum_{\substack{l\in\mathcal I_{\mathrm{pl}}\\1\le s\le n_l^{\mathrm{mb}}}}
\tilde{\mathbf P}_{l,s}^{\mathcal U}
\Big[\omega_{l,s}^\tau(\mathbf I_{MN}-\mathbf\Gamma_{l,s}\bar{\mathbf L})
\mathbf\Psi_\tau
+(1-\omega_{l,s}^\tau)\mathbf\Psi_{\tau-1}\Big].
\end{aligned}
\label{eq:pcsi_mean_graph_filter_recursion}
\end{equation}
After $T$ updates, this map relates the initial prior error to
the terminal mean data error as
\begin{equation}
\bar{\boldsymbol\epsilon}^T(k)
=\mathbf\Psi_T\bar{\mathbf e}(k+1\mid k).
\label{eq:pcsi_nominal_graph_filter}
\end{equation}
The matrix $\mathbf\Psi_T$ now connects communication to
posterior estimation. Substitute~\eqref{eq:pcsi_nominal_graph_filter}
and the mean of~\eqref{eq:pcsi_stacked_prior_error} into the
mean of~\eqref{eq:pcsi_stacked_posterior_error}.
Since $\mathbb E[\mathbf w(k)]=\mathbf0_N$, this gives
\begin{equation}
\bar{\mathbf e}(k+1)
=\mathbf F_0\bar{\mathbf e}(k)+\mathbf F_1\bar{\mathbf e}(k-1),
\quad k\geq1,
\label{eq:pcsi_mean_error_dynamics}
\end{equation}
where $\bar{\mathbf e}(k)=\mathbb E[\mathbf e(k)]$ and
\begin{equation}
\begin{aligned}
\mathbf F_0\triangleq{}&
\big[\tilde{\mathbf M}_0
+(\mathbf I_{MN}-\tilde{\mathbf M}_0)\mathbf\Psi_T\big]
(\tilde{\mathbf A}-\tilde{\mathbf K}\tilde{\mathbf C})
+\tilde{\mathbf M}_\beta,\\
\mathbf F_1\triangleq{}&-\tilde{\mathbf M}_\beta\tilde{\mathbf A}.
\end{aligned}
\label{eq:mean_error_transition_matrices}
\end{equation}
The coefficient $\mathbf F_0$ combines local prediction
$\tilde{\mathbf A}-\tilde{\mathbf K}\tilde{\mathbf C}$,
communication $\mathbf\Psi_T$, and posterior weights
$\tilde{\mathbf M}_0,\tilde{\mathbf M}_\beta$.
The coefficient $\mathbf F_1$ retains the preceding posterior
error $\bar{\mathbf e}(k-1)$.
For fixed gains, weights, and $T$, neither $\mathbf F_0$ nor
$\mathbf F_1$ depends on the SNR gain $\rho$.
The pilot-noise bias correction in~\eqref{eq:regularized_ota_term}
therefore removes the SNR dependence from the mean recursion,
although the individual errors $\mathbf e(k)$ remain random.

To determine the decay of $\bar{\mathbf e}(k)$, we use the
same coordinate assignment as the estimator. We first analyze
$\mathbf W_{l,s}^\top\bar{\mathbf e}_i(k)$ for $(l,s)\in\mathcal O_i$,
then the errors for $(l,s)\in\mathcal U_i$ driven by source data.

\subsection{Local Mean-Error Dynamics}
\label{subsec:mean_local_dynamics}
For $n_i^{\mathcal O}>0$, express the local mean error as
$\bar{\mathbf e}_i^{\mathrm J,\mathcal O}(k)
=(\mathbf W_i^{\mathcal O})^\top\bar{\mathbf e}_i(k)$.
This vector is empty when $n_i^{\mathcal O}=0$.
The communication update~\eqref{eq:update_z} retains the
$\mathcal O_i$ coordinates, so
$\mathbf P^{\mathcal O_i}\boldsymbol\epsilon_i^\tau(k)
=\mathbf P^{\mathcal O_i}\mathbf e_i(k+1\mid k)$ for $0\leq\tau\leq T$.
The posterior corrections also act only on $\mathcal U_i$,
as shown in Subsection~\ref{subsec:local_posterior_estimation}.
Thus communication and posterior correction leave the local prediction
unchanged on $\mathcal O_i$.

The separation identities~\eqref{eq:nodewise_separation_identities}
remove the network-assisted coordinates from both the retained-state
dynamics and the local innovation.
Moreover, the dual bases and $\mathbf K_i=\mathbf V_i^{\mathcal O}\mathbf G_i$
give $\mathbf P_{l,s}^{\mathcal U_i}\mathbf K_i=\mathbf0$ and
$\mathbf K_i\mathbf C_i\mathbf P_{l,s}^{\mathcal U_i}=\mathbf0$
for $(l,s)\in\mathcal U_i$.
Taking the expectation of the local prediction eliminates the zero-mean disturbance
$\mathbf w(k)$ and yields
\begin{equation}
\bar{\mathbf e}_i^{\mathrm J,\mathcal O}(k+1)
=\mathbf F_i^{\mathcal O}\bar{\mathbf e}_i^{\mathrm J,\mathcal O}(k),
\label{eq:pcsi_local_reduced_mean_error}
\end{equation}
where $\mathbf F_i^{\mathcal O}
=\mathbf J_i^{\mathcal O}-\mathbf G_i\mathbf C_i\mathbf V_i^{\mathcal O}$.
The chosen local gain satisfies $r(\mathbf F_i^{\mathcal O})<d$.
Hence $\bar{\mathbf e}_i^{\mathrm J,\mathcal O}(k)$ decays
strictly faster than $d^k$.
This local error also determines the accuracy of the source
estimates sent to nodes that cannot observe the same mini-block.
We next retain this source-error input when analyzing the
communicated coordinates.

\subsection{Mean Data Errors for a Communicated Jordan Mini-Block}
\label{subsec:mean_block_communication}
Fix a mini-block $(l,s)$ with $l\in\mathcal I_{\mathrm{pl}}$ and
$\mathcal V_{l,s}^{\mathcal U}\neq\varnothing$.
By~\eqref{eq:pcsi_modal_node_sets}, the non-source nodes
$\mathcal V_{l,s}^{\mathcal U}$ update segment $(l,s)$ by communication,
while the source nodes $\mathcal V_{l,s}^{\mathcal O}$ retain
its local prior. Proposition~\ref{prop:existence_of_modal_sources}
guarantees $\mathcal V_{l,s}^{\mathcal O}\neq\varnothing$.
To express the two groups of errors, let
$\mathbf E_{\mathcal V_{l,s}^{\mathcal U}}
\in\mathbb R^{|\mathcal V_{l,s}^{\mathcal U}|\times M}$ and
$\mathbf E_{\mathcal V_{l,s}^{\mathcal O}}
\in\mathbb R^{|\mathcal V_{l,s}^{\mathcal O}|\times M}$
select the corresponding node rows, and set
$n_{l,s,\mathcal U}=|\mathcal V_{l,s}^{\mathcal U}|$.
The required Jordan-coordinate errors are
\begin{equation}
\begin{aligned}
\bar{\boldsymbol\epsilon}_{l,s,\mathcal U}^{\mathrm J,\tau}(k)
&\triangleq(\mathbf E_{\mathcal V_{l,s}^{\mathcal U}}\otimes\mathbf W_{l,s}^\top)
\bar{\boldsymbol\epsilon}^\tau(k),\\
\bar{\boldsymbol\epsilon}_{l,s,\mathcal O}^{\mathrm J,\tau}(k)
&\triangleq(\mathbf E_{\mathcal V_{l,s}^{\mathcal O}}\otimes\mathbf W_{l,s}^\top)
\bar{\boldsymbol\epsilon}^\tau(k),\\
\bar{\mathbf e}_{l,s,\mathcal U}^{\mathrm J}(k)
&\triangleq\operatorname{col}_{i\in\mathcal V_{l,s}^{\mathcal U}}
(\mathbf W_{l,s}^\top\bar{\mathbf e}_i(k)),
\end{aligned}
\label{eq:pcsi_complete_modal_coordinates}
\end{equation}
Here $\bar{\boldsymbol\epsilon}_i^\tau(k)
=\mathbb E[\boldsymbol\epsilon_i^\tau(k)]$.
The vectors
$\bar{\boldsymbol\epsilon}_{l,s,\mathcal U}^{\mathrm J,\tau}(k)$ and
$\bar{\boldsymbol\epsilon}_{l,s,\mathcal O}^{\mathrm J,\tau}(k)$
are the non-source and source mean data errors.
$\bar{\mathbf e}_{l,s,\mathcal U}^{\mathrm J}(k)$ is the
non-source mean posterior error.
For non-source nodes, $\mathbf W_{l,s}^\top\mathbf K_i=\mathbf0$.
The chain dynamics~\eqref{eq:jordan_segment_dynamics} and
\eqref{eq:pcsi_stacked_prior_error} give the initialization
\begin{equation}
\begin{aligned}
\bar{\boldsymbol\epsilon}_{l,s,\mathcal U}^{\mathrm J,-1}(k)
&=\bar{\boldsymbol\epsilon}_{l,s,\mathcal U}^{\mathrm J,0}(k),\\
\bar{\boldsymbol\epsilon}_{l,s,\mathcal U}^{\mathrm J,0}(k)
&=(\mathbf I_{n_{l,s,\mathcal U}}\otimes\mathbf J_{l,s})
\bar{\mathbf e}_{l,s,\mathcal U}^{\mathrm J}(k)
+\mathbf f_{l,s}(k),\\
\bar{\boldsymbol\epsilon}_{l,s,\mathcal O}^{\mathrm J,\tau}(k)
&=\bar{\boldsymbol\epsilon}_{l,s,\mathcal O}^{\mathrm J,0}(k),
\quad 0\le\tau\le T.
\end{aligned}
\label{eq:pcsi_non_source_prior_modal_error}
\end{equation}
Here the source coordinates remain fixed during communication, and
\begin{equation}
\mathbf f_{l,s}(k)\triangleq
\operatorname{col}_{i\in\mathcal V_{l,s}^{\mathcal U}}
\left(\sum_{t>s}\mathbf J_{l,s,t}
\mathbf W_{l,t}^\top\bar{\mathbf e}_i(k)\right)
\label{eq:segment_prediction_input}
\end{equation}
collects the input from later segments of the same chain.
It can contain both locally retained errors and network-assisted errors,
but never an error from an earlier segment.
The source error
$\bar{\boldsymbol\epsilon}_{l,s,\mathcal O}^{\mathrm J,0}(k)$
is fixed over $\tau=0,\ldots,T$ within control cycle $k$.
Its value is the source nodes' local mean prediction error,
which~\eqref{eq:pcsi_local_reduced_mean_error} expresses as
\begin{equation}
\bar{\boldsymbol\epsilon}_{l,s,\mathcal O}^{\mathrm J,0}(k)
=\operatorname{col}_{i\in\mathcal V_{l,s}^{\mathcal O}}
\big(\mathbf W_{l,s}^\top\mathbf V_i^{\mathcal O}
\mathbf F_i^{\mathcal O}
\bar{\mathbf e}_i^{\mathrm J,\mathcal O}(k)\big).
\label{eq:pcsi_source_local_mean_errors}
\end{equation}
Because $r(\mathbf F_i^{\mathcal O})<d$,
$\bar{\boldsymbol\epsilon}_{l,s,\mathcal O}^{\mathrm J,0}(k)$
decays faster than $d^k$ across control cycles.
It can be nonzero at finite $k$ and must therefore remain
in the communication dynamics.
The graph matrix governing the non-source coordinates is
\begin{equation}
\mathbf L_{\mathcal V_{l,s}^{\mathcal U}}
\triangleq\mathbf E_{\mathcal V_{l,s}^{\mathcal U}}\mathbf L
\mathbf E_{\mathcal V_{l,s}^{\mathcal U}}^\top.
\label{eq:pcsi_grounded_laplacian}
\end{equation}
The principal submatrix $\mathbf L_{\mathcal V_{l,s}^{\mathcal U}}$
retains the full-graph degrees on its diagonal, including
edges to source nodes in $\mathcal V_{l,s}^{\mathcal O}$.
To see that it is positive definite, extend a nonzero
$\mathbf v\in\mathbb R^{n_{l,s,\mathcal U}}$ by zeros at the
source nodes to form $\mathbf u=\mathbf E_{\mathcal V_{l,s}^{\mathcal U}}^\top\mathbf v$
and
$\mathbf v^\top\mathbf L_{\mathcal V_{l,s}^{\mathcal U}}\mathbf v
=\mathbf u^\top\mathbf L\mathbf u>0$.
The strict inequality follows from graph connectivity and
$\mathcal V_{l,s}^{\mathcal O}\neq\varnothing$, because a nonzero
$\mathbf u$ with a zero source entry cannot be constant.

Set $\mathbf S_{l,s}=\mathbf I_{n_{l,s,\mathcal U}}
-\gamma_{l,s}\mathbf L_{\mathcal V_{l,s}^{\mathcal U}}$.
Extracting the coordinates~\eqref{eq:pcsi_complete_modal_coordinates}
from the mean of~\eqref{eq:pcsi_stacked_internal_error} yields
\begin{equation}
\begin{aligned}
\bar{\boldsymbol\epsilon}_{l,s,\mathcal U}^{\mathrm J,\tau+1}(k)
={}&\omega_{l,s}^\tau(\mathbf S_{l,s}\otimes\mathbf I_{m_{l,s}})\bar{\boldsymbol\epsilon}_{l,s,\mathcal U}^{\mathrm J,\tau}(k)+(1-\omega_{l,s}^\tau)
\bar{\boldsymbol\epsilon}_{l,s,\mathcal U}^{\mathrm J,\tau-1}(k)\\
&-\omega_{l,s}^\tau\gamma_{l,s}
\big(\mathbf E_{\mathcal V_{l,s}^{\mathcal U}}\mathbf L
\mathbf E_{\mathcal V_{l,s}^{\mathcal O}}^\top
\otimes\mathbf I_{m_{l,s}}\big)
\bar{\boldsymbol\epsilon}_{l,s,\mathcal O}^{\mathrm J,0}(k),
\end{aligned}
\label{eq:pcsi_source_driven_modal_recursion}
\end{equation}
for $0\leq\tau<T$, initialized by
\eqref{eq:pcsi_non_source_prior_modal_error}.
The terms involving
$\bar{\boldsymbol\epsilon}_{l,s,\mathcal U}^{\mathrm J,\tau}(k)$ and
$\bar{\boldsymbol\epsilon}_{l,s,\mathcal U}^{\mathrm J,\tau-1}(k)$
update the two stored non-source errors.
The term multiplying
$\bar{\boldsymbol\epsilon}_{l,s,\mathcal O}^{\mathrm J,0}(k)$
transmits the source errors through the source-to-non-source
edges. We first characterize the response to the non-source
initial errors and then add the response to this source input.

For fixed $k$, define the homogeneous response
$\bar{\boldsymbol\epsilon}_{l,s,\mathcal U,\mathrm h}^{\mathrm J,\tau}(k)$ by
\begin{equation}
\begin{aligned}
\bar{\boldsymbol\epsilon}_{l,s,\mathcal U,\mathrm h}^{\mathrm J,\tau+1}(k)
={}&\omega_{l,s}^\tau(\mathbf S_{l,s}\otimes\mathbf I_{m_{l,s}})
\bar{\boldsymbol\epsilon}_{l,s,\mathcal U,\mathrm h}^{\mathrm J,\tau}(k)\\
&+(1-\omega_{l,s}^\tau)
\bar{\boldsymbol\epsilon}_{l,s,\mathcal U,\mathrm h}^{\mathrm J,\tau-1}(k),
\quad 0\leq\tau<T,\\
\bar{\boldsymbol\epsilon}_{l,s,\mathcal U,\mathrm h}^{\mathrm J,-1}(k)
={}&\bar{\boldsymbol\epsilon}_{l,s,\mathcal U,\mathrm h}^{\mathrm J,0}(k)
=\bar{\boldsymbol\epsilon}_{l,s,\mathcal U}^{\mathrm J,0}(k).
\end{aligned}
\label{eq:pcsi_homogeneous_modal_initialization}
\end{equation}
The source-error input
$\bar{\boldsymbol\epsilon}_{l,s,\mathcal O}^{\mathrm J,0}(k)$
in~\eqref{eq:pcsi_source_driven_modal_recursion} is set to zero
in~\eqref{eq:pcsi_homogeneous_modal_initialization}. The two
non-source initial errors are unchanged.
Thus $\bar{\boldsymbol\epsilon}_{l,s,\mathcal U,\mathrm h}^{\mathrm J,T}(k)$
is the terminal response to the initial non-source errors alone.
It can differ from the actual terminal error
$\bar{\boldsymbol\epsilon}_{l,s,\mathcal U}^{\mathrm J,T}(k)$
when the source-error input is nonzero.

The homogeneous response can be evaluated without following
each communication update separately.
For $\omega_{l,s}^0=1$, define the scalar polynomials
\begin{equation}
\begin{aligned}
\psi_{0,l,s}(z)&=1,\quad\psi_{1,l,s}(z)=z,\\
\psi_{\tau+1,l,s}(z)
&=\omega_{l,s}^\tau z\psi_{\tau,l,s}(z)+(1-\omega_{l,s}^\tau)\psi_{\tau-1,l,s}(z),
\end{aligned}
\label{eq:pcsi_graph_polynomial_recursion}
\end{equation}
for $1\le\tau<T$. Replacing the scalar argument $z$ in
\eqref{eq:pcsi_graph_polynomial_recursion} by $\mathbf S_{l,s}$
reproduces the homogeneous update
\eqref{eq:pcsi_homogeneous_modal_initialization}.
Induction on $\tau$ therefore gives its terminal response as
\begin{equation}
\begin{aligned}
\bar{\boldsymbol\epsilon}_{l,s,\mathcal U,\mathrm h}^{\mathrm J,T}(k)
={}&[\psi_{T,l,s}(\mathbf S_{l,s})\otimes\mathbf I_{m_{l,s}}]\bar{\boldsymbol\epsilon}_{l,s,\mathcal U}^{\mathrm J,0}(k).
\end{aligned}
\label{eq:pcsi_homogeneous_graph_filter}
\end{equation}
Equation~\eqref{eq:pcsi_homogeneous_graph_filter} gives the
coefficient $\psi_{T,l,s}(\mathbf S_{l,s})\otimes\mathbf I_{m_{l,s}}$
relating the initial prior error
$\bar{\boldsymbol\epsilon}_{l,s,\mathcal U}^{\mathrm J,0}(k)$ to its
homogeneous terminal response. This prior error includes the chain input
$\mathbf f_{l,s}(k)$.
To recover the actual terminal error, we now include the
source-error input
$\bar{\boldsymbol\epsilon}_{l,s,\mathcal O}^{\mathrm J,0}(k)$.
For a fixed value of this input, the unique non-source error
at which the expected graph correction is zero is
\begin{equation}
\mathbf c_{l,s}(k)\triangleq
-\big[\mathbf L_{\mathcal V_{l,s}^{\mathcal U}}^{-1}
\mathbf E_{\mathcal V_{l,s}^{\mathcal U}}\mathbf L
\mathbf E_{\mathcal V_{l,s}^{\mathcal O}}^\top
\otimes\mathbf I_{m_{l,s}}\big]
\bar{\boldsymbol\epsilon}_{l,s,\mathcal O}^{\mathrm J,0}(k).
\label{eq:pcsi_source_error_equilibrium}
\end{equation}
Indeed, $\mathbf c_{l,s}(k)$ solves
$(\mathbf L_{\mathcal V_{l,s}^{\mathcal U}}\otimes\mathbf I_{m_{l,s}})\mathbf c_{l,s}(k)
+(\mathbf E_{\mathcal V_{l,s}^{\mathcal U}}\mathbf L
\mathbf E_{\mathcal V_{l,s}^{\mathcal O}}^\top\otimes\mathbf I_{m_{l,s}})
\bar{\boldsymbol\epsilon}_{l,s,\mathcal O}^{\mathrm J,0}(k)=\mathbf0$.
Thus $\mathbf c_{l,s}(k)$ is the communication equilibrium for
the source errors held fixed within control cycle $k$.
Its value changes with $k$ through
\eqref{eq:pcsi_source_local_mean_errors}.

Subtracting $\mathbf c_{l,s}(k)$ from
\eqref{eq:pcsi_source_driven_modal_recursion} removes the
source input. The shifted error
$\bar{\boldsymbol\epsilon}_{l,s,\mathcal U}^{\mathrm J,\tau}(k)-\mathbf c_{l,s}(k)$
then obeys~\eqref{eq:pcsi_homogeneous_modal_initialization}
with equal initial vectors
$\bar{\boldsymbol\epsilon}_{l,s,\mathcal U}^{\mathrm J,0}(k)-\mathbf c_{l,s}(k)$.
Applying the same polynomial $\psi_{T,l,s}$ to these shifted
initial errors gives
\begin{equation}
\begin{aligned}
\bar{\boldsymbol\epsilon}_{l,s,\mathcal U}^{\mathrm J,T}(k)
={}&[\psi_{T,l,s}(\mathbf S_{l,s})\otimes\mathbf I_{m_{l,s}}]\bar{\boldsymbol\epsilon}_{l,s,\mathcal U}^{\mathrm J,0}(k)\\
&+\big[(\mathbf I_{n_{l,s,\mathcal U}}-\psi_{T,l,s}(\mathbf S_{l,s}))
\otimes\mathbf I_{m_{l,s}}\big]\mathbf c_{l,s}(k).
\end{aligned}
\label{eq:pcsi_full_graph_filter}
\end{equation}
The term
$[\psi_{T,l,s}(\mathbf S_{l,s})\otimes\mathbf I_{m_{l,s}}]
\bar{\boldsymbol\epsilon}_{l,s,\mathcal U}^{\mathrm J,0}(k)$
is the homogeneous response. The term multiplying
$\mathbf c_{l,s}(k)$ is the response to the source errors.
Their sum is the terminal error required by the posterior update.

\subsection{Weight Design and Prescribed Mean-Error Rate}
\label{subsec:mean_chebyshev_design}
For a non-source node $i\in\mathcal V_{l,s}^{\mathcal U}$,
$\mathbf M_{0,i}$ and $\mathbf M_{\beta,i}$ have Jordan-coordinate
weights $-\beta_l\mathbf J_{l,s}^{-1}$ and $\beta_l\mathbf I_{m_{l,s}}$.
Substitute~\eqref{eq:pcsi_full_graph_filter} into the mean of
\eqref{eq:pcsi_stacked_posterior_error}, using the initialization
\eqref{eq:pcsi_non_source_prior_modal_error}.
The mean posterior error across control cycles then satisfies
\begin{equation}
\begin{aligned}
\bar{\mathbf e}_{l,s,\mathcal U}^{\mathrm J}(k+1)
={}&[\psi_{T,l,s}(\mathbf S_{l,s})
\otimes(\mathbf J_{l,s}+\beta_l\mathbf I_{m_{l,s}})]\bar{\mathbf e}_{l,s,\mathcal U}^{\mathrm J}(k)\\
&-(\mathbf I_{n_{l,s,\mathcal U}}\otimes\beta_l\mathbf J_{l,s})
\bar{\mathbf e}_{l,s,\mathcal U}^{\mathrm J}(k-1)\\
&+\big[(\mathbf I_{n_{l,s,\mathcal U}}-\psi_{T,l,s}(\mathbf S_{l,s}))
\otimes(\mathbf I_{m_{l,s}}+\beta_l\mathbf J_{l,s}^{-1})\big]\\
&\times[\mathbf c_{l,s}(k)-\mathbf f_{l,s}(k)]\\
&+\mathbf f_{l,s}(k)-\beta_l\mathbf f_{l,s}(k-1).
\end{aligned}
\label{eq:pcsi_non_source_posterior_recursion}
\end{equation}
Equation~\eqref{eq:pcsi_non_source_posterior_recursion} evolves
$\bar{\mathbf e}_{l,s,\mathcal U}^{\mathrm J}(k)$ across control
cycles after the $T$ communication updates are completed.
The history term contains
$\bar{\mathbf e}_{l,s,\mathcal U}^{\mathrm J}(k-1)$.
The source-error term contains $\mathbf c_{l,s}(k)$, a fixed linear
function of
$\{\bar{\mathbf e}_i^{\mathrm J,\mathcal O}(k)\}_{i\in\mathcal V_{l,s}^{\mathcal O}}$.
These local errors already decay faster than $d^k$.
The chain inputs $\mathbf f_{l,s}(k),\mathbf f_{l,s}(k-1)$ involve
only later segments of the same original chain.
Thus the network-assisted branches form a cascade when ordered from
later to earlier segments.
It remains to choose the weights so that the homogeneous current-and-previous
non-source error transition also has all eigenvalues inside
$|z|<d$.

For an eigenvalue $\mu$ of
$\mathbf L_{\mathcal V_{l,s}^{\mathcal U}}$, define the real scalar
$g_{l,s}(\mu)=\psi_{T,l,s}(1-\gamma_{l,s}\mu)$.
Since $\mathbf J_{l,s}$ has diagonal eigenvalue $\lambda_l$,
the corresponding homogeneous posterior-error transition has characteristic
polynomial $[q_{l,s,\mu}(z)]^{m_{l,s}}$, where
\begin{equation}
q_{l,s,\mu}(z)=z^2-(\lambda_l+\beta_l)g_{l,s}(\mu)z+d_-^2.
\label{eq:pcsi_modal_characteristic_polynomial}
\end{equation}
The roots of $q_{l,s,\mu}(z)$ govern the homogeneous decay across control
cycles $k$ of
$(\bar{\mathbf e}_{l,s,\mathcal U}^{\mathrm J}(k),
\bar{\mathbf e}_{l,s,\mathcal U}^{\mathrm J}(k-1))$.
The identity $\beta_l\lambda_l=d_-^2<d^2$ fixes the history
coefficient, while the communication attenuation
$g_{l,s}(\mu)$ controls the coefficient multiplying the current
posterior error. We therefore seek a bound on $|g_{l,s}(\mu)|$
that places both roots inside $|z|<d$.
Set $\delta=d_-/d\in(0,1)$ and $h_l=\lambda_l+\beta_l$.
Applying the root condition to
\eqref{eq:pcsi_modal_characteristic_polynomial} yields
the sufficient attenuation requirement
\begin{equation}
|\psi_{T,l,s}(1-\gamma_{l,s}\mu)|
<\alpha_l\triangleq
\frac{d(1-\delta^4)}{|h_l-\delta^2h_l^*|},\quad \mu\in\sigma(\mathbf L_{\mathcal V_{l,s}^{\mathcal U}}).
\label{eq:pcsi_modal_attenuation}
\end{equation}
For every $l\in\mathcal I_{\mathrm{pl}}$,
$|h_l-\delta^2h_l^*|\geq d(1-\delta^4)>0$.
Hence $0<\alpha_l\leq1$, with equality when $|\lambda_l|=d$.
The remaining task is to enforce the same bound $\alpha_l$
for every eigenvalue $\mu$ of the non-source graph matrix.
Let $\mu_{l,s}^{\mathrm{min}}>0$ and $\mu_{l,s}^{\mathrm{max}}$
be the smallest and largest eigenvalues of
$\mathbf L_{\mathcal V_{l,s}^{\mathcal U}}$, respectively.
The interval $[\mu_{l,s}^{\mathrm{min}},\mu_{l,s}^{\mathrm{max}}]$
contains every eigenvalue governing the homogeneous non-source errors.
To attenuate all these error components with one communication weight,
map this interval to an interval symmetric about zero by choosing
\begin{equation}
\gamma_{l,s}\triangleq\frac2{\mu_{l,s}^{\mathrm{max}}+\mu_{l,s}^{\mathrm{min}}},
\quad
\theta_{l,s}\triangleq
\frac{\mu_{l,s}^{\mathrm{max}}-\mu_{l,s}^{\mathrm{min}}}
{\mu_{l,s}^{\mathrm{max}}+\mu_{l,s}^{\mathrm{min}}}.
\label{eq:pcsi_chebyshev_step_size}
\end{equation}
Then $1-\gamma_{l,s}\mu\in[-\theta_{l,s},\theta_{l,s}]$ for every
$\mu\in[\mu_{l,s}^{\mathrm{min}},\mu_{l,s}^{\mathrm{max}}]$,
with $0\leq\theta_{l,s}<1$.
To reduce the largest value of
$|\psi_{T,l,s}(1-\gamma_{l,s}\mu)|$ on this interval,
initialize the weights by $\omega_{l,s}^0=1$ and
$\omega_{l,s}^1=2/(2-\theta_{l,s}^2)$. For $2\leq\tau<T$, use
$\omega_{l,s}^\tau=(1-\theta_{l,s}^2\omega_{l,s}^{\tau-1}/4)^{-1}$.
Only weights with $0\leq\tau<T$ are used. For $T=1$,
only $\omega_{l,s}^0$ is needed.
Substitution into~\eqref{eq:pcsi_graph_polynomial_recursion}
gives the normalized Chebyshev polynomial
\begin{equation}
\psi_{T,l,s}(z)=
\begin{cases}
\dfrac{\mathcal T_T(z/\theta_{l,s})}
{\mathcal T_T(1/\theta_{l,s})},&\theta_{l,s}>0,\\[1.5ex]
z^T,&\theta_{l,s}=0,
\end{cases}
\label{eq:pcsi_normalized_chebyshev_polynomial}
\end{equation}
Here $\mathcal T_T$ is the Chebyshev polynomial of the first
kind. For $\theta_{l,s}>0$, the transformed graph eigenvalue
$(1-\gamma_{l,s}\mu)/\theta_{l,s}$ lies in $[-1,1]$.
Since $\mathbf S_{l,s}$ is real symmetric and
$|\mathcal T_T(t)|\leq1$ on $[-1,1]$, the homogeneous error map satisfies
\begin{equation}
\|\psi_{T,l,s}(\mathbf S_{l,s})\|_2
\leq\frac1{\mathcal T_T(1/\theta_{l,s})},\quad\theta_{l,s}>0.
\label{eq:pcsi_graph_filter_norm_bound}
\end{equation}
The attenuation
$1/\mathcal T_T(1/\theta_{l,s})$ in
\eqref{eq:pcsi_graph_filter_norm_bound} decreases as
more communication updates are performed.
For $\theta_{l,s}=0$, $\mathbf S_{l,s}=\mathbf0$ and
$\psi_{T,l,s}(\mathbf S_{l,s})=\mathbf0$ for $T\geq1$,
so~\eqref{eq:pcsi_modal_attenuation} holds after one update.
Segments with $\mathcal V_{l,s}^{\mathcal U}=\varnothing$ require
no communication design. Conjugate Jordan mini-blocks share the
same node sets and graph spectrum and use the same real weights.

For $0<\theta_{l,s}<1$,
$\mathcal T_T(1/\theta_{l,s})=
\cosh(T\operatorname{arccosh}(1/\theta_{l,s}))$.
Solving $1/\mathcal T_T(1/\theta_{l,s})<\alpha_l$ therefore
gives the cycle-count requirement below.
Together with $r(\mathbf F_i^{\mathcal O})<d$, it ensures
that both the local and communication-dependent mean errors
decay faster than $d^k$.

\begin{theorem}[Prescribed Mean-Error Convergence]
\label{thm:pcsi_mean_convergence}
Consider Algorithm~\ref{alg:tracking_protocol} under
Assumptions~\ref{as:time_scales}--\ref{as:global_modal_source_observability}.
Fix $d\in(0,1)$ for which
Assumption~\ref{as:network_decoupled_local_detectability}
holds and select $d_-\in(0,d)$.
For each segment with $\mathcal V_{l,s}^{\mathcal U}\neq\varnothing$,
take $\gamma_{l,s}$ from \eqref{eq:pcsi_chebyshev_step_size} and use
the Chebyshev weights $\omega_{l,s}^\tau$.
Let the number of communication cycles per control cycle
be a positive integer $T$ satisfying
\begin{equation}
\max_{\substack{l\in\mathcal I_{\mathrm{pl}},\ 1\le s\le n_l^{\mathrm{mb}}\\
\mathcal V_{l,s}^{\mathcal U}\neq\varnothing,\ \theta_{l,s}>0}}
\frac{\operatorname{arccosh}(\alpha_l^{-1})}
{\operatorname{arccosh}(\theta_{l,s}^{-1})}<T,
\label{eq:pcsi_chebyshev_order_range}
\end{equation}
where an empty maximum is interpreted as zero.
Then Algorithm~\ref{alg:tracking_protocol} is asymptotically unbiased and
\begin{equation}
\limsup_{k\to\infty}\|\bar{\mathbf e}(k)\|^{1/k}<d,
\quad
\|\bar{\mathbf e}(k)\|=o(d^k).
\label{eq:pcsi_mean_decay_rate}
\end{equation}
\end{theorem}
\begin{proof}
Please see~\ref{app:proof_mean_convergence}.

\end{proof}
Theorem~\ref{thm:pcsi_mean_convergence} gives a finite
communication requirement for each admissible target $d$, because
there are finitely many segments, $\alpha_l>0$, and
$0<\theta_{l,s}<1$ for every segment included in
\eqref{eq:pcsi_chebyshev_order_range}.
Increasing $T$ sufficiently therefore makes the mean
posterior error $\bar{\mathbf e}(k)$ decay faster than $d^k$.
For each original chain, the non-source sets are nested:
$\mathcal V_{l,s}^{\mathcal U}\subseteq\mathcal V_{l,1}^{\mathcal U}$.
The corresponding grounded Laplacians are principal submatrices of
$\mathbf L_{\mathcal V_{l,1}^{\mathcal U}}$.
Their smallest eigenvalues cannot decrease and their largest eigenvalues
cannot increase. Hence $\theta_{l,s}\le\theta_{l,1}$, while $\alpha_l$ is
common to the chain. The largest cycle-count requirement within each
chain is therefore attained by its first segment, whose sources are
exactly the nodes with $r_{i,l}=1$.
This conclusion concerns $\mathbb E[\mathbf e(k)]$.
Section~\ref{sec:finite_snr_analysis} examines the fluctuations
around that mean and their contribution to the state estimation
MSE $\mathcal M(k)$.

\section{MSE Performance Analysis}
\label{sec:finite_snr_analysis}
Theorem~\ref{thm:pcsi_mean_convergence} establishes the decay of the mean
estimation error $\bar{\mathbf e}(k)$. To determine the state estimation
MSE $\mathcal M(k)$, we also retain the zero-mean communication
fluctuations. Fading and pilot estimation
errors multiply the data error $\boldsymbol\epsilon^\tau(k)$, while
pilot errors also introduce a term proportional to the plant state
$\mathbf x(k+1)$. We follow these terms through the $T$ communication
updates and the posterior update, then calculate their effects on
convergence to the MSE platform and on the platform itself.

\subsection{Stochastic Perturbations in OTA Fusion}
\label{subsec:finite_snr_ota_perturbation}
Substituting the aggregate CSI estimate \eqref{eq:aggregated_csi}
and the received-data model \eqref{eq:received_data} into
\eqref{eq:regularized_ota_term}, and using
\eqref{eq:pcsi_fusion_error_form} and
$\mathbf z_j^\tau(k)=\boldsymbol\epsilon_j^\tau(k)+\mathbf x(k+1)$,
yields
\begin{equation}
\begin{aligned}
\mathbf s_i^\tau(k)
={}&\bar{\mathbf s}_i^\tau(k)
-\sum_{j=1}^{M}
[\boldsymbol\Delta^\tau(k)]_{ij}
\boldsymbol\epsilon_j^\tau(k)\\
&-\sum_{j=1}^{M}
[\boldsymbol\Delta_N^\tau(k)]_{ij}
\big(\boldsymbol\epsilon_j^\tau(k)+\mathbf x(k+1)\big)
+\frac{1}{\sqrt\rho}\mathbf n_i^\tau(k).
\end{aligned}
\label{eq:finite_fusion_error_decomposition}
\end{equation}
The three departures from the conditional-mean signal
$\bar{\mathbf s}_i^\tau(k)$ are the fading term involving
$\boldsymbol\Delta^\tau(k)$, the pilot-error term involving
$\boldsymbol\Delta_N^\tau(k)$, and the receiver data-noise term
$\rho^{-1/2}\mathbf n_i^\tau(k)$. We specify these coefficients below.
Let $\mathbf H_i^\tau(k)=\sum_{j\in\mathcal N_i}\mathbf H_{ij}^\tau(k)$.
With exact channel sums, the channel products in
\eqref{eq:regularized_ota_term} give the matrix-weighted graph
Laplacian $\mathbf L^\tau(k)\in\mathbb R^{MN\times MN}$, with blocks
\begin{equation}
[\mathbf L^\tau(k)]_{ij}=
\begin{cases}
-\dfrac{(\mathbf H_i^\tau(k))^\top
\mathbf H_{ij}^\tau(k)}{N_r},
&j\in\mathcal N_i,\\[1.5ex]
-\displaystyle\sum_{m\in\mathcal N_i}
[\mathbf L^\tau(k)]_{im},
&j=i,\\[1.5ex]
\mathbf0_{N\times N},&\text{otherwise}.
\end{cases}
\label{eq:weighted_laplacian}
\end{equation}
The weights are channel correlation matrices, rather than scalar
edge weights. Each block row sums to zero, so
$\mathbf L^\tau(k)(\mathbf1_M\otimes\mathbf I_N)=\mathbf0$.
Thus exact channel sums do not couple a common state into the fusion signal.
Assumption~\ref{as:Wireless_Fading_Channels} gives
$\mathbb E[\mathbf L^\tau(k)]=\bar{\mathbf L}$ and hence
\begin{equation}
\mathbf L^\tau(k)
=\bar{\mathbf L}+\boldsymbol\Delta^\tau(k),
\quad
\mathbb E[\boldsymbol\Delta^\tau(k)]=\mathbf0,
\label{eq:weighted_laplacian_decomposition}
\end{equation}
Thus $\boldsymbol\Delta^\tau(k)$ measures the fluctuation around
the mean graph correction $\bar{\mathbf L}$.
The additive receiver-noise term in
\eqref{eq:finite_fusion_error_decomposition} is
\begin{equation}
\mathbf n_i^\tau(k)
\triangleq
\frac{1}{N_r}
\left[
(\mathbf H_i^\tau(k))^\top
+\frac{(\breve{\mathbf N}_{p,i}^\tau(k))^\top}{\sqrt\rho}
\right]
\boldsymbol\eta_i^\tau(k),
\label{eq:finite_equivalent_data_noise}
\end{equation}
The remaining perturbation $\boldsymbol\Delta_N^\tau(k)$ comes from
using the noisy aggregate CSI $\hat{\mathbf H}_i^\tau(k)$.
Expansion of~\eqref{eq:regularized_ota_term} gives its blocks as
\begin{equation}
[\boldsymbol\Delta_N^\tau(k)]_{ij}=
\begin{cases}
\displaystyle-\frac{1}{N_r\sqrt\rho}
(\breve{\mathbf N}_{p,i}^\tau(k))^\top
\mathbf H_{ij}^\tau(k),
&j\in\mathcal N_i,\\[2ex]
\displaystyle\frac{1}{N_r\sqrt\rho}
\bigg[(\breve{\mathbf N}_{p,i}^\tau(k))^\top
\mathbf H_i^\tau(k)\\[1.5ex]
\displaystyle{}
+(\mathbf H_i^\tau(k))^\top
\breve{\mathbf N}_{p,i}^\tau(k)\bigg]\\[1.5ex]
\displaystyle{}
+\frac{1}{N_r\rho}
\bigg[(\breve{\mathbf N}_{p,i}^\tau(k))^\top
\breve{\mathbf N}_{p,i}^\tau(k)\\[1.5ex]
\displaystyle{}
-N_r\mathbf X_p^{-\top}\mathbf X_p^{-1}\bigg],
&j=i,\\[2ex]
\mathbf0_{N\times N},&\text{otherwise}.
\end{cases}
\label{eq:finite_pilot_noise_perturbation}
\end{equation}
Subtracting $N_r\mathbf X_p^{-\top}\mathbf X_p^{-1}/\rho$
in~\eqref{eq:regularized_ota_term} centers the pilot-noise Gram
matrix in~\eqref{eq:finite_pilot_noise_perturbation}.
This bias correction and the independence of
$\boldsymbol\eta_i^\tau(k)$ yield
\begin{equation}
\mathbb E[\boldsymbol\Delta_N^\tau(k)]=\mathbf0,\;
\mathbb E[\mathbf n_i^\tau(k)\mid\mathbf H_i^\tau(k),
\breve{\mathbf N}_{p,i}^\tau(k)]=\mathbf0.
\label{eq:finite_primitive_perturbation_means}
\end{equation}
Unlike $\boldsymbol\Delta^\tau(k)$,
$\boldsymbol\Delta_N^\tau(k)$ need not have zero block-row sums.
It therefore acts on both $\boldsymbol\epsilon^\tau(k)$ and the
common state $\mathbf x(k+1)$ in
\eqref{eq:finite_fusion_error_decomposition}.
The receiver-noise vector $\mathbf n_i^\tau(k)$ depends on
$\mathbf H_i^\tau(k)$, $\breve{\mathbf N}_{p,i}^\tau(k)$, and
$\boldsymbol\eta_i^\tau(k)$, and is independent of
$\boldsymbol\epsilon^\tau(k)$ and $\mathbf x(k+1)$.
Thus the pilot-error term requires plant statistics in the MSE
calculation, whereas $\rho^{-1/2}\mathbf n_i^\tau(k)$ is an additive disturbance.

\subsection{Global Stochastic Error Dynamics}
\label{subsec:finite_snr_global_dynamics}
The perturbations in \eqref{eq:finite_fusion_error_decomposition}
act at each communication update. To obtain their accumulated effect
on the posterior error, we first follow the data error
$\boldsymbol\epsilon^\tau(k)$ through the $T$ updates.
Stack the receiver noises as
$\mathbf n^\tau(k)\triangleq
\operatorname{col}(\mathbf n_1^\tau(k),\ldots,\mathbf n_M^\tau(k))$
and write the common plant state as
$\tilde{\mathbf x}(k)\triangleq\mathbf1_M\otimes\mathbf x(k)$.
Then \eqref{eq:pcsi_stacked_internal_error} becomes
\begin{equation}
\begin{aligned}
\boldsymbol\epsilon^{\tau+1}(k)
={}&\tilde{\mathbf P}^{\mathcal O}\boldsymbol\epsilon^0(k)
+(\mathbf\Lambda_1^\tau+\mathbf E_{\epsilon}^\tau(k))
\boldsymbol\epsilon^\tau(k)\\
&+\mathbf\Lambda_2^\tau\boldsymbol\epsilon^{\tau-1}(k)
+\mathbf E_x^\tau(k)\tilde{\mathbf x}(k+1)\\
&+\rho^{-1/2}\mathbf n_{\mathrm{in}}^\tau(k).
\end{aligned}
\label{eq:finite_global_internal_error}
\end{equation}
Here $\boldsymbol\epsilon^{-1}(k)=\boldsymbol\epsilon^0(k)
=\mathbf e(k+1\mid k)$. The deterministic matrices multiplying
the current and preceding data errors are, respectively,
\begin{align*}
\mathbf\Lambda_1^\tau
&\triangleq\sum_{\substack{l\in\mathcal I_{\mathrm{pl}}\\1\le s\le n_l^{\mathrm{mb}}}}
\omega_{l,s}^\tau\tilde{\mathbf P}_{l,s}^{\mathcal U}
(\mathbf I_{MN}-\mathbf\Gamma_{l,s}\bar{\mathbf L}),\\
\mathbf\Lambda_2^\tau
&\triangleq\sum_{\substack{l\in\mathcal I_{\mathrm{pl}}\\1\le s\le n_l^{\mathrm{mb}}}}
(1-\omega_{l,s}^\tau)\tilde{\mathbf P}_{l,s}^{\mathcal U}.
\end{align*}
The perturbation multiplying the current data error is
\begin{equation*}
\begin{aligned}
\mathbf E_{\epsilon}^\tau(k)
\triangleq{}&-\sum_{\substack{l\in\mathcal I_{\mathrm{pl}}\\1\le s\le n_l^{\mathrm{mb}}}}
\omega_{l,s}^\tau\tilde{\mathbf P}_{l,s}^{\mathcal U}\mathbf\Gamma_{l,s}\\
&\times(\boldsymbol\Delta^\tau(k)+\boldsymbol\Delta_N^\tau(k))
\end{aligned}
\end{equation*}
and contains both fading and pilot fluctuations.
The plant-state coefficient
\begin{equation*}
\mathbf E_x^\tau(k)\triangleq
-\sum_{\substack{l\in\mathcal I_{\mathrm{pl}}\\1\le s\le n_l^{\mathrm{mb}}}}
\omega_{l,s}^\tau\tilde{\mathbf P}_{l,s}^{\mathcal U}
\mathbf\Gamma_{l,s}\boldsymbol\Delta_N^\tau(k)
\end{equation*}
contains only pilot errors and acts on the common plant state
$\tilde{\mathbf x}(k+1)$. The receiver data noise enters through
\begin{equation*}
\mathbf n_{\mathrm{in}}^\tau(k)\triangleq
\sum_{\substack{l\in\mathcal I_{\mathrm{pl}}\\1\le s\le n_l^{\mathrm{mb}}}}
\omega_{l,s}^\tau\tilde{\mathbf P}_{l,s}^{\mathcal U}
\mathbf\Gamma_{l,s}\mathbf n^\tau(k).
\end{equation*}
All selected segments use the same slot channels and pilot samples.
The stacked coefficients therefore retain their cross-segment correlations.
The Jordan-chain connections enter through the full prediction matrix
$\tilde{\mathbf A}-\tilde{\mathbf K}\tilde{\mathbf C}$.
Because \eqref{eq:finite_global_internal_error} is linear in the
initial error $\mathbf e(k+1\mid k)$, the fixed state
$\tilde{\mathbf x}(k+1)$, and receiver noises, its terminal value has
the form
\begin{equation}
\begin{aligned}
\boldsymbol\epsilon^T(k)
={}&[\mathbf\Psi_T+\mathbf R_T(k)]\mathbf e(k+1\mid k)\\
&+\mathbf Q_T(k)\tilde{\mathbf x}(k+1)
+\rho^{-1/2}\tilde{\mathbf n}_T(k).
\end{aligned}
\label{eq:finite_terminal_graph_filter}
\end{equation}
The mean coefficient $\mathbf\Psi_T$ is already defined in
\eqref{eq:pcsi_mean_graph_filter_recursion}.
The matrices $\mathbf R_T(k),\mathbf Q_T(k)\in\mathbb R^{MN\times MN}$
and vector $\tilde{\mathbf n}_T(k)\in\mathbb R^{MN}$
measure, respectively, the accumulated prior-error perturbation,
plant-state input caused by pilot errors, and receiver data noise.
To compute them, start with
$\mathbf R_{-1}=\mathbf R_0=\mathbf Q_{-1}=\mathbf Q_0=\mathbf0$
and $\tilde{\mathbf n}_{-1}=\tilde{\mathbf n}_0=\mathbf0$, and iterate
\begin{align} 
\mathbf R_{\tau+1}
& =(\mathbf\Lambda_1^\tau+\mathbf E_{\epsilon}^\tau)\mathbf R_\tau
+\mathbf\Lambda_2^\tau\mathbf R_{\tau-1}+\mathbf E_{\epsilon}^\tau\mathbf\Psi_\tau,
\label{eq:finite_graph_filter_perturbation}\\ 
\mathbf Q_{\tau+1}
& =(\mathbf\Lambda_1^\tau+\mathbf E_{\epsilon}^\tau)\mathbf Q_\tau
+\mathbf\Lambda_2^\tau\mathbf Q_{\tau-1}+\mathbf E_x^\tau,
\label{eq:finite_state_coupling_perturbation}\\ 
\tilde{\mathbf n}_{\tau+1}
& =(\mathbf\Lambda_1^\tau+\mathbf E_{\epsilon}^\tau)\tilde{\mathbf n}_\tau
+\mathbf\Lambda_2^\tau\tilde{\mathbf n}_{\tau-1}+\mathbf n_{\mathrm{in}}^\tau,
\label{eq:finite_accumulated_data_noise}
\end{align}
for $\tau=0,\ldots,T-1$, where the common argument $k$ is suppressed
in these three recursions. Substitution into
\eqref{eq:finite_global_internal_error} verifies
\eqref{eq:finite_terminal_graph_filter} at each communication cycle.

Independence across communication cycles and the zero means in
\eqref{eq:weighted_laplacian_decomposition} and
\eqref{eq:finite_primitive_perturbation_means} give
$\mathbb E[\mathbf R_T(k)]=\mathbb E[\mathbf Q_T(k)]
=\mathbb E[\tilde{\mathbf n}_T(k)]=\mathbf0$.
These identities recover the mean-error relation in
Section~\ref{sec:mean_convergence_analysis}. The accumulated
perturbations need not have zero second moments, so
$\mathbf R_T(k)\mathbf e(k+1\mid k)$ and
$\mathbf Q_T(k)\tilde{\mathbf x}(k+1)$ must be retained in the MSE analysis.

Let $\tilde{\mathbf w}(k)\triangleq\mathbf1_M\otimes\mathbf w(k)$
denote the replicated process disturbance. The stacked posterior-error
recursion is given by \eqref{eq:pcsi_stacked_posterior_error}.
Substituting \eqref{eq:finite_terminal_graph_filter},
$\mathbf e(k+1\mid k)
=(\tilde{\mathbf A}-\tilde{\mathbf K}\tilde{\mathbf C})\mathbf e(k)
-\tilde{\mathbf w}(k)$, and
$\tilde{\mathbf x}(k+1)=\tilde{\mathbf A}\tilde{\mathbf x}(k)
+\tilde{\mathbf w}(k)$ into \eqref{eq:pcsi_stacked_posterior_error}
yields
\begin{equation}
\begin{aligned}
\mathbf e(k+1)
={}&\mathbf F_0\mathbf e(k)
+\mathbf F_1\mathbf e(k-1)\\
&+(\mathbf I_{MN}-\tilde{\mathbf M}_0)
\mathbf R_T(k)
(\tilde{\mathbf A}-\tilde{\mathbf K}\tilde{\mathbf C})\mathbf e(k)\\
&+(\mathbf I_{MN}-\tilde{\mathbf M}_0)
\mathbf Q_T(k)
\tilde{\mathbf A}\tilde{\mathbf x}(k)+\boldsymbol\nu(k),
\end{aligned}
\label{eq:finite_master_error_dynamics}
\end{equation}
The terminal coefficients satisfy
$\mathbf\Psi_T\tilde{\mathbf w}(k)=\tilde{\mathbf w}(k)$ and
$\mathbf Q_T(k)\tilde{\mathbf w}(k)=\mathbf R_T(k)\tilde{\mathbf w}(k)$,
since $\tilde{\mathbf w}(k)=\mathbf1_M\otimes\mathbf w(k)$ lies in
the replicated-state subspace. Thus the random coefficients multiplying
the current process disturbance cancel.
With $\tilde{\mathbf M}_1\triangleq\mathbf I_{MN}-\tilde{\mathbf M}_0$
denoting the posterior communication weight, the remaining disturbance is
\begin{equation}
\boldsymbol\nu(k)
=-\tilde{\mathbf w}(k)+\tilde{\mathbf M}_\beta\tilde{\mathbf w}(k-1)
+\rho^{-1/2}\tilde{\mathbf M}_1\tilde{\mathbf n}_T(k).
\label{eq:finite_lumped_disturbance}
\end{equation}
The first two terms of~\eqref{eq:finite_master_error_dynamics} are
$\mathbf F_0\mathbf e(k)+\mathbf F_1\mathbf e(k-1)$ from the mean
recursion~\eqref{eq:pcsi_mean_error_dynamics}.
The next two contain, respectively, the prior-error perturbation
$\mathbf R_T(k)$ and plant-state coefficient
$\mathbf Q_T(k)$.
The disturbance $\boldsymbol\nu(k)$ collects the current and preceding
process disturbances and accumulated receiver data noise.

\subsection{Augmented Stochastic Error System}
\label{subsec:finite_snr_augmented_system}
The term $\tilde{\mathbf M}_\beta\tilde{\mathbf w}(k-1)$ in
\eqref{eq:finite_lumped_disturbance} contains the same past process
noise that also enters $\mathbf e(k)$. Keeping
$\tilde{\mathbf w}(k-1)$ in the state makes this dependence explicit.

For $k\geq1$, define the augmented error state as
\begin{equation}
\mathbf e^{\mathrm a}(k)
\triangleq
\begin{bmatrix}
\mathbf e(k)\\
\mathbf e(k-1)\\
\tilde{\mathbf w}(k-1)
\end{bmatrix}
\in\mathbb R^{3MN}.
\label{eq:finite_augmented_state}
\end{equation}
Then \eqref{eq:finite_master_error_dynamics} admits the first-order representation
\begin{equation}
\mathbf e^{\mathrm a}(k+1)
=\big(\mathbf F_{\mathrm a}+\mathbf U_{\mathrm a}(k)\big)
\mathbf e^{\mathrm a}(k)
+\mathbf U_x(k)\tilde{\mathbf x}(k)
+\boldsymbol\nu^{\mathrm a}(k).
\label{eq:finite_augmented_system}
\end{equation}
The deterministic augmented error-transition matrix $\mathbf F_{\mathrm a}$
combines the two-step mean-error coefficients $\mathbf F_0,\mathbf F_1$
and the stored process disturbance,
\begin{equation}
\mathbf F_{\mathrm a}
\triangleq
\begin{bmatrix}
\mathbf F_0
&\mathbf F_1
&\tilde{\mathbf M}_\beta\\
\mathbf I_{MN}&\mathbf0&\mathbf0\\
\mathbf0&\mathbf0&\mathbf0
\end{bmatrix}.
\label{eq:finite_nominal_augmented_matrix}
\end{equation}
The perturbation $\mathbf R_T(k)$ changes only the current-error equation,
so $\mathbf U_{\mathrm a}(k)\triangleq\operatorname{blkdiag}\big(
\tilde{\mathbf M}_1\mathbf R_T(k)
(\tilde{\mathbf A}-\tilde{\mathbf K}\tilde{\mathbf C}),
\mathbf0_{2MN\times2MN}\big)$.
The pilot-induced plant input enters the same block through
$\mathbf U_x(k)\triangleq\operatorname{col}\big(
\tilde{\mathbf M}_1\mathbf Q_T(k)\tilde{\mathbf A},
\mathbf0_{2MN\times MN}\big)$.
The disturbance is
$\boldsymbol\nu^{\mathrm a}(k)\triangleq\operatorname{col}\big(
-\tilde{\mathbf w}(k)+\rho^{-1/2}\tilde{\mathbf M}_1\tilde{\mathbf n}_T(k),
\mathbf0_{MN},\tilde{\mathbf w}(k)\big)$.
Thus the same current process noise $\tilde{\mathbf w}(k)$ enters
both the error and noise-storage blocks.

The block-triangular form of $\mathbf F_{\mathrm a}$ adds only zero eigenvalues
to the two-step mean-error transition. Hence
$r(\mathbf F_{\mathrm a})<d<1$ under
Theorem~\ref{thm:pcsi_mean_convergence}.

\subsection{Second-Moment Dynamics and Steady-State Estimation MSE}
\label{subsec:finite_snr_second_moment}
Write the augmented-error second moment as
$\boldsymbol\Sigma_{\mathrm a}(k)\triangleq
\mathbb E[\mathbf e^{\mathrm a}(k)(\mathbf e^{\mathrm a}(k))^\top]$.
To evaluate it from~\eqref{eq:finite_augmented_system}, the term
$\mathbf U_x(k)\tilde{\mathbf x}(k)$ introduces products of
$\tilde{\mathbf x}(k)$ with its transpose. We therefore also keep the
plant second moment
$\boldsymbol\Sigma_{\tilde x}(k)\triangleq
\mathbb E[\tilde{\mathbf x}(k)\tilde{\mathbf x}^\top(k)]$.
Moreover, $\mathbf U_{\mathrm a}(k)$ and $\mathbf U_x(k)$ use the same
channel and pilot samples. The expectation of their mixed quadratic
term therefore depends on the error--plant cross moment
$\boldsymbol\Sigma_{\mathrm a\tilde x}(k)\triangleq
\mathbb E[\mathbf e^{\mathrm a}(k)\tilde{\mathbf x}^\top(k)]$,
which records the correlation between the augmented error and plant state.

The disturbance contributions are described by
$\boldsymbol\Sigma_{\tilde w}\triangleq
\mathbb E[\tilde{\mathbf w}(k)\tilde{\mathbf w}^\top(k)]$ and
$\boldsymbol\Sigma_{\nu^{\mathrm a}}\triangleq
\mathbb E[\boldsymbol\nu^{\mathrm a}(k)(\boldsymbol\nu^{\mathrm a}(k))^\top]$.
The current process noise $\tilde{\mathbf w}(k)$ appears in the plant
recursion and in the first and last blocks of
$\boldsymbol\nu^{\mathrm a}(k)$. Their shared contribution is therefore
$\boldsymbol\Sigma_{\nu^{\mathrm a}\tilde w}\triangleq
\mathbb E[\boldsymbol\nu^{\mathrm a}(k)\tilde{\mathbf w}^\top(k)]
=\operatorname{col}\big(-\boldsymbol\Sigma_{\tilde w},
\mathbf0_{MN\times MN},\boldsymbol\Sigma_{\tilde w}\big)$.

The quadratic error, plant, and mixed products in
\eqref{eq:finite_augmented_system} are collected in the linear maps
\begin{equation}
\begin{aligned}
\mathcal H_{\rho,N_r}(\mathbf Z)
\triangleq{}&
\mathbf F_{\mathrm a}\mathbf Z\mathbf F_{\mathrm a}^\top
+\mathbb E[\mathbf U_{\mathrm a}(k)
\mathbf Z\mathbf U_{\mathrm a}^\top(k)],\\
\mathcal P_{\rho,N_r}(\mathbf Z)
\triangleq{}&
\mathbb E[\mathbf U_x(k)
\mathbf Z\mathbf U_x^\top(k)],\\
\mathcal C_{\rho,N_r}(\mathbf Z)
\triangleq{}&
\mathbb E[\mathbf U_{\mathrm a}(k)
\mathbf Z\mathbf U_x^\top(k)]+\mathbb E[\mathbf U_x(k)
\mathbf Z^\top\mathbf U_{\mathrm a}^\top(k)].
\end{aligned}
\label{eq:finite_second_moment_operators}
\end{equation}
The arguments of $\mathcal H_{\rho,N_r}$, $\mathcal P_{\rho,N_r}$,
and $\mathcal C_{\rho,N_r}$ have sizes $3MN\times3MN$,
$MN\times MN$, and $3MN\times MN$, respectively.
These maps propagate the augmented-error second moment, the plant
second moment, and their cross moment into the next error second moment.
In particular, $\mathcal C_{\rho,N_r}$ retains the mixed contribution
because the two random coefficients need not be independent.
Independence from $\mathbf e^{\mathrm a}(k),\tilde{\mathbf x}(k)$
and zero means eliminate terms linear in either coefficient, but not
the product of $\mathbf U_{\mathrm a}(k)$ and $\mathbf U_x(k)$.
The matrix representation of
$\mathcal H_{\rho,N_r}$ under vectorization is
\begin{equation}
\mathbf\Xi(\rho,N_r)
\triangleq
\mathbf F_{\mathrm a}\otimes\mathbf F_{\mathrm a}
+\mathbb E[\mathbf U_{\mathrm a}(k)
\otimes\mathbf U_{\mathrm a}(k)],
\label{eq:finite_covariance_transition_matrix}
\end{equation}
so that $\operatorname{vec}(\mathcal H_{\rho,N_r}(\mathbf Z))
=\mathbf\Xi(\rho,N_r)\operatorname{vec}(\mathbf Z)$.
Removing the plant input $\mathbf U_x(k)\tilde{\mathbf x}(k)$
and disturbance $\boldsymbol\nu^{\mathrm a}(k)$ from
\eqref{eq:finite_augmented_system} gives the homogeneous response
$\mathbf h(k+1)=[\mathbf F_{\mathrm a}+\mathbf U_{\mathrm a}(k)]\mathbf h(k)$.
Its second moment $\mathbf Z(k)\triangleq\mathbb E[\mathbf h(k)\mathbf h^\top(k)]$
satisfies $\mathbf Z(k+1)=\mathcal H_{\rho,N_r}(\mathbf Z(k))$.
Write $\mathcal I$ for the identity operator on real
$3MN\times3MN$ matrices. Powers of $\mathcal H_{\rho,N_r}$ denote
repeated composition, with $\mathcal H_{\rho,N_r}^{0}=\mathcal I$.
Together with the plant recursion, these maps give a closed evolution
for $\boldsymbol\Sigma_{\tilde x}(k)$,
$\boldsymbol\Sigma_{\mathrm a\tilde x}(k)$, and
$\boldsymbol\Sigma_{\mathrm a}(k)$.
The next theorem states these recursions and obtains the state
estimation MSE from the leading block of
$\boldsymbol\Sigma_{\mathrm a}(k)$.

\begin{theorem}[Exact Evolution of the Estimation-Error Second Moments]
\label{thm:finite_second_moment}
Under Assumptions~\ref{as:time_scales}--\ref{as:statistical_independence},
the moments of~\eqref{eq:finite_augmented_system} satisfy,
for $k\geq1$,
\begin{subequations}
\label{eq:finite_second_moment_solutions}
\begin{align}
\boldsymbol\Sigma_{\tilde x}(k+1)
&= \tilde{\mathbf A}\boldsymbol\Sigma_{\tilde x}(k)
   \tilde{\mathbf A}^{\top}
   +\boldsymbol\Sigma_{\tilde w},
\label{eq:finite_plant_second_moment_solution}\\
\boldsymbol\Sigma_{\mathrm a\tilde x}(k+1)
&= \mathbf F_{\mathrm a}
   \boldsymbol\Sigma_{\mathrm a\tilde x}(k)
   \tilde{\mathbf A}^{\top}
   +\boldsymbol\Sigma_{\nu^{\mathrm a}\tilde w},
\label{eq:finite_cross_second_moment_solution}\\
\boldsymbol\Sigma_{\mathrm a}(k+1)
&= \mathcal H_{\rho,N_r}(\boldsymbol\Sigma_{\mathrm a}(k))
   +\boldsymbol\Sigma_{\nu^{\mathrm a}}
\nonumber\\
&\quad
   +\mathcal P_{\rho,N_r}(\boldsymbol\Sigma_{\tilde x}(k))
   +\mathcal C_{\rho,N_r}
   (\boldsymbol\Sigma_{\mathrm a\tilde x}(k)).
\label{eq:finite_augmented_second_moment_solution}
\end{align}
\end{subequations}
Consequently, the global state estimation MSE is
\begin{equation}
\mathbb E[\|\mathbf e(k)\|^2]
=\operatorname{tr}[(\boldsymbol\Sigma_{\mathrm a}(k))_{11}],
\label{eq:finite_tracking_mse}
\end{equation}
where $(\boldsymbol\Sigma_{\mathrm a}(k))_{11}$ is the leading
$MN\times MN$ diagonal block.
\end{theorem}

\begin{proof}
Please see~\ref{app:proof_second_moment}.
\end{proof}

Because $r(\tilde{\mathbf A})<1$ and
$r(\mathbf F_{\mathrm a})<d<1$ under
Theorem~\ref{thm:pcsi_mean_convergence}, the sequences in
\eqref{eq:finite_plant_second_moment_solution} and
\eqref{eq:finite_cross_second_moment_solution} converge. Denote their
limits by
\begin{equation}
\boldsymbol\Sigma_{\tilde x}^\star
\triangleq\sum_{j=0}^{\infty}
\tilde{\mathbf A}^{j}\boldsymbol\Sigma_{\tilde w}(\tilde{\mathbf A}^{\top})^{j},
\quad
\boldsymbol\Sigma_{\mathrm a\tilde x}^\star
\triangleq\sum_{j=0}^{\infty}
\mathbf F_{\mathrm a}^{j}\boldsymbol\Sigma_{\nu^{\mathrm a}\tilde w}(\tilde{\mathbf A}^{\top})^{j}.
\label{eq:finite_stationary_plant_cross_moments}
\end{equation}
Convergence of the augmented-error second moment is then ensured by
stability of the homogeneous second-moment operator
$\mathcal H_{\rho,N_r}$, equivalently
$r(\mathbf\Xi(\rho,N_r))<1$.

\begin{corollary}[Mean-Square Boundedness and Steady-State MSE]
\label{cor:finite_steady_state_mse}
Under the conditions of Theorems~\ref{thm:pcsi_mean_convergence} and
\ref{thm:finite_second_moment}, if
\begin{equation}
r(\mathbf\Xi(\rho,N_r))<1,
\label{eq:finite_mean_square_stability_condition}
\end{equation}
then $\mathbf e^{\mathrm a}(k)$ in~\eqref{eq:finite_augmented_system}
is mean-square bounded, and $\boldsymbol\Sigma_{\mathrm a}(k)$
converges to
\begin{equation}
\boldsymbol\Sigma_{\mathrm a}^\star
 =(\mathcal I-\mathcal H_{\rho,N_r})^{-1}\big[
\boldsymbol\Sigma_{\nu^{\mathrm a}}
+\mathcal P_{\rho,N_r}(\boldsymbol\Sigma_{\tilde x}^\star)
+\mathcal C_{\rho,N_r}(\boldsymbol\Sigma_{\mathrm a\tilde x}^\star)\big].
\label{eq:finite_steady_state_augmented_covariance}
\end{equation}
Consequently, the steady-state MSE is
\begin{equation}
\lim_{k\to\infty}\mathbb E[\|\mathbf e(k)\|^2]
=\operatorname{tr}[(\boldsymbol\Sigma_{\mathrm a}^\star)_{11}].
\label{eq:finite_steady_state_tracking_mse}
\end{equation}
\end{corollary}

\begin{proof}
Please see~\ref{app:proof_steady_state_mse}.

\end{proof}

To describe departures from stationarity, use
$\mathbf D_{\mathrm a}(k)\triangleq\boldsymbol\Sigma_{\mathrm a}(k)-\boldsymbol\Sigma_{\mathrm a}^\star$
for the augmented-error moment,
$\mathbf D_x(k)\triangleq\boldsymbol\Sigma_{\tilde x}(k)-\boldsymbol\Sigma_{\tilde x}^\star$
for the plant moment, and
$\mathbf D_{\mathrm a x}(k)\triangleq\boldsymbol\Sigma_{\mathrm a\tilde x}(k)-\boldsymbol\Sigma_{\mathrm a\tilde x}^\star$
for their cross moment.

\begin{theorem}[Exact Decomposition of the MSE Deviation]
\label{thm:finite_exact_centered_mse_decomposition}
Under Theorems~\ref{thm:pcsi_mean_convergence} and
\ref{thm:finite_second_moment} and condition
\eqref{eq:finite_mean_square_stability_condition}, the state
estimation MSE deviation satisfies, for $k\geq1$,
\begin{equation}
\mathbb E[\|\mathbf e(k)\|^2]
-\operatorname{tr}[(\boldsymbol\Sigma_{\mathrm a}^\star)_{11}]
=H(k)+P(k)+C(k),
\label{eq:finite_exact_centered_mse_decomposition}
\end{equation}
where the initial-error, plant-moment, and cross-moment responses are
\begin{equation}
\begin{aligned}
H(k)&\triangleq
\operatorname{tr}\!\left[
\bigl(\mathcal H_{\rho,N_r}^{k-1}(\mathbf D_{\mathrm a}(1))\bigr)_{11}
\right],\\
P(k)&\triangleq
\operatorname{tr}\!\left[\left(
\sum_{t=1}^{k-1}\mathcal H_{\rho,N_r}^{k-1-t}
\bigl(\mathcal P_{\rho,N_r}(\mathbf D_x(t))\bigr)
\right)_{11}\right],\\
C(k)&\triangleq
\operatorname{tr}\!\left[\left(
\sum_{t=1}^{k-1}\mathcal H_{\rho,N_r}^{k-1-t}
\bigl(\mathcal C_{\rho,N_r}(\mathbf D_{\mathrm a x}(t))\bigr)
\right)_{11}\right].
\end{aligned}
\label{eq:finite_exact_mse_components}
\end{equation}
\end{theorem}
\begin{proof}
Please see
\ref{app:proof_exact_centered_mse_decomposition}.

\end{proof}
The scalar $H(k)$ is the response of the initial error-moment
deviation under $\mathcal H_{\rho,N_r}$. The terms $P(k)$ and $C(k)$
are the responses to deviations of
$\boldsymbol\Sigma_{\tilde x}(k)$ and
$\boldsymbol\Sigma_{\mathrm a\tilde x}(k)$ from stationarity.
Thus~\eqref{eq:finite_exact_centered_mse_decomposition} separates
the estimator's own transient from the plant statistics entering
through pilot-induced coupling.

\subsection{SNR-Dependent Transient and Steady-State Performance}
\label{subsec:finite_snr_transient_contributions}
We first bound the decay rates of $H(k),P(k),C(k)$, then
quantify how the SNR gain $\rho$ changes their magnitudes.
The steady-state analysis subsequently varies both $\rho$ and
the process-noise covariance $\mathbf W_w$.

\begin{corollary}[Convergence Rates of the MSE Deviation Terms]
\label{cor:finite_componentwise_transient_rates}
Under Assumptions~\ref{as:time_scales}--\ref{as:statistical_independence},
Theorem~\ref{thm:pcsi_mean_convergence}, and
\eqref{eq:finite_mean_square_stability_condition},
\begin{equation}
\begin{aligned}
\limsup_{k\to\infty}|H(k)|^{1/(2k)}
&\leq\sqrt{r(\mathbf\Xi(\rho,N_r))},\\
\limsup_{k\to\infty}|P(k)|^{1/(2k)}
&\leq\max\!\left\{
\sqrt{r(\mathbf\Xi(\rho,N_r))},r(\tilde{\mathbf A})\right\},\\
\limsup_{k\to\infty}|C(k)|^{1/(2k)}
&\leq\max\!\left\{
\sqrt{r(\mathbf\Xi(\rho,N_r))},
\sqrt{r(\mathbf F_{\mathrm a})r(\tilde{\mathbf A})}\right\}.
\end{aligned}
\label{eq:finite_componentwise_root_mse_rates}
\end{equation}
Consequently,
\begin{equation}
\begin{aligned}
&\limsup_{k\to\infty}
\left|
\mathbb E[\|\mathbf e(k)\|^2]
-\operatorname{tr}[(\boldsymbol\Sigma_{\mathrm a}^\star)_{11}]
\right|^{1/(2k)}\\
&\leq
\max\!\left\{
\sqrt{r(\mathbf\Xi(\rho,N_r))},r(\tilde{\mathbf A}),
\sqrt{r(\mathbf F_{\mathrm a})r(\tilde{\mathbf A})}\right\}.
\end{aligned}
\label{eq:finite_root_mse_convergence_rate}
\end{equation}
\end{corollary}

\begin{proof}
Please see~\ref{app:proof_componentwise_transient_rates}.

\end{proof}

Corollary~\ref{cor:finite_componentwise_transient_rates} distinguishes
the estimator's second-moment decay $\sqrt{r(\mathbf\Xi(\rho,N_r))}$
from the plant decay $r(\tilde{\mathbf A})$. A small but nonzero
plant input may therefore affect the long-term MSE rate.
To quantify its magnitude, hold $N_r,T$ fixed and increase $\rho$.
In~\eqref{eq:finite_pilot_noise_perturbation}, every block of
$\boldsymbol\Delta_N^\tau(k)$ contains a factor $\rho^{-1/2}$
or $\rho^{-1}$, and the data noise in
\eqref{eq:finite_fusion_error_decomposition} has the factor
$\rho^{-1/2}$. These terms vanish in mean square, whereas
$\boldsymbol\Delta^\tau(k)$ does not depend on $\rho$.
Consequently, pilot-based CSI estimation errors and receiver data
noise disappear at high SNR, while fading remains at fixed $N_r$.

Set $\boldsymbol\Delta_N^\tau(k)=\mathbf0$ in
\eqref{eq:finite_global_internal_error} and retain the same
channel realizations. Let $\mathbf U_{\mathrm a,\infty}(k)$
be the resulting limit of $\mathbf U_{\mathrm a}(k)$.
For the homogeneous error system
$\mathbf h(k+1)=[\mathbf F_{\mathrm a}+
\mathbf U_{\mathrm a,\infty}(k)]\mathbf h(k)$,
the second-moment operator maps $\mathbf Z$ to
$\mathbb E[(\mathbf F_{\mathrm a}+\mathbf U_{\mathrm a,\infty}(k))
\mathbf Z(\mathbf F_{\mathrm a}+\mathbf U_{\mathrm a,\infty}(k))^\top]$.
Its matrix under vectorization is
\begin{equation}
\mathbf\Xi_\infty(N_r)
\triangleq{}
\mathbf F_{\mathrm a}\otimes\mathbf F_{\mathrm a}
+\mathbb E[\mathbf U_{\mathrm a,\infty}(k)
\otimes\mathbf U_{\mathrm a,\infty}(k)].
\label{eq:finite_pcsi_second_moment_transition}
\end{equation}
The finite recursions
\eqref{eq:finite_graph_filter_perturbation}--\eqref{eq:finite_accumulated_data_noise}
pass these communication-cycle scalings to the terminal coefficients.
Every term in $\mathbf Q_T(k)$ contains a pilot-error factor.
Consequently,
$\lim_{\rho\to\infty}\mathbf\Xi(\rho,N_r)=\mathbf\Xi_\infty(N_r)$.
For $\mathbf Z$ of the corresponding dimensions in
\eqref{eq:finite_second_moment_operators}, the bounds
$\|\mathcal P_{\rho,N_r}(\mathbf Z)\|_{\mathrm F}
=\mathcal O(\rho^{-1})\|\mathbf Z\|_{\mathrm F}$ and
$\|\mathcal C_{\rho,N_r}(\mathbf Z)\|_{\mathrm F}
=\mathcal O(\rho^{-1/2})\|\mathbf Z\|_{\mathrm F}$
hold as $\rho\to\infty$.

The mean-square stability condition for the high-SNR limiting
homogeneous system is
\begin{equation}
r(\mathbf\Xi_\infty(N_r))<1.
\label{eq:finite_pcsi_mean_square_stability}
\end{equation}
Since $\mathbf\Xi(\rho,N_r)\to\mathbf\Xi_\infty(N_r)$ as
$\rho\to\infty$, continuity of the spectral radius implies that
\eqref{eq:finite_pcsi_mean_square_stability} ensures
$r(\mathbf\Xi(\rho,N_r))<1$ for all sufficiently large $\rho$.
If $r(\mathbf\Xi_\infty(N_r))<d^2$, the limiting homogeneous error
$\mathbf h(k)$ decays in root mean square faster than $d^k$.
The input from the plant can still retain the rate
$r(\tilde{\mathbf A})$ identified in
Corollary~\ref{cor:finite_componentwise_transient_rates}.
The next bound combines these decay rates with the
$\mathcal O(\rho^{-1})$ plant-moment input and
$\mathcal O(\rho^{-1/2})$ cross-moment input.

\begin{corollary}[Uniform High-SNR Bound on the MSE Deviation]
\label{cor:finite_snr_transient_envelope}
Under the conditions of Theorems~\ref{thm:pcsi_mean_convergence} and
\ref{thm:finite_second_moment}, suppose that
\eqref{eq:finite_pcsi_mean_square_stability} holds.
Choose a fixed estimator decay bound $\gamma_{\mathrm e}$ and
plant decay bound $\gamma_{\mathrm x}$ satisfying
\begin{equation}
\begin{gathered}
\max\{d,\sqrt{r(\mathbf\Xi_\infty(N_r))}\}<\gamma_{\mathrm e}<1,\\
r(\tilde{\mathbf A})<\gamma_{\mathrm x}<1,\qquad \gamma_{\mathrm x}\ne \gamma_{\mathrm e}.
\end{gathered}
\label{eq:finite_snr_transient_auxiliary_rates}
\end{equation}
Then there exist finite constants
$\rho_0,C_{\mathrm b},C_{\mathrm x},C_{\mathrm c}>0$, independent of
$k$ and $\rho$, such that
for every $\rho\geq\rho_0$ and $k\geq1$,
\begin{equation}
\begin{aligned}
\left|
\mathbb E[\|\mathbf e(k)\|^2]
-\operatorname{tr}[(\boldsymbol\Sigma_{\mathrm a}^\star)_{11}]
\right|
\leq{}&
C_{\mathrm b} \gamma_{\mathrm e}^{2k}
+\frac{C_{\mathrm x}}{\rho}\gamma_{\mathrm x}^{2k}
+\frac{C_{\mathrm c}}{\sqrt\rho}(\gamma_{\mathrm e}\gamma_{\mathrm x})^k.
\end{aligned}
\label{eq:finite_snr_transient_bound}
\end{equation}
\end{corollary}

\begin{proof}
Please see~\ref{app:proof_snr_transient_envelope}.

\end{proof}

\begin{remark}[Effect of SNR on the MSE Transient]
\label{rem:finite_snr_transient_contributions}
In~\eqref{eq:finite_snr_transient_bound}, the three bounds decay as
$\gamma_{\mathrm e}^{2k}$, $\gamma_{\mathrm x}^{2k}$, and $(\gamma_{\mathrm e}\gamma_{\mathrm x})^k$.
When $\gamma_{\mathrm x}>\gamma_{\mathrm e}$, the plant-moment bound has the slowest decay,
while the cross-moment rate $\sqrt{\gamma_{\mathrm e}\gamma_{\mathrm x}}$ lies between the two.
Increasing $\rho$ reduces the latter two amplitudes by
$\rho^{-1}$ and $\rho^{-1/2}$, respectively. It does not change
$\gamma_{\mathrm x}$ at fixed $N_r$.
If $r(\tilde{\mathbf A})<d$ and
$r(\mathbf\Xi(\rho,N_r))<d^2$, Corollary~\ref{cor:finite_componentwise_transient_rates}
gives the strict prescribed rate in Definition~\ref{def:converge_speed}.
At fixed $N_r$, increasing $\rho$ alone need not place
$r(\mathbf\Xi_\infty(N_r))$ below $d^2$.
\end{remark}

We now turn from the MSE transient to its platform.
Write $\mathbf W_w=\alpha\mathbf W_0$, where
$\mathbf W_0\succeq\mathbf0$ is fixed and $\alpha\geq0$ measures
the process-noise magnitude. With the segment partition, gains, fusion weights, topology,
$T$, and $\mathbf X_p$ fixed, display the platform's dependence on
$\alpha$ and the SNR gain $\rho$ as
\begin{equation}
\mathcal M_\star(\alpha,\rho;N_r)
\triangleq\lim_{k\to\infty}\mathbb E[\|\mathbf e(k)\|^2],
\label{eq:ssmse:metric}
\end{equation}
whenever the limit exists. Process noise $\mathbf w(k)$ and receiver
data noise $\boldsymbol\eta_i^\tau(k)$ are independent, so their
stationary second-moment contributions add. The former is proportional
to $\alpha$, whereas the latter retains the factor $1/\rho$ from
receiver data noise.

\begin{theorem}[Steady-State MSE Dependence]\label{thm:ssmse}
Under Theorem~\ref{thm:pcsi_mean_convergence} and
\eqref{eq:finite_pcsi_mean_square_stability}, fix $N_r$. There exist
$\rho_0\geq1$ and $C_w,C_n\geq0$, independent of $\alpha,\rho$,
such that, for $\alpha\geq0$ and $\rho\geq\rho_0$, the following hold.

(i) The limit in \eqref{eq:ssmse:metric} exists and is finite. It satisfies
\begin{equation}\label{eq:ssmse:affine}
 \mathcal M_\star(\alpha,\rho;N_r)=\alpha a(\rho,N_r)+b(\rho,N_r),
\end{equation}
where $a(\rho,N_r)$ and $b(\rho,N_r)$ are finite and nonnegative. The steady-state MSE is nondecreasing in $\alpha$ and nonincreasing in $\rho$. 

(ii) The uniform bound
\begin{equation}\label{eq:ssmse:bound}
 0\le\mathcal M_\star(\alpha,\rho;N_r)\le C_w\alpha+\frac{C_n}{\rho}
\end{equation}
holds. Moreover, there exist finite $a_\infty(N_r),c_\infty(N_r)\ge0$ such that
\begin{equation}\label{eq:ssmse:limits}
 \lim_{\rho\to\infty}a(\rho,N_r)=a_\infty(N_r),\quad
 \lim_{\rho\to\infty}\rho b(\rho,N_r)=c_\infty(N_r).
\end{equation}
In the absence of process noise,
\begin{equation}\label{eq:ssmse:noiseless}
 \mathcal M_\star(0,\rho;N_r)=\frac{c_\infty(N_r)}{\rho}+o(\rho^{-1}).
\end{equation}
(iii) As the process-noise magnitude tends to zero and the
SNR gain tends to infinity,
\begin{equation}\label{eq:ssmse:jointlimit}
\lim_{\substack{\alpha\to0^+\\\rho\to\infty}}
\mathcal M_\star(\alpha,\rho;N_r)=0.
\end{equation}
For each fixed $\alpha$, the high-SNR limit is $\alpha a_\infty(N_r)$. 
\end{theorem}
\begin{proof}
Please see~\ref{app:ssmse}.

\end{proof}
The steady-state limit in Theorem~\ref{thm:ssmse} is independent of the initial second moments. A fixed receive-array size is sufficient for a vanishing steady-state MSE under~\eqref{eq:finite_pcsi_mean_square_stability} as $\alpha\to0$ and $\rho\to\infty$. The fading fluctuations remain in $\mathbf\Xi_\infty(N_r)$ and affect the constants in the bound. The transient convergence behavior continues to be characterized by Corollary~\ref{cor:finite_snr_transient_envelope}.

\subsection{Receive-Array Size and the Ideal Communication Limit}
\label{subsec:finite_receive_array_hardening}
A high SNR reduces pilot and receiver noise, but leaves fading in
$\mathbf\Xi_\infty(N_r)$. Increasing $N_r$ instead averages the
channel products themselves. We now fix $\rho>0$
and $T$ and follow this averaging from the slot perturbations to
the MSE platform and convergence rate.

Each normalized channel product in \eqref{eq:weighted_laplacian} and
\eqref{eq:finite_pilot_noise_perturbation} is an average of $N_r$
independent receive-antenna contributions. The bias-compensation term in
\eqref{eq:finite_pilot_noise_perturbation} removes the mean of the
pilot-noise Gram matrix. Consequently, at every communication cycle
$\tau$,
\begin{equation}
\mathbb E\!\left[\|\boldsymbol\Delta^\tau(k)\|_{\mathrm F}^2
+\|\boldsymbol\Delta_N^\tau(k)\|_{\mathrm F}^2
+\|\mathbf n^\tau(k)\|^2\right]=\mathcal O(N_r^{-1}).
\label{eq:finite_receive_array_primitive_scaling}
\end{equation}
The constants hidden in \eqref{eq:finite_receive_array_primitive_scaling}
may depend on $\rho$ and the fixed system parameters but not on $N_r$.
Since $T$ is finite, every term in
$\mathbf R_T(k)$ contains at least one
$\boldsymbol\Delta^\tau(k)$ or
$\boldsymbol\Delta_N^\tau(k)$, while every term in
$\mathbf Q_T(k)$ contains at least one
$\boldsymbol\Delta_N^\tau(k)$. Similarly, each term in the accumulated
noise $\tilde{\mathbf n}_T(k)$ contains a factor
$\mathbf n^\tau(k)$. Slot independence passes the bounds in
\eqref{eq:finite_receive_array_primitive_scaling} through these products.
The required fourth moments are finite under the stated Gaussian
model. Therefore
\begin{equation}
\mathbb E\!\left[\|\mathbf U_{\mathrm a}(k)\|_{\mathrm F}^2
+\|\mathbf U_x(k)\|_{\mathrm F}^2
+\|\tilde{\mathbf n}_T(k)\|^2\right]=\mathcal O(N_r^{-1}).
\label{eq:finite_receive_array_terminal_scaling}
\end{equation}
By \eqref{eq:finite_covariance_transition_matrix},
\eqref{eq:finite_receive_array_terminal_scaling} gives
\begin{equation}
\left\|\mathbf\Xi(\rho,N_r)
-\mathbf F_{\mathrm a}\otimes\mathbf F_{\mathrm a}\right\|_{\mathrm F}
=\mathcal O(N_r^{-1}).
\label{eq:finite_receive_array_operator_scaling}
\end{equation}
Thus, the matrix difference
$\mathbf\Xi(\rho,N_r)-\mathbf F_{\mathrm a}\otimes\mathbf F_{\mathrm a}$ decays at the rate
$1/N_r$. By continuity of the spectral radius and
Theorem~\ref{thm:pcsi_mean_convergence},
\begin{equation}
\lim_{N_r\to\infty}\sqrt{r(\mathbf\Xi(\rho,N_r))}=r(\mathbf F_{\mathrm a})<d.
\label{eq:finite_receive_array_spectral_limit}
\end{equation}
Therefore $r(\mathbf\Xi(\rho,N_r))<d^2<1$ for all sufficiently
large $N_r$. A sufficiently large receive array thus ensures
mean-square boundedness and a homogeneous root-MSE factor
$\sqrt{r(\mathbf\Xi(\rho,N_r))}$ strictly smaller than $d$.
Equation~\eqref{eq:finite_receive_array_terminal_scaling} also shows
that $\mathbf U_x(k)$ and
$\tilde{\mathbf n}_T(k)$ vanish in mean square as $N_r\to\infty$.
Accordingly, $\boldsymbol\nu^{\mathrm a}(k)$ in
\eqref{eq:finite_augmented_system} converges in mean square to
$\boldsymbol\nu_{\mathrm w}^{\mathrm a}(k)$, and its second moment converges to
$\boldsymbol\Sigma_{\nu_{\mathrm w}}$, where
\begin{equation}
\begin{aligned}
\boldsymbol\nu^{\mathrm a}(k)
&\xrightarrow[N_r\to\infty]{L^2}
\boldsymbol\nu_{\mathrm w}^{\mathrm a}(k)
\triangleq
\begin{bmatrix}
-\tilde{\mathbf w}(k)\\
\mathbf0_{MN}\\
\tilde{\mathbf w}(k)
\end{bmatrix},\\
\boldsymbol\Sigma_{\nu_{\mathrm w}}
&\triangleq
\mathbb E\!\left[
\boldsymbol\nu_{\mathrm w}^{\mathrm a}(k)
(\boldsymbol\nu_{\mathrm w}^{\mathrm a}(k))^\top
\right].
\end{aligned}
\label{eq:finite_process_only_augmented_disturbance}
\end{equation}
Combining
\eqref{eq:finite_receive_array_terminal_scaling} with the uniform
stability for all sufficiently large $N_r$ and the fixed-point formula
\eqref{eq:finite_steady_state_augmented_covariance} gives
\begin{equation}
\lim_{N_r\to\infty}\boldsymbol\Sigma_{\mathrm a}^\star
=\sum_{j=0}^{\infty}
\mathbf F_{\mathrm a}^{j}\boldsymbol\Sigma_{\nu_{\mathrm w}}
(\mathbf F_{\mathrm a}^{\top})^{j}.
\label{eq:finite_receive_array_stationary_limit}
\end{equation}
Thus, all communication-induced contributions to
$\operatorname{tr}[(\boldsymbol\Sigma_{\mathrm a}^\star)_{11}]$ vanish, and the
limiting value of $\operatorname{tr}[(\boldsymbol\Sigma_{\mathrm a}^\star)_{11}]$
is determined solely by $\boldsymbol\Sigma_{\nu_{\mathrm w}}$,
equivalently by $\boldsymbol\Sigma_{\tilde w}$.

\begin{corollary}[Prescribed MSE Convergence in the Ideal Communication Limit]
\label{cor:finite_joint_ideal_mse_convergence}
Under the design conditions of
Theorem~\ref{thm:pcsi_mean_convergence}, consider the joint
ideal-communication limit in which $\rho\to\infty$ and
$N_r\to\infty$. The limiting augmented-error system, whose homogeneous
second-moment transition matrix is $\mathbf F_{\mathrm a}\otimes\mathbf F_{\mathrm a}$, has mean-square-stable homogeneous dynamics, and its state estimation MSE deviation satisfies
\begin{equation}
\limsup_{k\to\infty}
\left|
\mathbb E[\|\mathbf e(k)\|^2]
-\lim_{N_r\to\infty}\operatorname{tr}[(\boldsymbol\Sigma_{\mathrm a}^\star)_{11}]
\right|^{1/(2k)}<d.
\label{eq:finite_joint_ideal_mse_decay}
\end{equation}
\end{corollary}

\begin{proof}
Please see~\ref{app:proof_ideal_mse_convergence}.

\end{proof}
If the process noise is absent, then
$\boldsymbol\Sigma_{\tilde w}=\boldsymbol\Sigma_{\nu_{\mathrm w}}=\mathbf0$.
The limit in \eqref{eq:finite_receive_array_stationary_limit} is then
zero, and the limiting state estimation MSE itself decays as
$o(d^{2k})$.

\section{Numerical Results}
\label{sec:numerical_results}

This section evaluates the proposed distributed state estimator
over wireless MIMO fading channels.
The first experiment validates the transient and steady-state
estimation-MSE predictions for the proposed protocol and compares its
performance with representative centralized and distributed estimators.
The second experiment quantifies the receive-array gain predicted by the
channel-hardening analysis. The third experiment evaluates the SNR-dependent
onset of persistent plant-state dominance in the exact centered-MSE transient.

\subsection{Simulation Setup and Performance Metric}
\label{subsec:numerical_setup}

We consider a high-order discrete-time linear benchmark with $M=8$ network
nodes and state dimension $N=20$. The closed-loop matrix $\hat{\mathbf A}$ in
\eqref{eq:plant_closed_loop} has two dominant Jordan blocks associated with
the eigenvalues $0.98$ and $0.95$, while the remaining blocks have eigenvalue
magnitudes between $0.10$ and $0.60$. This separation captures a system in
which a small number of slow dynamics determine the estimation transient,
whereas the other internal dynamics are sufficiently damped. Selecting
$d=0.7$ therefore places exactly the two dominant Jordan blocks in
$\mathcal I_{\mathrm{pl}}$ and provides a nontrivial test of the
prescribed-rate design in a higher-dimensional system.
The process disturbance is Gaussian with covariance
$\mathbf W_w=10^{-2}\mathbf I_{20}$. The local sensing matrices satisfy
Assumptions~\ref{as:global_modal_source_observability} and
\ref{as:network_decoupled_local_detectability}, and the observer gains
are selected as specified in
Subsection~\ref{subsec:local_prediction}. Unless otherwise stated,
$\mathbf x(0)\sim\mathcal N(\mathbf0,\mathbf I_{20})$ and
$\hat{\mathbf x}_i(-1)=\hat{\mathbf x}_i(0)=\mathbf0$.

\begin{figure}[t]
\centering
\includegraphics[width=\columnwidth]{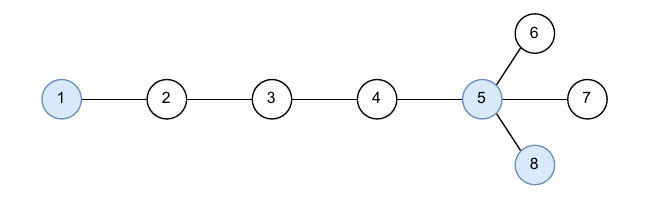}
\caption{Relay-and-branch communication topology and sensing configuration
used in the simulations. The shaded nodes $1$, $5$, and $8$ are modal-source
nodes for both performance-limiting Jordan blocks.}
\label{fig:numerical_topology}
\end{figure}

The network nodes are connected through the sparse relay-and-branch topology
shown in Fig.~\ref{fig:numerical_topology}. Nodes $1$--$5$ form a multi-hop
backbone, with node~$5$ connecting the peripheral nodes $6$--$8$. This
configuration represents a communication-constrained network in which
information from a remote source node must traverse multiple relays before
reaching the local branch. The shaded nodes $1$, $5$, and $8$ are
modal-source nodes for both performance-limiting Jordan blocks and use direct
measurements, whereas the unshaded nodes have no direct measurements. Although
the overall sensing network is globally observable, the closed neighborhood of
node~$3$, consisting of nodes $2$--$4$, contains no modal-source node for
either block and therefore provides no measurement of either
performance-limiting Jordan-coordinate component. The proposed protocol
must consequently propagate the required modal information over multiple
hops.

The wireless channels and receiver noises follow the distributions in
Section~\ref{sec:WC_Network}, with an independent channel realization in every
fast communication cycle. The common numerical parameters are listed in
Table~\ref{tab:numerical_parameters}. The proposed protocol uses $T=3$, which
satisfies the communication-order condition in
Theorem~\ref{thm:pcsi_mean_convergence}.

\begin{table}[t]
\caption{Common Numerical Parameters}
\label{tab:numerical_parameters}
\centering
\small
\begin{tabular}{lcc}
\hline
\textbf{Description}&\textbf{Symbol}&\textbf{Value}\\ \hline
Number of nodes&$M$&$8$\\
State/transmit dimension&$N=N_t$&$20$\\
Prescribed decay factor&$d$&$0.7$\\
Internal decay factor&$d_-$&$0.6993$\\
Communication cycles per control cycle&$T$&$3$\\
Process-noise covariance&$\mathbf W_w$&$10^{-2}\mathbf I_{20}$\\
Common pilot matrix&$\mathbf X_p$&$\mathbf I_{20}$\\
Default receive antennas&$N_r$&$64$\\
Default SNR gain&$10\log_{10}\rho$&$10$~dB\\ \hline
\end{tabular}
\end{table}

For each operating pair $(\rho,N_r)$, we report the network-averaged
state estimation MSE $\mathcal M(k)/M$. When
$r(\mathbf\Xi(\rho,N_r))<1$, its steady-state value is
$\mathcal M_\star/M$, with $\mathcal M_\star$ given by
\eqref{eq:finite_steady_state_tracking_mse}. The expectations in the
second-moment recursion are evaluated by independent channel samples, whereas
the trajectory curves use independent realizations of the plant disturbance,
channels, and receiver noises.

The proposed protocol is compared with the following benchmarks.
\begin{itemize}
\item \emph{Ideal centralized Kalman filter (CKF-Ideal)} receives all source
measurements through error-free links and provides the centralized performance reference.

\item \emph{Centralized Kalman filter with imperfect CSI (CKF-IPCSI)} retains
centralized processing but recovers the source measurements over fading links using estimated CSI, and serves as the centralized wireless reference.

\item \emph{Distributed Kalman filter with imperfect CSI (DKF-IPCSI)} implements the distributed information-form Kalman filter proposed in \cite{olfatisaber2007dkf}, in which each node combines local Kalman updates with information received from its neighbors over the wireless links. It is the principal comparison with a fully distributed Kalman-consensus architecture.

\item \emph{Gradient-tracking distributed state estimator with imperfect CSI
(GT-DSE-IPCSI)} implements Algorithm~1 in \cite{yang2025network}. Local observers reconstruct the observable quotient states, which are fused through gradient-tracking-based distributed optimization with $3$ communication iterations per control cycle.

\item \emph{Reduced-communication distributed Kalman filter with imperfect CSI (DKF-RC-IPCSI)} implements the filter in \cite{battilotti2025reduced}. It reduces the communication dimension by exchanging virtual outputs instead of full state estimates. To satisfy the condition of Theorem~2 in \cite{battilotti2025reduced}, it uses $5$ communication iterations per control cycle.
\end{itemize}
All methods use the same plant, sensing configuration, process disturbance,
initial-state distribution, and wireless-channel statistics. For the Monte
Carlo baselines, the empirical steady-state MSE is evaluated by averaging the
last ten samples. GT-DSE-IPCSI is treated separately because its time trajectory
does not reach a steady region over the reported horizon.

\subsection{Transient and Steady-State Estimation MSE}
\label{subsec:numerical_mse_time}

\begin{figure}[t]
\centering
\includegraphics[width=\columnwidth]{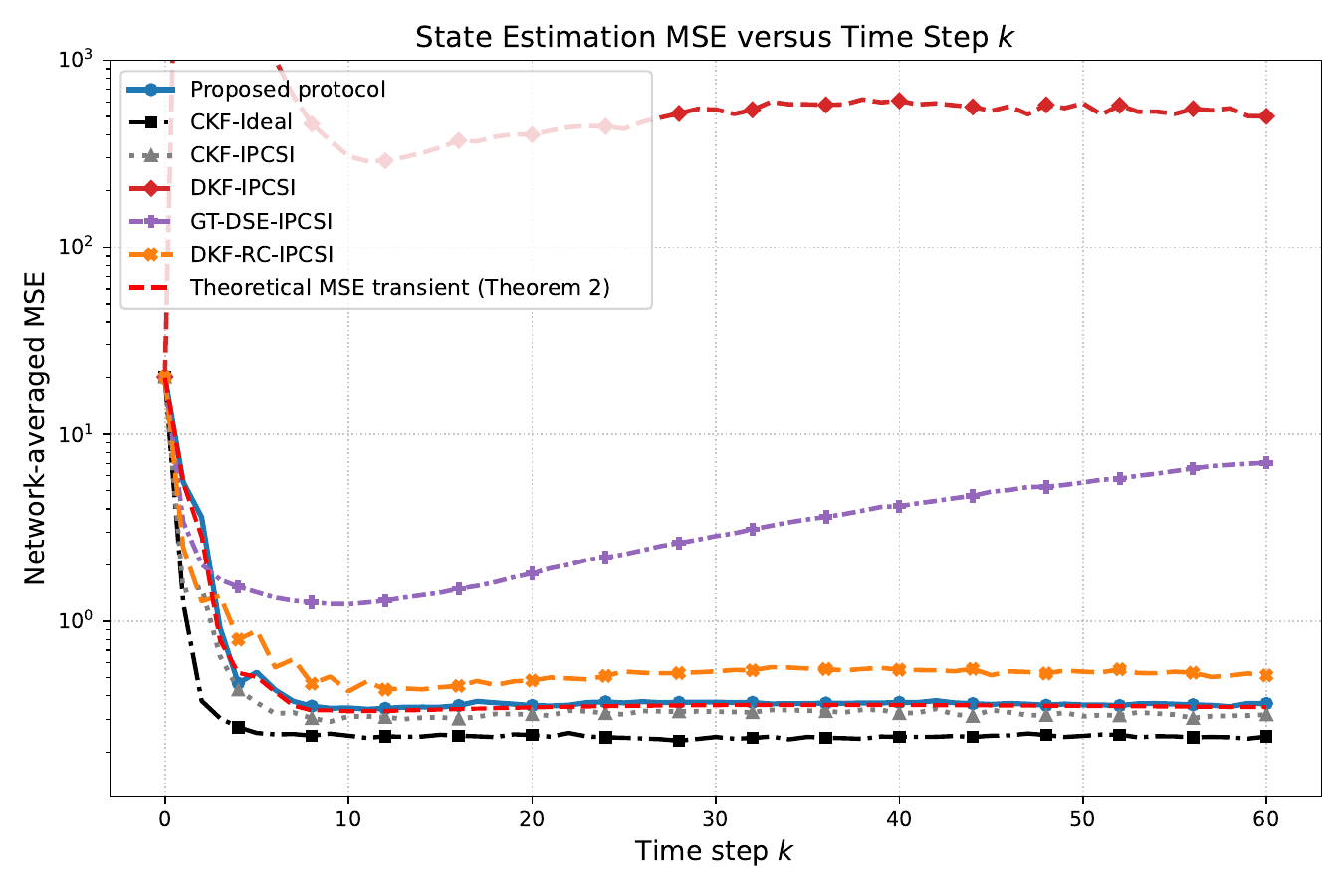}
\caption{Network-averaged state estimation MSE versus $k$ at $N_r=64$ and an SNR of
$10$~dB. Each Monte Carlo curve is averaged over $500$ independent trajectories, and the theoretical MSE transient from
Theorem~\ref{thm:finite_second_moment} is computed using $500$ independent joint channel samples. The ordinate is truncated at $10^3$ to resolve the lower
curves; DKF-IPCSI reaches $2.45\times10^5$ at $k=1$.}
\label{fig:numerical_mse_time}
\end{figure}

Figure~\ref{fig:numerical_mse_time} shows the MSE trajectories for the default
configuration in Table~\ref{tab:numerical_parameters}. The red dashed curve
represents the theoretical MSE transient $\mathcal M(k)/M$ computed from
the second-moment recursions in Theorem~\ref{thm:finite_second_moment}.
The theoretical steady-state MSE of the proposed protocol is $0.3357$, while the average over its final ten samples is $0.3406$, a relative difference of $1.5\%$.
This agreement supports the steady-state covariance in
Corollary~\ref{cor:finite_steady_state_mse} and the predicted mean-square
boundedness under $r(\mathbf\Xi(\rho,N_r))<1$.

The steady-state MSEs of CKF-Ideal, CKF-IPCSI, DKF-IPCSI, and DKF-RC-IPCSI are
$0.2420$, $0.3154$, $531.2$, and $0.5270$, respectively. The centralized
filters serve as the expected lower-MSE references. Among the distributed
baselines, DKF-RC-IPCSI performs best, and the proposed protocol achieves a $35.4\%$ lower steady-state MSE than
DKF-RC-IPCSI.

GT-DSE-IPCSI initially decreases but subsequently grows to $7.055$ at $k=60$.
Since its gradient-tracking states are designed for reliable numerical
exchange, imperfect-CSI distortions are repeatedly propagated and accumulated.
DKF-RC-IPCSI settles to a bounded empirical error floor with five communication iterations per control cycle, but
the uncompensated wireless iterations raise that floor. In contrast, the
proposed protocol
combines the $T=3$ Chebyshev recursion with mode-selective fusion, historical
compensation, and imperfect-CSI bias compensation. The observed behavior is
consistent with the stochastic error dynamics and mean-square stability
condition in Corollary~\ref{cor:finite_steady_state_mse}.

\subsection{Receive-Array Gain}
\label{subsec:numerical_mse_receive_array}

\begin{figure}[t]
\centering
\includegraphics[width=\columnwidth]{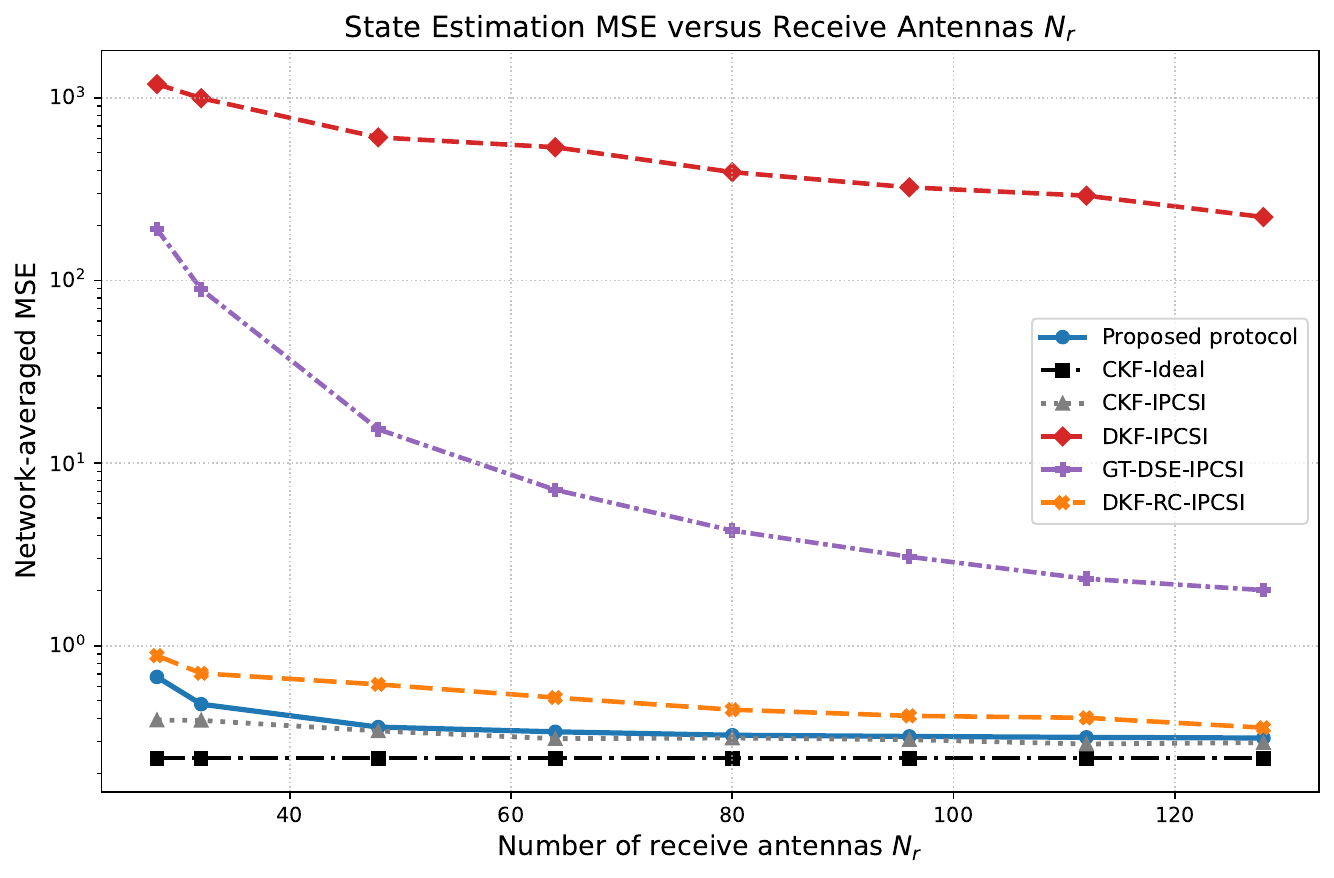}
\caption{Network-averaged state estimation MSE versus $N_r$ at an SNR of $10$~dB and
$T=3$. Each theoretical steady-state point for the proposed protocol uses
$300$ independent joint channel samples. Each Monte Carlo point averages
$300$ trajectories, while CKF-Ideal is evaluated by its deterministic
centralized Riccati recursion.}
\label{fig:numerical_mse_receive_array}
\end{figure}

We next fix the SNR at $10$~dB and set $T=3$ while varying $N_r$ from $28$ to
$128$. Figure~\ref{fig:numerical_mse_receive_array}
plots the theoretical steady-state value $\mathcal M_\star/M$ for the proposed protocol; the baseline values follow the evaluation described above.

For the proposed protocol, the mean-square stability condition is satisfied
throughout the reported antenna range: the estimated
$r(\mathbf\Xi(\rho,N_r))$ decreases from $0.8936$ at $N_r=28$ to $0.5248$ at
$N_r=128$. Its theoretical steady-state MSE correspondingly decreases from
$0.6747$ to $0.3124$, a
$53.7\%$ reduction. The corresponding Monte Carlo estimate decreases from
$0.8022$ to $0.3227$ and approaches the theoretical value as $N_r$ increases.
The observed spectral and MSE trends are consistent with
\eqref{eq:finite_receive_array_spectral_limit} and
\eqref{eq:finite_receive_array_stationary_limit}, respectively.

DKF-RC-IPCSI shows the clearest baseline array gain, decreasing from $0.8825$
to $0.3559$. At $N_r=128$, the proposed theoretical value is $12.2\%$ below
the DKF-RC-IPCSI steady-state MSE. CKF-IPCSI decreases from $0.3915$ to
$0.2949$, while CKF-Ideal remains $0.2419$. DKF-IPCSI also improves as the
receive array grows. GT-DSE-IPCSI decreases with $N_r$, although its values
represent a growing transient rather than a steady-state error floor.

\subsection{SNR-Dependent Plant-State Dominance}
\label{subsec:numerical_snr_transient}

The third experiment evaluates the exact network-averaged centered
root-MSE transient
\begin{equation}
\sqrt{\frac{|\mathcal M(k)-\mathcal M_\star|}{M}}
=\frac{|H(k)+P(k)+C(k)|^{1/2}}{\sqrt M},
\label{eq:numerical_exact_root_mse_transient}
\end{equation}
where the equality follows from
Theorem~\ref{thm:finite_exact_centered_mse_decomposition}. We set
$N_r=96$, use a displayed horizon $K=100$, and initialize
\begin{equation}
\hat{\mathbf x}_i(-1)=\hat{\mathbf x}_i(0),
\
\hat{\mathbf x}_i(0)\sim\mathcal N(\mathbf0,9\mathbf I_{20}).
\label{eq:numerical_transient_initialization}
\end{equation}
For the finite-$N_r$ operating points, the second-moment operators are evaluated
using $500$ independent channel samples for each SNR.

For the numerical classification only, set $\kappa=50$ and define the
finite-horizon onset of persistent plant-state dominance as
\begin{equation}
\begin{aligned}
k_{\mathrm P,K}^{(\kappa)}(\rho,N_r)
&\triangleq
\min\Bigl\{k_0\in\{1,\ldots,K\}:
\ \forall\,n\in\{k_0,\ldots,K\},\\[-1pt]
&\qquad
|H(n)|+|C(n)|
\le \frac{|P(n)|}{\kappa}
\Bigr\}.
\end{aligned}
\label{eq:numerical_persistent_plant_dominance}
\end{equation}
If the defining set is empty, the onset is left undefined. For every
$k\geq k_{\mathrm P,K}^{(\kappa)}$ within the displayed horizon,
\begin{equation}
\begin{aligned}
\sqrt{1-\kappa^{-1}}\sqrt{\frac{|P(k)|}{M}}
&\leq\sqrt{\frac{|\mathcal M(k)-\mathcal M_\star|}{M}}\leq\sqrt{1+\kappa^{-1}}\sqrt{\frac{|P(k)|}{M}}.
\end{aligned}
\label{eq:numerical_plant_dominance_tightness}
\end{equation}
Thus, for $\kappa=50$, the centered root-MSE transient differs from
$\sqrt{|P(k)|/M}$ by at most $1.01\%$ in relative magnitude over this
segment. The joint-ideal curve is displayed as the finite-horizon tail
maximum of this quantity over $n\in\{k,\ldots,160\}$. The coefficient $C_d$ in
the $C_d d^k$ slope reference is selected so that the reference meets
this tail maximum at $k=100$.

\begin{figure}[t]
\centering
\includegraphics[width=\columnwidth]{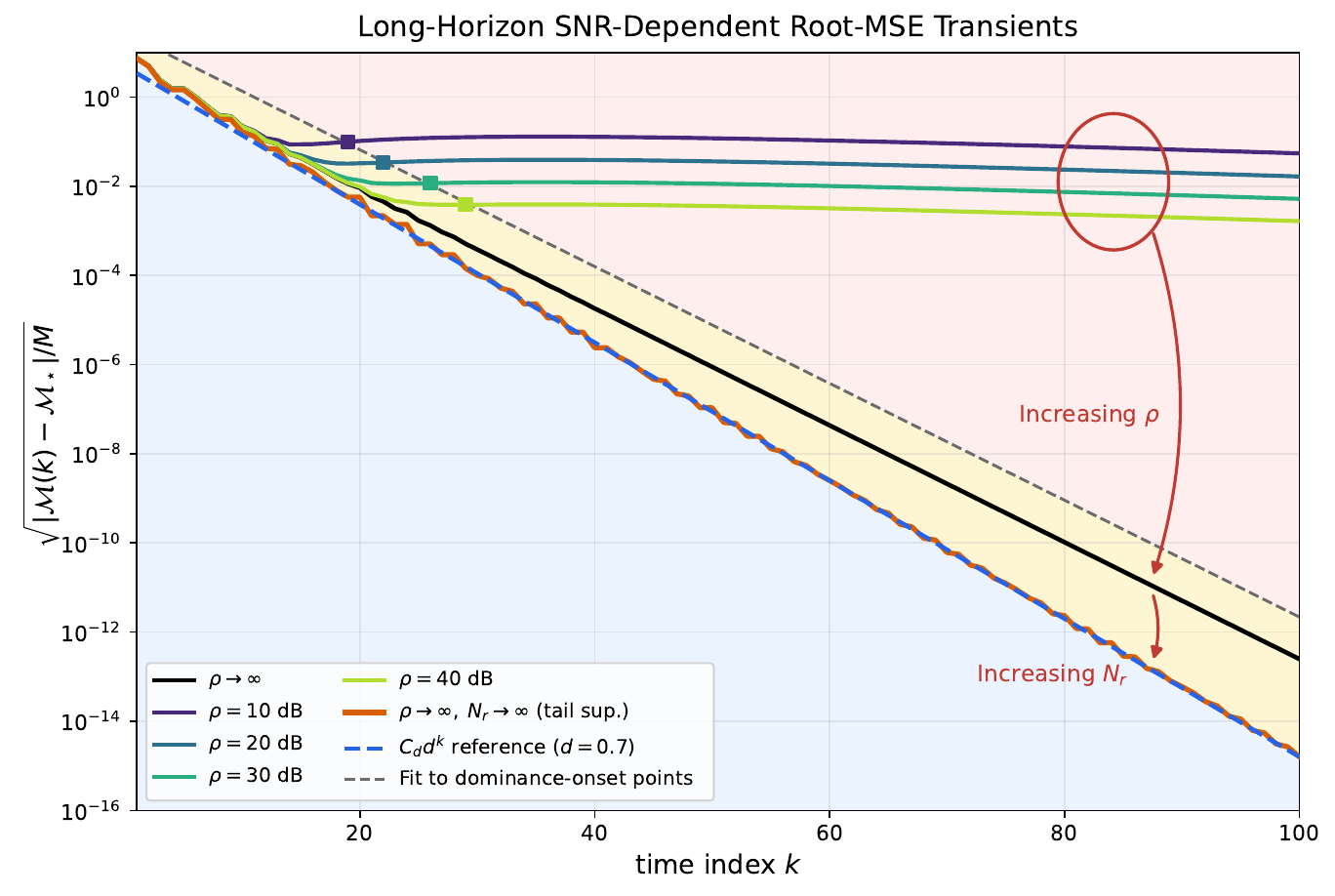}
\caption{Exact network-averaged centered root-MSE transients at $N_r=96$. Each square
marks the onset of persistent plant-state dominance for one finite-SNR curve,
using a $100$-step horizon and a dominance ratio of $50$. The gray dashed line is a spectral-rate-constrained
fit to these numerical onset points, and the background shading is a
visual guide. The joint-ideal curve is shown through its tail maximum
over $n\leq160$ and compared with a $C_d d^k$ reference matched to it
at $k=100$.}
\label{fig:numerical_snr_transient}
\end{figure}

For SNRs of $10$, $20$, $30$, and $40$~dB, the persistent-dominance
onsets are $19$, $22$, $26$, and $29$, respectively. Hence, the onset
moves to later times as $\rho$ increases, consistently with the
high-SNR scaling of the plant-state operator established in
Subsection~\ref{subsec:finite_snr_transient_contributions}. The square markers certify
the dominance condition only over the finite numerical horizon; the
fitted gray line is not a theoretical boundary.

As $\rho\to\infty$ with finite $N_r$, the pilot-noise-induced coupling
vanishes while instantaneous fading remains, yielding the limiting factor
$\sqrt{r(\mathbf\Xi_\infty(96))}=0.740$. In both idealized cases,
$P(k)=C(k)=0$, so no plant-dominance marker is assigned. Under the joint
limit $\rho\to\infty$ and $N_r\to\infty$, the finite-horizon tail
supremum of the orange curve follows the prescribed $d^k$ reference with
$d=0.70$, consistently with
Corollary~\ref{cor:finite_joint_ideal_mse_convergence}.

\section{Conclusion}
\label{sec:conclusion}
This paper investigated distributed state estimation over wireless MIMO fading
channels with imperfect aggregate CSI when locally unobservable
performance-limiting modes require multi-hop information propagation.
A distributed estimator was developed by integrating bias-compensated
over-the-air fusion, mode-selective Chebyshev filtering, and historical
compensation. The bias-compensated fusion rule renders the mean-error dynamics
independent of the SNR under the adopted channel model, and a feasible
communication order guarantees asymptotic unbiasedness with a mean error
decaying as $o(d^k)$. An augmented stochastic error system was further
established to characterize the transient and steady-state second moments.
The analysis provided a mean-square stability condition in terms of
$r(\mathbf\Xi(\rho,N_r))$, decomposed the centered MSE exactly into
homogeneous, plant-state, and error--plant cross-moment contributions, and
characterized the SNR-dependent onset of persistent plant-state dominance.
It also quantified the channel-hardening benefit of increasing the
receive-array size. In the ideal-communication limit, the root of the
absolute MSE deviation attains the prescribed decay rate. At a fixed
receive-array size satisfying the high-SNR mean-square stability condition,
the steady-state MSE vanishes as the process-noise covariance tends to zero
and the SNR gain tends to infinity. Numerical results corroborate the
predicted transient and steady-state behavior and the receive-array gain.


\appendix
\section{Proof of Proposition~\ref{prop:existence_of_modal_sources}}
\label{app:proof_existence_of_modal_sources}
Fix $l\in\mathcal{I}_{\mathrm{pl}}$ and $1\le s\le n_l^{\mathrm{mb}}$. Since $\mathbf{J}_l$ is a single Jordan block, its eigenspace at $\lambda_l$ is one-dimensional. Choose $\mathbf{v}_l$ as the first coordinate vector in $\ker(\lambda_l\mathbf{I}_{m_l}-\mathbf{J}_l)$.

Suppose, for a contradiction, that
$\mathcal{V}_{l,s}^{\mathcal{O}}=\varnothing$.
Then $(l,s)\in\mathcal{U}_i$ for every $i\in\mathcal{V}$, so
$a_{l,s}<r_{i,l}$ at every node.
Since $a_{l,s}\ge1$, the first column of $\mathbf C_i\mathbf V_l$ is zero.
Hence $\mathbf{C}_i\mathbf{V}_l\mathbf{v}_l=\mathbf{0}$ for every $i\in\mathcal{V}$. Set $\mathbf{v}=\mathbf{V}_l\mathbf{v}_l$.
The columns of $\mathbf{V}_l$ are linearly
independent, so $\mathbf{v}\neq\mathbf{0}$.
Moreover,
\begin{equation}
\hat{\mathbf{A}}\mathbf{v}
=\mathbf{V}_l\mathbf{J}_l\mathbf{v}_l
=\lambda_l\mathbf{v},\
\mathbf{C}\mathbf{v}
=\operatorname{col}_{i\in\mathcal{V}}
(\mathbf{C}_i\mathbf{V}_l\mathbf{v}_l)
=\mathbf{0}.
\end{equation}
Thus $\mathbf{v}$ is a nonzero eigenvector of $\hat{\mathbf{A}}$ that is annihilated by the stacked observation matrix $\mathbf{C}$.
This contradicts the PBH observability test for $(\hat{\mathbf{A}},\mathbf{C})$ under Assumption~\ref{as:global_modal_source_observability}. Therefore $\mathcal{V}_{l,s}^{\mathcal{O}}\neq\varnothing$.
Since $(l,s)$ was arbitrary, the conclusion holds for every performance-limiting mini-block. \hfill $\blacksquare$

\section{Polynomial Representation of the Communication Filter}
\label{app:proof_homogeneous_graph_filter}
Fix $(l,s)$ with $l\in\mathcal I_{\mathrm{pl}}$ and
$\mathcal V_{l,s}^{\mathcal U}\neq\varnothing$ and use the matrix
$\mathbf S_{l,s}$ defined in Section~\ref{subsec:mean_block_communication}.
The homogeneous recurrence and its two initial vectors are given in
\eqref{eq:pcsi_homogeneous_modal_initialization}. Since $\omega_{l,s}^0=1$,
the first update gives
\begin{equation}
\bar{\boldsymbol\epsilon}_{l,s,\mathcal U,\mathrm h}^{\mathrm J,1}(k)
=(\mathbf S_{l,s}\otimes\mathbf I_{m_{l,s}})
\bar{\boldsymbol\epsilon}_{l,s,\mathcal U}^{\mathrm J,0}(k).
\label{eq:pcsi_proof_first_homogeneous_update}
\end{equation}
This agrees with \eqref{eq:pcsi_homogeneous_graph_filter} at $T=1$,
because $\psi_{1,l,s}(z)=z$.
Suppose the representation holds at $\nu=\tau-1,\tau$:
\begin{equation}
\begin{aligned}
\bar{\boldsymbol\epsilon}_{l,s,\mathcal U,\mathrm h}^{\mathrm J,\nu}(k)
={}&[\psi_{\nu,l,s}(\mathbf S_{l,s})\otimes\mathbf I_{m_{l,s}}]\\
&\times\bar{\boldsymbol\epsilon}_{l,s,\mathcal U}^{\mathrm J,0}(k).
\end{aligned}
\label{eq:pcsi_proof_polynomial_induction_hypothesis}
\end{equation}
Substitution into
\eqref{eq:pcsi_homogeneous_modal_initialization} yields
\begin{equation}
\begin{aligned}
\bar{\boldsymbol\epsilon}_{l,s,\mathcal U,\mathrm h}^{\mathrm J,\tau+1}(k)
={}&\Big[\big(\omega_{l,s}^\tau\mathbf S_{l,s}
\psi_{\tau,l,s}(\mathbf S_{l,s})\\
&+(1-\omega_{l,s}^\tau)
\psi_{\tau-1,l,s}(\mathbf S_{l,s})\big)
\otimes\mathbf I_{m_{l,s}}\Big]\\
&\times\bar{\boldsymbol\epsilon}_{l,s,\mathcal U}^{\mathrm J,0}(k)\\
={}&[\psi_{\tau+1,l,s}(\mathbf S_{l,s})\otimes\mathbf I_{m_{l,s}}]\\
&\times\bar{\boldsymbol\epsilon}_{l,s,\mathcal U}^{\mathrm J,0}(k).
\end{aligned}
\label{eq:pcsi_proof_polynomial_induction_step}
\end{equation}
where the last equality follows from
\eqref{eq:pcsi_graph_polynomial_recursion}. The initialization in
\eqref{eq:pcsi_homogeneous_modal_initialization} supplies the
$\nu=0$ case with $\psi_{0,l,s}(z)=1$, so induction proves
\eqref{eq:pcsi_homogeneous_graph_filter}.

To verify \eqref{eq:pcsi_normalized_chebyshev_polynomial},
first suppose $\theta_{l,s}>0$.
The Chebyshev recurrence evaluated at $1/\theta_{l,s}$ gives
\begin{equation}
\begin{gathered}
\frac{2\mathcal T_1(1/\theta_{l,s})}
{\theta_{l,s}\mathcal T_2(1/\theta_{l,s})}
=\frac{2}{2-\theta_{l,s}^2},\\
\frac{2\mathcal T_\tau(1/\theta_{l,s})}
{\theta_{l,s}\mathcal T_{\tau+1}(1/\theta_{l,s})}
=\left[
1-\frac{\theta_{l,s}^2}{4}
\frac{2\mathcal T_{\tau-1}(1/\theta_{l,s})}
{\theta_{l,s}\mathcal T_\tau(1/\theta_{l,s})}
\right]^{-1},\ \tau\geq2.
\end{gathered}
\label{eq:pcsi_proof_chebyshev_weight_recursion}
\end{equation}
These ratios equal $\omega_{l,s}^\tau$ in
Subsection~\ref{subsec:mean_chebyshev_design}.
The normalized-Chebyshev expression agrees with
$\psi_{0,l,s}(z)=1$ and $\psi_{1,l,s}(z)=z$.
Assume it holds at degrees $\tau-1$ and $\tau$.
Substituting these expressions into
\eqref{eq:pcsi_graph_polynomial_recursion}
and applying the Chebyshev recurrence yields
\begin{equation}
\begin{aligned}
\psi_{\tau+1,l,s}(z)
={}&\omega_{l,s}^\tau z
\frac{\mathcal T_\tau(z/\theta_{l,s})}
{\mathcal T_\tau(1/\theta_{l,s})}
+(1-\omega_{l,s}^\tau)
\frac{\mathcal T_{\tau-1}(z/\theta_{l,s})}
{\mathcal T_{\tau-1}(1/\theta_{l,s})}\\
={}&\frac{\mathcal T_{\tau+1}(z/\theta_{l,s})}
{\mathcal T_{\tau+1}(1/\theta_{l,s})}.
\end{aligned}
\label{eq:pcsi_proof_normalized_chebyshev_recursion}
\end{equation}
Induction proves \eqref{eq:pcsi_normalized_chebyshev_polynomial}.
If $\theta_{l,s}=0$, all used weights equal one, and
\eqref{eq:pcsi_graph_polynomial_recursion}
gives $\psi_{\tau,l,s}(z)=z^\tau$.
\hfill$\blacksquare$

\section{Proof of Theorem~\ref{thm:pcsi_mean_convergence}}
\label{app:proof_mean_convergence}
For each retained tail of an original chain, the eigendirection at
$\lambda_l$ is its first coordinate.
The corresponding output vector is
$\mathbf C_i\mathbf v_{l,r_{i,l}}$.
At a fixed eigenvalue with $|\lambda|\ge d$,
Assumption~\ref{as:network_decoupled_local_detectability}
makes these output vectors linearly independent.
The PBH criterion therefore gives
\eqref{eq:effective_anchored_detectability}.
All undetectable retained eigenvalues have modulus strictly below $d$.
Hence one can choose $\mathbf G_i$ with
$r(\mathbf J_i^{\mathcal O}
-\mathbf G_i\mathbf C_i\mathbf V_i^{\mathcal O})<d$.
The separation identities~\eqref{eq:nodewise_separation_identities}
and the retained-coordinate support of $\mathbf K_i$ make
\eqref{eq:pcsi_local_reduced_mean_error} autonomous.
If $n_i^{\mathcal O}=0$, the choice $\mathbf K_i=\mathbf0$ is admissible.
Conjugate-consistent partitions and paired gains yield a real $\mathbf K_i$.

For the network-assisted coordinates,
\eqref{eq:pcsi_non_source_posterior_recursion} gives a two-step
homogeneous transition together with source and chain inputs.
The source equilibrium $\mathbf c_{l,s}(k)$ depends only on
locally retained errors.
By \eqref{eq:segment_prediction_input}, the other inputs involve
only segments $(l,t)$ with $t>s$, at the current or preceding time.
To characterize the homogeneous transition,
fix $(l,s)$ with $l\in\mathcal I_{\mathrm{pl}}$ and
$\mathcal V_{l,s}^{\mathcal U}\neq\varnothing$.
Since the grounded Laplacian is symmetric positive definite,
choose an orthogonal matrix $\mathbf Q_{l,s}$ such that
\begin{equation}
\mathbf L_{\mathcal V_{l,s}^{\mathcal U}}
=\mathbf Q_{l,s}\operatorname{diag}
(\mu_1^{l,s},\ldots,\mu_{n_{l,s,\mathcal U}}^{l,s})\mathbf Q_{l,s}^\top.
\label{eq:pcsi_grounded_laplacian_eigendecomposition}
\end{equation}
Premultiplication by $\mathbf Q_{l,s}^\top\otimes\mathbf I_{m_{l,s}}$
decouples the homogeneous transition into graph modes.
For $1\leq j\leq n_{l,s,\mathcal U}$, the current-and-previous
posterior-error transition of the $j$th graph mode is
\begin{equation}
\mathbf A_{l,s,j}^{\mathcal U,a}
\triangleq
\begin{bmatrix}
g_{l,s}(\mu_j^{l,s})(\mathbf J_{l,s}+\beta_l\mathbf I_{m_{l,s}})
&-\beta_l\mathbf J_{l,s}\\
\mathbf I_{m_{l,s}}&\mathbf0
\end{bmatrix},
\label{eq:pcsi_complete_jordan_graph_mode_companion}
\end{equation}
where $g_{l,s}(\mu)$ is the graph-filter gain defined in
Subsection~\ref{subsec:mean_chebyshev_design}.
For $z\neq0$, taking the Schur complement of the lower-right
block $z\mathbf I_{m_{l,s}}$ gives
\begin{equation}
\begin{aligned}
&\det(z\mathbf I_{2m_{l,s}}-\mathbf A_{l,s,j}^{\mathcal U,a})\\
&=\det\!\left[
z^2\mathbf I_{m_{l,s}}
-g_{l,s}(\mu_j^{l,s})(\mathbf J_{l,s}+\beta_l\mathbf I_{m_{l,s}})z
+\beta_l\mathbf J_{l,s}
\right].
\end{aligned}
\label{eq:pcsi_schur_complement_characteristic_determinant}
\end{equation}
Both sides are polynomials in $z$, so the equality extends
to $z=0$.
Writing $\mathbf J_{l,s}=\lambda_l\mathbf I_{m_{l,s}}+\mathbf N_{l,s}$,
where $\mathbf N_{l,s}$ is strictly upper triangular, and using
$\beta_l\lambda_l=d_-^2$, we obtain
\begin{equation}
\begin{aligned}
&\det(z\mathbf I_{2m_{l,s}}-\mathbf A_{l,s,j}^{\mathcal U,a})\\
&=\det\!\left[
q_{l,s,\mu_j^{l,s}}(z)\mathbf I_{m_{l,s}}
+\big(\beta_l-g_{l,s}(\mu_j^{l,s})z\big)\mathbf N_{l,s}
\right]\\
&=[q_{l,s,\mu_j^{l,s}}(z)]^{m_{l,s}}.
\end{aligned}
\label{eq:pcsi_complete_jordan_characteristic_determinant}
\end{equation}
Here $q_{l,s,\mu}(z)$ is defined in
\eqref{eq:pcsi_modal_characteristic_polynomial}.
Strict upper triangularity of $\mathbf N_{l,s}$ gives the last
equality. Hence, the diagonal Jordan subblock contributes only
the zeros of $q_{l,s,\mu_j^{l,s}}(z)$, with their multiplicities
multiplied by $m_{l,s}$.

Let $\hat z=z/d$. The equation $q_{l,s,\mu_j^{l,s}}(z)=0$ becomes
\begin{equation}
\hat z^2-\frac{h_l}{d}g_{l,s}(\mu_j^{l,s})\hat z+\delta^2=0.
\label{eq:pcsi_normalized_characteristic_equation}
\end{equation}
For a monic quadratic $z^2+a_1z+a_2$, the Schur--Cohn
criterion~\cite{iserles2009first} is
\begin{equation}
\begin{aligned}
|a_2|&<1,\\
|a_1-a_2a_1^*|&<1-|a_2|^2.
\end{aligned}
\label{eq:pcsi_schur_cohn_criterion}
\end{equation}
In \eqref{eq:pcsi_normalized_characteristic_equation},
$a_2=\delta^2<1$.
Since $g_{l,s}(\mu_j^{l,s})$ is real, the second inequality is
equivalent to
\begin{equation}
|g_{l,s}(\mu_j^{l,s})|<\alpha_l,
\label{eq:pcsi_modal_schur_condition}
\end{equation}
with $\alpha_l$ given by \eqref{eq:pcsi_modal_attenuation}.

We verify that the threshold in
\eqref{eq:pcsi_modal_attenuation} is well defined and does not
exceed one, including when $\lambda_l$ is nonreal. Write
$\lambda_l=|\lambda_l|e^{\mathrm i\varphi_l}$. From
$\beta_l=d_-^2/\lambda_l$ and the definitions of $\delta$ and $h_l$ in Subsection~\ref{subsec:mean_chebyshev_design},
\begin{equation}
\begin{aligned}
h_l-\delta^2h_l^*
={}&\left(
|\lambda_l|-\frac{d_-^4}{d^2|\lambda_l|}
\right)e^{\mathrm i\varphi_l}\\
&-\left(
\frac{d_-^2|\lambda_l|}{d^2}
-\frac{d_-^2}{|\lambda_l|}
\right)e^{-\mathrm i\varphi_l}.
\end{aligned}
\label{eq:pcsi_complex_modal_attenuation_expansion}
\end{equation}
Because $|\lambda_l|\geq d>d_-$, the two coefficients in parentheses
in \eqref{eq:pcsi_complex_modal_attenuation_expansion} are nonnegative,
and the first is no smaller than the second. The reverse triangle
inequality therefore gives
\begin{equation}
\begin{aligned}
|h_l-\delta^2h_l^*|
\geq{}&|\lambda_l|
-\frac{d_-^4}{d^2|\lambda_l|}
-\frac{d_-^2|\lambda_l|}{d^2}
+\frac{d_-^2}{|\lambda_l|}\\
={}&(1-\delta^2)
\left(|\lambda_l|+\frac{d_-^2}{|\lambda_l|}\right)\\
\geq{}&(1-\delta^2)
\left(d+\frac{d_-^2}{d}\right)\\
={}&d(1-\delta^4)>0.
\end{aligned}
\label{eq:pcsi_complex_modal_attenuation_lower_bound}
\end{equation}
The second inequality uses the fact that
$r+d_-^2/r$ is strictly increasing for $r\geq d>d_-$. Therefore,
$0<\alpha_l\leq1$, with equality if and only if
$|\lambda_l|=d$.
For $\theta_{l,s}>0$, using
$\mathcal T_T(x)=\cosh(T\operatorname{arccosh}(x))$ for $x>1$,
\eqref{eq:pcsi_chebyshev_order_range}, and
\eqref{eq:pcsi_normalized_chebyshev_polynomial} gives
\begin{equation}
\max_{\mu\in[\mu_{l,s}^{\mathrm{min}},\mu_{l,s}^{\mathrm{max}}]}
\left|\psi_{T,l,s}(1-\gamma_{l,s}\mu)\right|<\alpha_l.
\label{eq:pcsi_interval_modal_schur_condition}
\end{equation}
For $\theta_{l,s}=0$, the same strict inequality holds because the
attenuation is zero. Therefore, \eqref{eq:pcsi_modal_schur_condition}
holds for every grounded-Laplacian eigenvalue, and
$r(\mathbf A_{l,s,j}^{\mathcal U,a})<d$ for all $j$.

For the global mean-error recurrence
\eqref{eq:pcsi_mean_error_dynamics}, define
$\bar{\mathbf e}_{\mathrm{mean}}^a(k)
=\operatorname{col}(\bar{\mathbf e}(k),\bar{\mathbf e}(k-1))$.
Its first-order representation is
\begin{equation}
\bar{\mathbf e}_{\mathrm{mean}}^a(k+1)
=\mathbf A_{\mathrm{mean}}\bar{\mathbf e}_{\mathrm{mean}}^a(k),
\quad k\geq1,
\label{eq:pcsi_first_order_mean_error_dynamics}
\end{equation}
where
\begin{equation}
\mathbf A_{\mathrm{mean}}\triangleq
\begin{bmatrix}
\mathbf F_0&\mathbf F_1\\
\mathbf I_{MN}&\mathbf0_{MN\times MN}
\end{bmatrix}
\label{eq:pcsi_global_mean_companion}
\end{equation}
is the companion matrix of the second-order recurrence.

The local recursion \eqref{eq:pcsi_local_reduced_mean_error}
is autonomous.
Every Jordan coordinate belongs either to a locally retained
vector or to exactly one non-source segment vector.
Their basis rows partition $\mathbf I_M\otimes\mathbf V^{-1}$.
The transformations $\mathbf Q_{l,s}^\top$ within the non-source
node sets are orthogonal, so the complete coordinate transformation
is invertible, also when applied to current and preceding errors.

Order the locally retained coordinates and their histories first.
For each original chain, order its network-assisted segment groups
by decreasing $s$, keeping each group's current and preceding errors
together.
The source inputs then arise from earlier groups.
The chain inputs in \eqref{eq:segment_prediction_input} also arise
only from earlier groups, since they involve $t>s$.
Consequently, the global transition is block lower triangular.
Its diagonal blocks are $\mathbf F_i^{\mathcal O}$,
zero blocks of size $n_i^{\mathcal O}$, and the
graph-mode companion matrices $\mathbf A_{l,s,j}^{\mathcal U,a}$.
The zero blocks are the redundant histories of the first-order
local recursions.
Thus
\begin{equation}
\begin{aligned}
&\det(z\mathbf I_{2MN}-\mathbf A_{\mathrm{mean}})\\
&=\prod_{i\in\mathcal V}
\left[z^{n_i^{\mathcal O}}
\det(z\mathbf I_{n_i^{\mathcal O}}-\mathbf F_i^{\mathcal O})\right]\\
&\phantom{=}\times
\prod_{\substack{l\in\mathcal I_{\mathrm{pl}},\ 1\le s\le n_l^{\mathrm{mb}}\\
\mathcal V_{l,s}^{\mathcal U}\neq\varnothing}}
\prod_{j=1}^{n_{l,s,\mathcal U}}
[q_{l,s,\mu_j^{l,s}}(z)]^{m_{l,s}}.
\end{aligned}
\label{eq:pcsi_global_characteristic_determinant_factorization}
\end{equation}
An empty local factor is interpreted as one.
For every nonempty local branch,
\eqref{eq:pcsi_local_reduced_mean_error}
and the gain selection give $r(\mathbf F_i^{\mathcal O})<d$.
The graph-mode analysis establishes the same root bound
for each $q_{l,s,\mu_j^{l,s}}(z)$.
Consequently, every root of the global characteristic
polynomial lies inside $|z|<d$, including the additional
roots at zero, and
\begin{equation}
r(\mathbf A_{\mathrm{mean}})<d.
\label{eq:pcsi_global_mean_spectral_radius_bound}
\end{equation}
The admissible initializations in
Section~\ref{sec:local_info_and_state_esti}
ensure that $\bar{\mathbf e}_{\mathrm{mean}}^a(1)$ is finite.
Choose an auxiliary decay factor $d_{\mathrm{aux}}$ such that
$r(\mathbf A_{\mathrm{mean}})<d_{\mathrm{aux}}<d$. A standard
finite-dimensional matrix-power bound gives a constant
$C_{\mathrm{aux}}>0$ such that
\begin{equation}
\left\|\mathbf A_{\mathrm{mean}}^k\right\|
\leq C_{\mathrm{aux}}d_{\mathrm{aux}}^k,
\qquad k\in\mathbb Z_{\geq0}.
\label{eq:pcsi_mean_matrix_power_bound}
\end{equation}
Iterating \eqref{eq:pcsi_first_order_mean_error_dynamics} from $k=1$
and using \eqref{eq:pcsi_mean_matrix_power_bound} give, for $k\geq1$,
\begin{equation}
\begin{aligned}
\|\bar{\mathbf e}(k)\|
&\leq\|\bar{\mathbf e}_{\mathrm{mean}}^a(k)\|\\
&\leq C_{\mathrm{aux}}d_{\mathrm{aux}}^{k-1}
\|\bar{\mathbf e}_{\mathrm{mean}}^a(1)\|.
\end{aligned}
\label{eq:pcsi_augmented_mean_error_power_bound}
\end{equation}
Taking $k$th roots in
\eqref{eq:pcsi_augmented_mean_error_power_bound} and, separately,
dividing it by $d^k$ yield
\begin{equation}
\begin{gathered}
\limsup_{k\to\infty}
\|\bar{\mathbf e}(k)\|^{1/k}
\le d_{\mathrm{aux}}<d,\\
\frac{\|\bar{\mathbf e}(k)\|}{d^k}
\le
\frac{C_{\mathrm{aux}}}{d_{\mathrm{aux}}}
\|\bar{\mathbf e}_{\mathrm{mean}}^a(1)\|
\left(\frac{d_{\mathrm{aux}}}{d}\right)^k
\to 0.
\end{gathered}
\label{eq:pcsi_mean_decay_from_power_bound}
\end{equation}
This proves \eqref{eq:pcsi_mean_decay_rate}.
\hfill$\blacksquare$

\section{Schur Stability of the Deterministic Augmented Transition}
\label{app:proof_nominal_augmented_stability}
Grouping the first two block components of
$\mathbf e^{\mathrm a}(k)$ makes
$\mathbf F_{\mathrm a}$ in
\eqref{eq:finite_nominal_augmented_matrix} block upper triangular.
Its diagonal blocks are $\mathbf A_{\mathrm{mean}}$ in
\eqref{eq:pcsi_global_mean_companion} and
$\mathbf0_{MN\times MN}$. Hence,
\begin{equation}
\sigma(\mathbf F_{\mathrm a})
=\sigma(\mathbf A_{\mathrm{mean}})\cup\{0\},
\quad
r(\mathbf F_{\mathrm a})=r(\mathbf A_{\mathrm{mean}}).
\label{eq:finite_augmented_spectral_identity}
\end{equation}
The bound \eqref{eq:pcsi_global_mean_spectral_radius_bound}
proved in \ref{app:proof_mean_convergence} therefore yields
\begin{equation}
r(\mathbf F_{\mathrm a})<d<1.
\label{eq:finite_nominal_augmented_radius}
\end{equation}
\hfill$\blacksquare$

\section{Cross-Moment Properties of the Stochastic Components}
\label{app:proof_cross_moments}
The variables $\mathbf e^{\mathrm a}(k)$ and
$\tilde{\mathbf x}(k)$ depend on initial conditions and innovations
up to control cycle $k-1$.
The current coefficients $\mathbf U_{\mathrm a}(k)$ and
$\mathbf U_x(k)$ depend only on the channels and pilot noises
within control cycle $k$. They are therefore jointly independent
of these historical variables. The zero-mean terminal coefficients
in \eqref{eq:finite_graph_filter_perturbation}
and \eqref{eq:finite_state_coupling_perturbation} give
$\mathbb E[\mathbf U_{\mathrm a}(k)]
=\mathbb E[\mathbf U_x(k)]=\mathbf0$.
Consequently,
\begin{equation}
\begin{aligned}
&\mathbb E[\mathbf U_{\mathrm a}(k)\mathbf e^{\mathrm a}(k)
(\mathbf e^{\mathrm a}(k))^\top\mathbf F_{\mathrm a}^\top]
=\mathbf0,\\
&\mathbb E[\mathbf F_{\mathrm a}\mathbf e^{\mathrm a}(k)
\tilde{\mathbf x}^\top(k)\mathbf U_x^\top(k)]
=\mathbf0.
\end{aligned}
\label{eq:finite_vanishing_cross_moments}
\end{equation}
The process-noise simplification in
\eqref{eq:finite_lumped_disturbance} follows directly from the
terminal recursions. For any
$\mathbf v=\mathbf1_M\otimes\mathbf v_0$,
$\bar{\mathbf L}\mathbf v=\boldsymbol\Delta^\tau(k)\mathbf v=\mathbf0$.
The completeness identity
$\mathbf P^{\mathcal O_i}+
\sum_{(l,s)\in\mathcal U_i}\mathbf P_{l,s}^{\mathcal U_i}=\mathbf I_N$
and the unit sum of the two communication weights imply,
by induction in \eqref{eq:pcsi_mean_graph_filter_recursion},
$\mathbf\Psi_\tau\mathbf v=\mathbf v$.
Moreover, the coefficients in
\eqref{eq:finite_global_internal_error} satisfy
$\mathbf E_{\epsilon}^\tau(k)\mathbf v=\mathbf E_x^\tau(k)\mathbf v$.
The identical initializations of $\mathbf R_\tau$ and
$\mathbf Q_\tau$, together with
\eqref{eq:finite_graph_filter_perturbation} and
\eqref{eq:finite_state_coupling_perturbation}, therefore imply
$\mathbf R_\tau(k)\mathbf v=\mathbf Q_\tau(k)\mathbf v$.
Thus the coefficients of $\tilde{\mathbf w}(k)$ in the posterior
disturbance reduce to $-\mathbf I_{MN}$.

For compactness, use the fixed disturbance-injection matrices
\begin{equation}
\begin{aligned}
\mathbf L_w&\triangleq
\operatorname{col}(-\mathbf I_{MN},\mathbf0_{MN\times MN},\mathbf I_{MN}),\\
\mathbf J_\eta&\triangleq
\operatorname{col}(\tilde{\mathbf M}_1,\mathbf0_{MN\times MN},
\mathbf0_{MN\times MN}).
\end{aligned}
\label{eq:finite_noise_injection_matrices}
\end{equation}
Then $\boldsymbol\nu^{\mathrm a}(k)
=\mathbf L_w\tilde{\mathbf w}(k)
+\rho^{-1/2}\mathbf J_\eta\tilde{\mathbf n}_T(k)$.
Conditioning on the historical variables and all current
channels and pilot noises leaves only the independent,
zero-mean process and receiver data noises. Hence
\begin{equation}
\mathbb E[\boldsymbol\nu^{\mathrm a}(k)
\mid \mathbf e^{\mathrm a}(k),\tilde{\mathbf x}(k),
\{\mathbf H_{ij}^\tau(k),\mathbf N_{p,i}^\tau(k)\}_{i,j,\tau}]
=\mathbf0.
\label{eq:finite_conditional_augmented_disturbance}
\end{equation}
The tower property eliminates the cross terms between
$\boldsymbol\nu^{\mathrm a}(k)$ and the historical or
multiplicatively perturbed states. Its remaining moments are
\begin{equation}
\begin{aligned}
\boldsymbol\Sigma_{\nu^{\mathrm a}}
={}&\mathbf L_w\boldsymbol\Sigma_{\tilde w}\mathbf L_w^\top\\
&+\rho^{-1}\mathbf J_\eta
\mathbb E[\tilde{\mathbf n}_T(k)\tilde{\mathbf n}_T^\top(k)]
\mathbf J_\eta^\top,\\
\boldsymbol\Sigma_{\nu^{\mathrm a}\tilde w}
={}&\mathbf L_w\boldsymbol\Sigma_{\tilde w}.
\end{aligned}
\label{eq:finite_disturbance_second_moments}
\end{equation}
The coefficients $\mathbf U_{\mathrm a}(k)$ and $\mathbf U_x(k)$
share channel and pilot samples and need not be mutually independent.
In particular,
\begin{equation}
\begin{aligned}
&\mathbb E[\mathbf U_{\mathrm a}(k)\mathbf e^{\mathrm a}(k)
\tilde{\mathbf x}^\top(k)\mathbf U_x^\top(k)]\\
&=\mathbb E[\mathbf U_{\mathrm a}(k)
\boldsymbol\Sigma_{\mathrm a\tilde x}(k)\mathbf U_x^\top(k)],
\end{aligned}
\label{eq:finite_nonzero_cross_moment_factorization}
\end{equation}
which need not vanish. This is the mixed contribution retained
by $\mathcal C_{\rho,N_r}$.
\hfill$\blacksquare$

\section{Proof of Theorem~\ref{thm:finite_second_moment}}
\label{app:proof_second_moment}
The replicated plant dynamics are
$\tilde{\mathbf x}(k+1)=\tilde{\mathbf A}\tilde{\mathbf x}(k)
+\tilde{\mathbf w}(k)$.
Independence and the zero mean of $\tilde{\mathbf w}(k)$
give \eqref{eq:finite_plant_second_moment_solution}.
Multiplying \eqref{eq:finite_augmented_system} by
$\tilde{\mathbf x}^\top(k+1)$ and taking expectations eliminates
terms linear in $\mathbf U_{\mathrm a}(k)$ or $\mathbf U_x(k)$.
The conditional-mean identity
\eqref{eq:finite_conditional_augmented_disturbance} also gives
$\mathbb E[\boldsymbol\nu^{\mathrm a}(k)\tilde{\mathbf x}^\top(k)]
=\mathbf0$.
The remaining process-noise contribution is
$\boldsymbol\Sigma_{\nu^{\mathrm a}\tilde w}$ in
\eqref{eq:finite_disturbance_second_moments}, which establishes
\eqref{eq:finite_cross_second_moment_solution}.

Expanding the quadratic product of
\eqref{eq:finite_augmented_system} and using
\ref{app:proof_cross_moments} yields
\begin{equation}
\begin{aligned}
\boldsymbol\Sigma_{\mathrm a}(k+1)
={}&\mathbf F_{\mathrm a}\boldsymbol\Sigma_{\mathrm a}(k)
\mathbf F_{\mathrm a}^\top+\boldsymbol\Sigma_{\nu^{\mathrm a}}\\
&+\mathbb E[\mathbf U_{\mathrm a}(k)
\boldsymbol\Sigma_{\mathrm a}(k)\mathbf U_{\mathrm a}^\top(k)]\\
&+\mathbb E[\mathbf U_x(k)
\boldsymbol\Sigma_{\tilde x}(k)\mathbf U_x^\top(k)]\\
&+\mathbb E[\mathbf U_{\mathrm a}(k)
\boldsymbol\Sigma_{\mathrm a\tilde x}(k)\mathbf U_x^\top(k)]\\
&+\mathbb E[\mathbf U_x(k)
\boldsymbol\Sigma_{\mathrm a\tilde x}^\top(k)
\mathbf U_{\mathrm a}^\top(k)].
\end{aligned}
\label{eq:finite_augmented_covariance_recursion}
\end{equation}
The maps in \eqref{eq:finite_second_moment_operators}
convert this expression into
\eqref{eq:finite_augmented_second_moment_solution}.
The vectorization identity
$\operatorname{vec}(\mathbf A\mathbf B\mathbf C)
=(\mathbf C^\top\otimes\mathbf A)\operatorname{vec}(\mathbf B)$
gives the homogeneous transition
$\mathbf\Xi(\rho,N_r)$ in
\eqref{eq:finite_covariance_transition_matrix}.
Finally, the first block of $\mathbf e^{\mathrm a}(k)$ is
$\mathbf e(k)$, so
$\operatorname{tr}[(\boldsymbol\Sigma_{\mathrm a}(k))_{11}]
=\mathbb E[\|\mathbf e(k)\|^2]$.
This proves \eqref{eq:finite_tracking_mse}.
\hfill$\blacksquare$

\section{Proof of Corollary~\ref{cor:finite_steady_state_mse}}
\label{app:proof_steady_state_mse}
The plant recursion \eqref{eq:finite_plant_second_moment_solution}
has a unique stationary second moment because
$r(\tilde{\mathbf A})<1$.
The vectorized cross-moment recursion
\eqref{eq:finite_cross_second_moment_solution}
has transition $\tilde{\mathbf A}\otimes\mathbf F_{\mathrm a}$,
with spectral radius
$r(\tilde{\mathbf A})r(\mathbf F_{\mathrm a})<1$ by
\eqref{eq:finite_nominal_augmented_radius}.
Their homogeneous responses vanish, and their matrix series
converge absolutely to
\eqref{eq:finite_stationary_plant_cross_moments}.

Linearity of $\mathcal P_{\rho,N_r}$ and $\mathcal C_{\rho,N_r}$
therefore makes the forcing in
\eqref{eq:finite_augmented_second_moment_solution} converge to
$\boldsymbol\Sigma_{\nu^{\mathrm a}}
+\mathcal P_{\rho,N_r}(\boldsymbol\Sigma_{\tilde x}^\star)
+\mathcal C_{\rho,N_r}(\boldsymbol\Sigma_{\mathrm a\tilde x}^\star)$.
The homogeneous transition is $\mathbf\Xi(\rho,N_r)$,
which is Schur stable by
\eqref{eq:finite_mean_square_stability_condition}.
The variation-of-constants formula then gives a unique finite
limit of $\boldsymbol\Sigma_{\mathrm a}(k)$.
Taking the limit yields
\eqref{eq:finite_steady_state_augmented_covariance}.
Consequently,
\begin{equation}
\sup_{k\geq1}\mathbb E[\|\mathbf e^{\mathrm a}(k)\|^2]
=\sup_{k\geq1}\operatorname{tr}(\boldsymbol\Sigma_{\mathrm a}(k))
<\infty.
\label{eq:finite_augmented_mean_square_boundedness}
\end{equation}
Taking the leading-block trace proves
\eqref{eq:finite_steady_state_tracking_mse}.

Theorem~\ref{thm:pcsi_mean_convergence} and the zero-mean process
noise give $\mathbb E[\mathbf e^{\mathrm a}(k)]\to\mathbf0$,
while $r(\tilde{\mathbf A})<1$ gives
$\mathbb E[\tilde{\mathbf x}(k)]\to\mathbf0$.
Thus the limiting second moments also equal their stationary
covariance and cross-covariance matrices.
\hfill$\blacksquare$

\section{Proof of Theorem~\ref{thm:finite_exact_centered_mse_decomposition}}
\label{app:proof_exact_centered_mse_decomposition}
The stationary solution is given by
\eqref{eq:finite_steady_state_augmented_covariance}.
Subtract its defining fixed-point equation from
\eqref{eq:finite_augmented_second_moment_solution}.
With the deviations $\mathbf D_{\mathrm a}(k)$,
$\mathbf D_x(k)$, and $\mathbf D_{\mathrm a x}(k)$ defined
before the theorem, linearity gives
\begin{equation}
\begin{aligned}
\mathbf D_{\mathrm a}(k+1)
={}&\mathcal H_{\rho,N_r}(\mathbf D_{\mathrm a}(k))\\
&+\mathcal P_{\rho,N_r}(\mathbf D_x(k))
+\mathcal C_{\rho,N_r}(\mathbf D_{\mathrm a x}(k)).
\end{aligned}
\label{eq:finite_centered_operator_covariance_recursion}
\end{equation}
Iterating from $k=1$ yields
\begin{equation}
\begin{aligned}
\mathbf D_{\mathrm a}(k)
={}&\mathcal H_{\rho,N_r}^{k-1}(\mathbf D_{\mathrm a}(1))\\
&+\sum_{t=1}^{k-1}\mathcal H_{\rho,N_r}^{k-1-t}
\bigl(\mathcal P_{\rho,N_r}(\mathbf D_x(t))\bigr)\\
&+\sum_{t=1}^{k-1}\mathcal H_{\rho,N_r}^{k-1-t}
\bigl(\mathcal C_{\rho,N_r}(\mathbf D_{\mathrm a x}(t))\bigr).
\end{aligned}
\label{eq:finite_centered_covariance_solution}
\end{equation}
The sums are empty for $k=1$.
Taking the leading-block trace gives the three terms
$H(k),P(k),C(k)$ in \eqref{eq:finite_exact_mse_components},
and
\begin{equation}
\operatorname{tr}[(\mathbf D_{\mathrm a}(k))_{11}]
=\mathbb E[\|\mathbf e(k)\|^2]
-\operatorname{tr}[(\boldsymbol\Sigma_{\mathrm a}^\star)_{11}].
\label{eq:finite_centered_mse_block_extraction}
\end{equation}
This proves \eqref{eq:finite_exact_centered_mse_decomposition}.
The deviations are differences of moments, not covariance
matrices themselves; hence the three contributions can be signed.
\hfill$\blacksquare$

\section{Proof of
Corollary~\ref{cor:finite_componentwise_transient_rates}}
\label{app:proof_componentwise_transient_rates}
Subtracting the stationary relations
\eqref{eq:finite_stationary_plant_cross_moments}
from the plant and cross-moment recursions
\eqref{eq:finite_second_moment_solutions}, and iterating, gives
\begin{equation}
\begin{aligned}
\mathbf D_x(k)
={}&\tilde{\mathbf A}^k
\mathbf D_x(0)
(\tilde{\mathbf A}^\top)^k,\\
\mathbf D_{\mathrm a x}(k)
={}&\mathbf F_{\mathrm a}^{k-1}
\mathbf D_{\mathrm a x}(1)
(\tilde{\mathbf A}^\top)^{k-1}.
\end{aligned}
\label{eq:finite_transient_driving_deviations}
\end{equation}
Choose $q_\Xi$, $q_x$, and $q_a$ such that
$r(\mathbf\Xi(\rho,N_r))<q_\Xi<1$,
$r(\tilde{\mathbf A})<q_x<1$, and
$r(\mathbf F_{\mathrm a})<q_a<1$. The standard finite-dimensional matrix-power
bound implies that the powers of the three matrices are bounded by
constant multiples of $q_\Xi^k$, $q_x^k$, and $q_a^k$, respectively.
These bounds absorb any polynomial factors associated with nontrivial
Jordan blocks. Moreover, for fixed $(\rho,N_r)$,
$\mathcal P_{\rho,N_r}$ and $\mathcal C_{\rho,N_r}$ have finite induced
norms, while leading-block extraction followed by the trace is a
bounded linear functional. Combining these facts with
\eqref{eq:finite_covariance_transition_matrix},
\eqref{eq:finite_transient_driving_deviations}, and
\eqref{eq:finite_exact_mse_components} gives finite
constants $K_{\mathrm H},K_{\mathrm P},K_{\mathrm C}>0$ such that
\begin{equation}
\begin{aligned}
|H(k)|
&\leq K_{\mathrm H}q_\Xi^{k-1},\\
|P(k)|
&\leq K_{\mathrm P}\sum_{t=1}^{k-1}
q_\Xi^{k-1-t}q_x^{2t},\\
|C(k)|
&\leq K_{\mathrm C}\sum_{t=1}^{k-1}
q_\Xi^{k-1-t}(q_aq_x)^{t-1}.
\end{aligned}
\label{eq:finite_componentwise_convolution_bounds}
\end{equation}
For any $a,b\in(0,1)$, the $k$th-root upper limit of a geometric
convolution with terms $a^{k-1-t}b^t$ or
$a^{k-1-t}b^{t-1}$ does not exceed $\max\{a,b\}$. When $a=b$, the sum
contains the additional factor $k-1$, which does not affect this rate
because $(k-1)^{1/k}\to1$. Applying this fact to
\eqref{eq:finite_componentwise_convolution_bounds} gives
\begin{equation}
\begin{aligned}
\limsup_{k\to\infty}|H(k)|^{1/(2k)}
&\leq\sqrt{q_\Xi},\\
\limsup_{k\to\infty}|P(k)|^{1/(2k)}
&\leq\max\{\sqrt{q_\Xi},q_x\},\\
\limsup_{k\to\infty}|C(k)|^{1/(2k)}
&\leq\max\{\sqrt{q_\Xi},\sqrt{q_aq_x}\}.
\end{aligned}
\label{eq:finite_auxiliary_componentwise_root_rates}
\end{equation}
Letting $q_\Xi$, $q_x$, and $q_a$ decrease to their corresponding
spectral radii proves
\eqref{eq:finite_componentwise_root_mse_rates}.

The exact decomposition
\eqref{eq:finite_exact_centered_mse_decomposition} and the triangle
inequality give
\begin{equation}
\begin{aligned}
&\left|
\mathbb E[\|\mathbf e(k)\|^2]
-\operatorname{tr}[(\boldsymbol\Sigma_{\mathrm a}^\star)_{11}]
\right|\\
&\leq |H(k)|+|P(k)|+|C(k)|.
\end{aligned}
\label{eq:finite_centered_mse_triangle_bound}
\end{equation}
Taking $2k$th roots and upper limits in
\eqref{eq:finite_centered_mse_triangle_bound}, the resulting rate is no
greater than the largest of the three componentwise rates. Since
$r(\mathbf\Xi(\rho,N_r))<1$, $r(\tilde{\mathbf A})<1$, and
$r(\mathbf F_{\mathrm a})<1$, this largest rate is strictly less than one, which
proves \eqref{eq:finite_root_mse_convergence_rate}.
\hfill$\blacksquare$

\section{Proof of Corollary~\ref{cor:finite_snr_transient_envelope}}
\label{app:proof_snr_transient_envelope}
We first establish the high-SNR operator bounds stated after
\eqref{eq:finite_pcsi_second_moment_transition}.
Each term in the finite expansions of
$\mathbf U_x(k)$ and
$\mathbf U_{\mathrm a}(k)-
\mathbf U_{\mathrm a,\infty}(k)$
contains at least one pilot-noise perturbation
$\boldsymbol\Delta_N^\tau(k)$. For
fixed $T$, the terminal recursions
\eqref{eq:finite_graph_filter_perturbation}--\eqref{eq:finite_accumulated_data_noise}
and the finite Gaussian moments of the channel and pilot-noise variables
imply
\begin{equation}
\begin{aligned}
\mathbb E[\|\mathbf U_x(k)\|_{\mathrm F}^2]
&=\mathcal O(\rho^{-1}),\\
\mathbb E[\|\mathbf U_{\mathrm a}(k)
-\mathbf U_{\mathrm a,\infty}(k)\|_{\mathrm F}^2]
&=\mathcal O(\rho^{-1}),\\
\sup_{\rho\geq1}
\mathbb E[\|\mathbf U_{\mathrm a}(k)\|_{\mathrm F}^2]
&<\infty.
\end{aligned}
\label{eq:finite_terminal_perturbation_moment_bounds}
\end{equation}
Indeed, every term containing pilot noise has at least one factor
scaled by $\rho^{-1/2}$, while terms containing two such factors have
order at most $\rho^{-1}$. Since $T$ is finite, these expansions contain only finitely
many products. Submultiplicativity of the Frobenius norm
and independence across communication cycles bound the
second moment of each product by a product of per-cycle
second-moment bounds.

For every $\mathbf Z\in\mathbb R^{3MN\times MN}$, the
Cauchy--Schwarz inequality and
\eqref{eq:finite_terminal_perturbation_moment_bounds} give
\begin{equation}
\begin{aligned}
&\left\|\mathbb E[\mathbf U_{\mathrm a}(k)
\mathbf Z\mathbf U_x^\top(k)]\right\|_{\mathrm F}\\
&\leq
\Bigl(
\mathbb E[\|\mathbf U_{\mathrm a}(k)\|_{\mathrm F}^2]
\mathbb E[\|\mathbf U_x(k)\|_{\mathrm F}^2]
\Bigr)^{1/2}\|\mathbf Z\|_{\mathrm F}\\
&=\mathcal O(\rho^{-1/2})\|\mathbf Z\|_{\mathrm F},
\end{aligned}
\label{eq:finite_cross_operator_scaling}
\end{equation}
where the hidden constant is independent of $\rho$ and $\mathbf Z$.
Applying the same inequality to two factors of
$\mathbf U_x(k)$ gives the
$\mathcal O(\rho^{-1})$ scaling of $\mathcal P_{\rho,N_r}$; adding the
transposed cross term in
\eqref{eq:finite_second_moment_operators} gives the stated scaling of
$\mathcal C_{\rho,N_r}$.

To establish the convergence of $\mathbf\Xi(\rho,N_r)$, write
$\mathbf U_{\mathrm a}(k)$ as
$\mathbf U_{\mathrm a,\infty}(k)$ plus their difference and
expand the Kronecker product in
\eqref{eq:finite_covariance_transition_matrix}. The last two bounds in
\eqref{eq:finite_terminal_perturbation_moment_bounds} and another
application of the Cauchy--Schwarz inequality show that every term
containing this difference converges to zero. This proves the high-SNR operator limits and bounds stated in
Subsection~\ref{subsec:finite_snr_transient_contributions}.

We next prove the uniform transient envelope. By the definition of
$\gamma_{\mathrm e}$ in \eqref{eq:finite_snr_transient_auxiliary_rates},
$r(\mathbf\Xi_\infty(N_r))<\gamma_{\mathrm e}^2<1$. Hence, there exists a compatible
matrix norm $\|\cdot\|_\star$ such that
$\|\mathbf\Xi_\infty(N_r)\|_\star<\gamma_{\mathrm e}^2$. The convergence in
the high-SNR operator bounds permits $\rho_0$ to be
chosen sufficiently large that
$\|\mathbf\Xi(\rho,N_r)\|_\star\leq \gamma_{\mathrm e}^2$ for every
$\rho\geq\rho_0$. Norm equivalence then gives a constant
$K_\Xi>0$, independent of $k$ and $\rho$, such that
\begin{equation}
\sup_{\rho\geq\rho_0}
\|\mathbf\Xi^k(\rho,N_r)\|_{\mathrm F}
\leq K_\Xi \gamma_{\mathrm e}^{2k},\qquad k\geq0.
\label{eq:finite_uniform_covariance_power_bound}
\end{equation}
The stationary plant-state and cross moments in
\eqref{eq:finite_stationary_plant_cross_moments} are independent of
$\rho$ under the adopted normalization. The same finite-$T$ moment
argument shows that the disturbance moments in
\eqref{eq:finite_disturbance_second_moments} are uniformly bounded for
$\rho\geq\rho_0$. For a fixed initialization, the initial moment
$\boldsymbol\Sigma_{\mathrm a}(1)$ is uniformly bounded as well. Together with \eqref{eq:finite_uniform_covariance_power_bound} and the fixed-point
formula \eqref{eq:finite_steady_state_augmented_covariance}, these facts
give
\begin{equation}
\begin{aligned}
\sup_{\rho\geq\rho_0}
\|\boldsymbol\Sigma_{\mathrm a}^\star\|_{\mathrm F}&<\infty,\\
\sup_{\rho\geq\rho_0}
\|\mathbf D_{\mathrm a}(1)\|_{\mathrm F}&<\infty,\\
\sup_{\rho\geq\rho_0}
\|\mathbf D_{\mathrm a x}(1)\|_{\mathrm F}&<\infty.
\end{aligned}
\label{eq:finite_uniform_initial_moment_bounds}
\end{equation}
By Theorem~\ref{thm:pcsi_mean_convergence} and \ref{app:proof_nominal_augmented_stability}, the powers
of $\mathbf F_{\mathrm a}$ are bounded by a constant multiple of $d^k$, whereas
the powers of $\tilde{\mathbf A}$ are bounded by a constant multiple of
$\gamma_{\mathrm x}^k$. Since $d<\gamma_{\mathrm e}$,
\eqref{eq:finite_transient_driving_deviations} and
\eqref{eq:finite_uniform_initial_moment_bounds} therefore bound the
plant-state and error--plant driving deviations, uniformly in
$\rho\geq\rho_0$, by constant multiples of $\gamma_{\mathrm x}^{2k}$ and
$(\gamma_{\mathrm e}\gamma_{\mathrm x})^k$, respectively.

Substitution of these bounds and
the high-SNR operator bounds into
\eqref{eq:finite_exact_mse_components} produces two
geometric convolutions. Since $\gamma_{\mathrm x}\neq \gamma_{\mathrm e}$, they satisfy
\begin{equation}
\begin{aligned}
&\sum_{t=1}^{k-1}
\gamma_{\mathrm e}^{2(k-1-t)}\gamma_{\mathrm x}^{2t}
=\gamma_{\mathrm x}^2\frac{\gamma_{\mathrm e}^{2(k-1)}-\gamma_{\mathrm x}^{2(k-1)}}
{\gamma_{\mathrm e}^2-\gamma_{\mathrm x}^2},\\
&\sum_{t=1}^{k-1}
\gamma_{\mathrm e}^{2(k-1-t)}(\gamma_{\mathrm e}\gamma_{\mathrm x})^{t-1}
=\frac{\gamma_{\mathrm e}^{2(k-1)}-(\gamma_{\mathrm e}\gamma_{\mathrm x})^{k-1}}
{\gamma_{\mathrm e}^2-\gamma_{\mathrm e}\gamma_{\mathrm x}}.
\end{aligned}
\label{eq:finite_high_snr_geometric_convolutions}
\end{equation}
Consequently, $H(k)$, $P(k)$, and $C(k)$ are bounded by
$\rho$-independent constant multiples of $\gamma_{\mathrm e}^{2k}$,
$\rho^{-1}(\gamma_{\mathrm e}^{2k}+\gamma_{\mathrm x}^{2k})$, and
$\rho^{-1/2}(\gamma_{\mathrm e}^{2k}+(\gamma_{\mathrm e}\gamma_{\mathrm x})^k)$, respectively. For
$\rho\geq\rho_0$, all terms with factor $\gamma_{\mathrm e}^{2k}$ can be absorbed into
one constant $C_{\mathrm b}$. Applying the triangle inequality in
\eqref{eq:finite_centered_mse_triangle_bound} then gives
\eqref{eq:finite_snr_transient_bound}.
\hfill$\blacksquare$

\section{Proof of Theorem~\ref{thm:ssmse}}\label{app:ssmse}
For this proof, let $n=MN$,
$\mathbf J_w=\mathbf1_M\otimes\mathbf I_N$, and
$\boldsymbol\Omega_0=\mathbf J_w\mathbf W_0\mathbf J_w^\top$,
so $\boldsymbol\Sigma_{\tilde w}=\alpha\boldsymbol\Omega_0$.
The receive-array size $N_r$ remains fixed. The common segment partition, local gains, communication weights, $T$, and pilot matrix are fixed throughout the comparisons in $\alpha$ and $\rho$.
The operators are those in \eqref{eq:finite_second_moment_operators}.
The control-cycle index is suppressed, and dependence on $\rho$
is displayed when needed.

\subsection{Disturbance Decomposition and High-SNR Moment Limits}
\label{app:mse:radio}
Use the fixed injection matrices $\mathbf L_w$ and $\mathbf J_\eta$
in \eqref{eq:finite_noise_injection_matrices}, and define
\begin{equation}
\mathbf R_{\eta,T}(\rho)\triangleq
\mathbb E[\tilde{\mathbf n}_T(\rho)\tilde{\mathbf n}_T^\top(\rho)].
\label{eq:ssmse:noise_matrices}
\end{equation}
The disturbance in \eqref{eq:finite_augmented_system} is
\begin{equation}
\boldsymbol\nu^{\mathrm a}(\rho)
=\mathbf L_w\tilde{\mathbf w}
+\rho^{-1/2}\mathbf J_\eta\tilde{\mathbf n}_T(\rho).
\label{eq:ssmse:noise_split}
\end{equation}
The data noise has zero conditional mean given channels and pilot
noises and is independent of the process noise.
Equation~\eqref{eq:finite_disturbance_second_moments} therefore gives
\begin{equation}
\begin{aligned}
\boldsymbol\Sigma_{\nu^{\mathrm a}}
={}&\alpha\mathbf L_w\boldsymbol\Omega_0\mathbf L_w^\top
+\rho^{-1}\mathbf J_\eta\mathbf R_{\eta,T}(\rho)\mathbf J_\eta^\top,\\
\boldsymbol\Sigma_{\nu^{\mathrm a}\tilde w}
={}&\alpha\mathbf L_w\boldsymbol\Omega_0.
\end{aligned}
\label{eq:ssmse:noise_covariances}
\end{equation}
We next bound the accumulated receiver noise.
At communication cycle $\tau$, \eqref{eq:finite_equivalent_data_noise}
gives
\begin{equation}
\mathbf n_i^\tau(\rho)
=\frac1{N_r}
\big[(\mathbf H_i^\tau)^\top
+\rho^{-1/2}(\breve{\mathbf N}_{p,i}^\tau)^\top\big]
\boldsymbol\eta_i^\tau.
\label{eq:ssmse:slotnoise}
\end{equation}
Let $d_i=|\mathcal N_i|$ and
$\mathbf R_p=\mathbf X_p^{-\top}\mathbf X_p^{-1}$.
Independence and zero means remove the channel--pilot cross terms,
so
\begin{equation}
\mathbb E[\mathbf n_i^\tau(\rho)(\mathbf n_i^\tau(\rho))^\top]
=\frac{d_i}{N_r}\mathbf I_N+\frac1{\rho N_r}\mathbf R_p.
\label{eq:ssmse:slotcov}
\end{equation}
The covariance of the actual data-noise term
$\rho^{-1/2}\mathbf n_i^\tau(\rho)$ is $\rho^{-1}$ times this matrix.

To handle the two-step communication recursion, retain
$(\boldsymbol\epsilon^{\mathrm a})^\tau
=\operatorname{col}(\boldsymbol\epsilon^\tau,
\boldsymbol\epsilon^{\tau-1},\boldsymbol\epsilon^0,
\tilde{\mathbf x}(k+1))$.
Equation~\eqref{eq:finite_global_internal_error} gives its transition
\begin{equation}
\mathbf\Lambda^\tau(\rho)=
\begin{bmatrix}
\mathbf\Lambda_1^\tau+\mathbf E_{\epsilon}^\tau
&\mathbf\Lambda_2^\tau&\tilde{\mathbf P}^{\mathcal O}
&\mathbf E_x^\tau\\
\mathbf I_n&\mathbf0&\mathbf0&\mathbf0\\
\mathbf0&\mathbf0&\mathbf I_n&\mathbf0\\
\mathbf0&\mathbf0&\mathbf0&\mathbf I_n
\end{bmatrix}.
\label{eq:ssmse:slot_transition}
\end{equation}
The pilot terms in \eqref{eq:finite_pilot_noise_perturbation}
yield the expansion
\begin{equation}
\mathbf\Lambda^\tau(\rho)
=\mathbf F_{\tau,0}+\rho^{-1/2}\mathbf F_{\tau,1}
+\rho^{-1}\mathbf F_{\tau,2}.
\label{eq:ssmse:lambda_expansion}
\end{equation}
Here $\mathbf F_{\tau,0}$ depends only on fading,
$\mathbf F_{\tau,1}$ is linear in the centered pilot noise,
and $\mathbf F_{\tau,2}$ contains its centered quadratic terms.
The segment projectors and weights are deterministic, so the Gaussian degree and centering of these coefficients are preserved. Their second moments are finite. Thus
$\mathbf\Lambda^\tau(\rho)$ is uniformly bounded in mean square
for $\rho\geq1$ and converges in mean square to $\mathbf F_{\tau,0}$.

The augmented receiver-noise injection is
\begin{equation}
\mathbf N^\tau(\rho)\triangleq
\operatorname{col}(\mathbf n_{\mathrm{in}}^\tau(\rho),\mathbf0_{3n}).
\label{eq:ssmse:slotinjection}
\end{equation}
Unrolling the noise recursion in \eqref{eq:finite_accumulated_data_noise}
gives
\begin{equation}
\tilde{\mathbf n}_T(\rho)
=\left[
\sum_{\tau=0}^{T-1}
\left(\prod_{m=T-1}^{\tau+1}\mathbf\Lambda^m(\rho)\right)
\mathbf N^\tau(\rho)
\right]_1,
\label{eq:ssmse:accumulated}
\end{equation}
where $[\cdot]_1$ extracts the leading $n$ entries.
Products have larger indices on the left, and an empty product
is $\mathbf I_{4n}$.
Set
\begin{equation}
\begin{aligned}
u_m&\triangleq\sup_{\rho\geq1}
\mathbb E[\|\mathbf\Lambda^m(\rho)\|_{\mathrm F}^2],\\
v_\tau&\triangleq\sup_{\rho\geq1}
\mathbb E[\|\mathbf N^\tau(\rho)\|^2].
\end{aligned}
\label{eq:ssmse:fastconstants}
\end{equation}
These constants are finite by the Gaussian moment bounds.
The inequality $\|\sum_\tau\mathbf z_\tau\|^2
\leq T\sum_\tau\|\mathbf z_\tau\|^2$ and independence across cycles yield
\begin{equation}
\begin{aligned}
\mathbb E[\|\tilde{\mathbf n}_T(\rho)\|^2]
&\leq T\sum_{\tau=0}^{T-1}v_\tau
\prod_{m=\tau+1}^{T-1}u_m
\triangleq C_T<\infty,\\
\|\mathbf R_{\eta,T}(\rho)\|_{\mathrm F}
&\leq\operatorname{tr}\mathbf R_{\eta,T}(\rho)\leq C_T.
\end{aligned}
\label{eq:ssmse:RTbound}
\end{equation}
No same-cycle independence between $\mathbf\Lambda^\tau$
and $\mathbf N^\tau$ is needed: each noise injection is multiplied
only by transitions from later cycles.

Finite products converge in mean square by a telescoping expansion:
each difference term contains one convergent factor and factors from
distinct other cycles with uniformly bounded second moments.
This applies to the terminal coefficient recursions
\eqref{eq:finite_graph_filter_perturbation}--\eqref{eq:finite_accumulated_data_noise}.
Consequently,
\begin{equation}
\begin{aligned}
\tilde{\mathbf n}_T(\rho)
&\longrightarrow\tilde{\mathbf n}_{T,\infty}\quad\text{in mean square},\\
\mathbf R_{\eta,T}(\rho)&\longrightarrow\mathbf R_{\eta,T,\infty},\\
\mathbf U_{\mathrm a}(\rho)
&\longrightarrow\mathbf U_{\mathrm a,\infty}\quad\text{in mean square},\\
\mathbf U_x(\rho)&\longrightarrow\mathbf0\quad\text{in mean square}.
\end{aligned}
\label{eq:ssmse:radio_limits}
\end{equation}
The zero plant-input limit follows because $\mathbf E_x^\tau$
contains only pilot-error terms.
For random matrices converging in mean square with bounded second
moments, Cauchy--Schwarz also gives convergence of their quadratic
and Kronecker moments. For example,
\begin{equation}
\begin{aligned}
&\|\mathbb E[\mathbf U_\rho\otimes\mathbf U_\rho
-\mathbf U_\infty\otimes\mathbf U_\infty]\|_2\\
&\leq
\big(\mathbb E\|\mathbf U_\rho-\mathbf U_\infty\|_{\mathrm F}^2\big)^{1/2}
\Big[
\big(\mathbb E\|\mathbf U_\rho\|_{\mathrm F}^2\big)^{1/2}\\
&\hspace{24mm}
+\big(\mathbb E\|\mathbf U_\infty\|_{\mathrm F}^2\big)^{1/2}
\Big]\longrightarrow0.
\end{aligned}
\label{eq:ssmse:moment_continuity}
\end{equation}
Applied to \eqref{eq:ssmse:radio_limits}, this establishes
\begin{equation}
\begin{aligned}
\mathbf\Xi(\rho,N_r)&\longrightarrow\mathbf\Xi_\infty(N_r),\\
\mathcal P_{\rho,N_r}&\longrightarrow0,\qquad
\mathcal C_{\rho,N_r}\longrightarrow0
\end{aligned}
\label{eq:ssmse:operator_limits}
\end{equation}
in finite-dimensional operator norm.
The same finite-product bounds apply uniformly to
$\mathbf U_{\mathrm a}(\rho)$ and $\mathbf U_x(\rho)$.
The process-noise injection matrix $\mathbf L_w$ is fixed.

\subsection{Mean-Square Stability and a Uniform Inverse Bound}\label{app:mse:inverse}
By \eqref{eq:finite_pcsi_mean_square_stability}, $\mathbf I-\mathbf\Xi_\infty(N_r)$ is invertible. Define
\begin{equation}
\begin{aligned}
\mathbf R_{\Xi,\infty}
&\triangleq(\mathbf I_{9n^2}-\mathbf\Xi_\infty(N_r))^{-1},\\
\mathbf R_\Xi(\rho)
&\triangleq(\mathbf I_{9n^2}-\mathbf\Xi(\rho,N_r))^{-1}
\end{aligned}
\label{eq:ssmse:inverse_def}
\end{equation}
whenever the latter inverse exists. Continuity of the spectral radius and \eqref{eq:ssmse:operator_limits} imply $r(\mathbf\Xi(\rho,N_r))<1$ for all sufficiently large $\rho$. Increase $\rho_0\ge1$, if necessary, so that also
\begin{equation}\label{eq:ssmse:resolvent_condition}
 \left\lVert \mathbf R_{\Xi,\infty}
 \bigl(\mathbf\Xi(\rho,N_r)-\mathbf\Xi_\infty(N_r)\bigr)
 \right\rVert_2\le\tfrac12
\end{equation}
for every $\rho\ge\rho_0$. The factorization
\begin{equation}
\begin{aligned}
\mathbf I-\mathbf\Xi(\rho,N_r)
={}&(\mathbf I-\mathbf\Xi_\infty(N_r))\\
&\times[\mathbf I-\mathbf R_{\Xi,\infty}
(\mathbf\Xi(\rho,N_r)-\mathbf\Xi_\infty(N_r))]
\end{aligned}
\label{eq:ssmse:factorization}
\end{equation}
and the convergent Neumann series give
\begin{equation}\label{eq:ssmse:inversebound}
 \begin{aligned}
 \left\lVert \mathbf R_\Xi(\rho)\right\rVert_2
 &\le2\left\lVert \mathbf R_{\Xi,\infty}\right\rVert_2,\\
 \mathbf R_\Xi(\rho)&\longrightarrow \mathbf R_{\Xi,\infty}.
 \end{aligned}
\end{equation}
This inverse bound also holds when $\mathbf\Xi(\rho,N_r)$ is nonnormal.

\subsection{Stationary Moments and Closed-Form Coefficients}\label{app:mse:stationary}
Define $\mathbf X_0\in\mathbb{R}^{n\times n}$ and $\mathbf Y_0\in\mathbb{R}^{3n\times n}$ by
\begin{equation}
\begin{aligned}
\operatorname{vec}\mathbf X_0
={}&(\mathbf I_{n^2}-\tilde{\mathbf A}\otimes\tilde{\mathbf A})^{-1}
\operatorname{vec}\boldsymbol\Omega_0,\\
\operatorname{vec}\mathbf Y_0
={}&(\mathbf I_{3n^2}-\tilde{\mathbf A}\otimes\mathbf F_{\mathrm a})^{-1}\\
&\times\operatorname{vec}(\mathbf L_w\boldsymbol\Omega_0).
\end{aligned}
\label{eq:ssmse:xy}
\end{equation}
Substituting \eqref{eq:ssmse:noise_covariances} into
\eqref{eq:finite_plant_second_moment_solution} and
\eqref{eq:finite_cross_second_moment_solution}
shows that their stationary forcings are
$\alpha\boldsymbol\Omega_0$ and
$\alpha\mathbf L_w\boldsymbol\Omega_0$, respectively.
The mixed contribution $\mathcal C_{\rho,N_r}$
remains in the augmented-error recursion
\eqref{eq:finite_augmented_second_moment_solution}.

The transition matrices of the first two vectorized recursions are $\tilde{\mathbf A}\otimes\tilde{\mathbf A}$ and $\tilde{\mathbf A}\otimes \mathbf F_{\mathrm a}$. Their spectral radii are $r(\hat{\mathbf A})^2<1$ and $r(\hat{\mathbf A})r(\mathbf F_{\mathrm a})<1$. Thus their unique limits are
\begin{equation}\label{eq:ssmse:xy_limits}
 \boldsymbol\Sigma_{\tilde x}^\star=\alpha \mathbf X_0,\qquad
 \boldsymbol\Sigma_{\mathrm a\tilde x}^\star=\alpha \mathbf Y_0.
\end{equation}
The fixed-point forcing can now be written using
\begin{equation}
\begin{aligned}
\mathbf G_w(\rho)
\triangleq{}&\mathbf L_w\boldsymbol\Omega_0\mathbf L_w^\top\\
&+\mathcal P_{\rho,N_r}(\mathbf X_0)
+\mathcal C_{\rho,N_r}(\mathbf Y_0),\\
\mathbf G_\eta(\rho)
\triangleq{}&\mathbf J_\eta\mathbf R_{\eta,T}(\rho)
\mathbf J_\eta^\top.
\end{aligned}
\label{eq:ssmse:forcing}
\end{equation}
After including the third recursion, the stacked vectorized moment system is affine with a block triangular transition matrix. Its diagonal blocks are the two stable matrices just identified and $\mathbf\Xi(\rho,N_r)$, which is stable by \ref{app:mse:inverse}. All its eigenvalues are therefore inside the unit disk. Powers of this transition matrix converge to zero, and its affine recursion converges to a unique fixed point for every finite initial second moment.

At this fixed point, substitute \eqref{eq:ssmse:noise_covariances} and \eqref{eq:ssmse:xy_limits} into \eqref{eq:finite_augmented_second_moment_solution} to obtain
\begin{equation}\label{eq:ssmse:sigma_solution}
 \begin{aligned}
 \operatorname{vec}\boldsymbol\Sigma_{\mathrm a}^\star
 ={}&\mathbf R_\Xi(\rho)\\
 &\times\operatorname{vec}[\alpha\mathbf G_w(\rho)
 +\rho^{-1}\mathbf G_\eta(\rho)].
 \end{aligned}
\end{equation}
Define $\boldsymbol\ell_e=\operatorname{vec}\operatorname{diag}(\mathbf I_n,\mathbf0_{n\times n},\mathbf0_{n\times n})$, so
$\boldsymbol\ell_e^\top\operatorname{vec}\boldsymbol\Sigma_{\mathrm a}^\star=\operatorname{tr}[(\boldsymbol\Sigma_{\mathrm a}^\star)_{11}]$.
The leading-block trace is selected by $\boldsymbol\ell_e$, so \eqref{eq:ssmse:sigma_solution} yields
\begin{equation}
\begin{aligned}
\mathcal M_\star(\alpha,\rho;N_r)
={}&\boldsymbol\ell_e^\top\mathbf R_\Xi(\rho)\\
&\times\operatorname{vec}[\alpha\mathbf G_w(\rho)
+\rho^{-1}\mathbf G_\eta(\rho)].
\end{aligned}
\label{eq:ssmse:closed}
\end{equation}
Consequently, the coefficients in \eqref{eq:ssmse:affine} are
\begin{equation}\label{eq:ssmse:coefficients}
 \begin{aligned}
 a(\rho,N_r)&=\boldsymbol\ell_e^\top \mathbf R_\Xi(\rho)\operatorname{vec} \mathbf G_w(\rho),\\
 b(\rho,N_r)&=\rho^{-1}\boldsymbol\ell_e^\top \mathbf R_\Xi(\rho)\operatorname{vec} \mathbf G_\eta(\rho).
 \end{aligned}
\end{equation}
For fixed $T$, all expectations in \eqref{eq:ssmse:closed} are finite Gaussian polynomial moments determined by the specified channel and noise statistics. This is a finite-dimensional closed-form representation of the steady-state MSE. It is the sum of the sensors' MSEs, as in \eqref{eq:finite_tracking_mse}; division by $M$ gives the network average.

\subsection{Process-Noise Dependence and Nonnegativity}\label{app:mse:process}
Linearity in \eqref{eq:ssmse:sigma_solution} gives \eqref{eq:ssmse:affine}. Its coefficients are nonnegative for a physical reason that does not require $\mathcal C_{\rho,N_r}(\mathbf Y_0)$ to be positive semidefinite. Take zero initial states, which does not change the steady-state limit. Conditional on the entire sequence of channel and pilot samples, the plant and estimator are linear in the process innovations and the receiver data-noise innovations. Decompose the error into the responses generated by these two innovation groups. Their conditional cross moment is zero, because the groups are independent and have zero means. The unconditional MSE is consequently the sum of their two nonnegative squared-error energies. Scaling $\mathbf W_w$ from $\mathbf W_0$ to $\alpha \mathbf W_0$ multiplies the process-generated second moment by $\alpha$, whereas the receiver-noise contribution is unchanged. Their limits are precisely $a(\rho,N_r)$ and $b(\rho,N_r)$ in \eqref{eq:ssmse:affine}.

For general $\mathbf W_w^{(2)}\succeq \mathbf W_w^{(1)}$, realize the larger-covariance innovation as the sum of an innovation with covariance $\mathbf W_w^{(1)}$ and an independent zero-mean innovation with covariance $\mathbf W_w^{(2)}-\mathbf W_w^{(1)}$. Conditional linearity makes the additional error energy nonnegative. Hence the steady-state MSE cannot decrease. This proves the process-noise assertions in part (i).

\subsection{SNR Monotonicity for a Fixed Design}\label{app:mse:snr}
The scalar powers of $\rho$ in the physical slot recursion permit an order comparison that is stronger than a norm bound. Condition on the current channel realization in \eqref{eq:ssmse:lambda_expansion}. Centering of the quadratic pilot term by the bias compensation gives
\begin{equation}\label{eq:ssmse:orthogonality}
 \begin{aligned}
 \mathbb{E}_p[\mathbf F_{\tau,1}]&=\mathbb{E}_p[\mathbf F_{\tau,2}]=0,\\
 \mathbb{E}_p[\mathbf F_{\tau,1}\mathbf Z\mathbf F_{\tau,2}^\top]&=0
 \end{aligned}
\end{equation}
for every deterministic compatible $\mathbf Z$, where $\mathbb{E}_p$ integrates the current Gaussian pilot noises. The second identity follows because its entries are sums of odd-degree Gaussian moments: $\mathbf F_{\tau,1}$ is linear and $\mathbf F_{\tau,2}$ is centered quadratic. Cross terms with $\mathbf F_{\tau,0}$ vanish by the first identity. Therefore, for $\mathbf Z\succeq0$,
\begin{equation}
\begin{aligned}
&\mathbb E[\mathbf\Lambda^\tau(\rho)\mathbf Z
\mathbf\Lambda^\tau(\rho)^\top]\\
&=\mathcal K_{\tau,0}(\mathbf Z)
+\rho^{-1}\mathcal K_{\tau,1}(\mathbf Z)
+\rho^{-2}\mathcal K_{\tau,2}(\mathbf Z),
\end{aligned}
\label{eq:ssmse:positive_maps}
\end{equation}
where $\mathcal K_{\tau,j}(\mathbf Z)=\mathbb{E}[\mathbf F_{\tau,j}\mathbf Z\mathbf F_{\tau,j}^\top]\succeq0$.
Likewise the actual slot-noise vector has the form $\rho^{-1/2}\mathbf g_{\tau,0}+\rho^{-1}\mathbf g_{\tau,1}$, with $\mathbf g_{\tau,0}$ containing channel times data noise and $\mathbf g_{\tau,1}$ containing pilot times data noise. Their cross covariance is zero by the centered pilot noise, so the additive covariance is
\begin{equation}\label{eq:ssmse:positive_additive}
 \rho^{-1}\mathbf Q_{\tau,0}+\rho^{-2}\mathbf Q_{\tau,1},\qquad \mathbf Q_{\tau,j}\succeq0.
\end{equation}
There are no cross terms between this additive noise and the preceding joint state: conditional on the channel and pilot samples, the current data noise has zero mean.

For the comparison, retain the complete joint state during each slot. In addition to the fast-time state $(\boldsymbol\epsilon^{\mathrm a})^\tau$, retain
\begin{equation}
\begin{aligned}
\mathbf v(k)&\triangleq\operatorname{col}\big(
\mathbf e(k),\mathbf e(k-1),\\
&\tilde{\mathbf w}(k-1),\tilde{\mathbf w}(k)\big),\\
\boldsymbol\zeta^\tau(k)&\triangleq
\operatorname{col}((\boldsymbol\epsilon^{\mathrm a})^\tau(k),\mathbf v(k)).
\end{aligned}
\label{eq:ssmse:joint_memory}
\end{equation}
The last plant-state component of $(\boldsymbol\epsilon^{\mathrm a})^\tau$ already contains $\tilde{\mathbf x}(k+1)$. The retained variables make the posterior update \eqref{eq:pcsi_stacked_posterior_error} and the shift to $\mathbf e^{\mathrm a}(k+1)$ fixed linear maps of the terminal joint state. During communication, $\mathbf v(k)$ remains fixed. The enlarged transition is $\operatorname{diag}(\mathbf\Lambda^\tau(\rho),\mathbf I_{4n})$ and has the form \eqref{eq:ssmse:lambda_expansion} and satisfies \eqref{eq:ssmse:orthogonality}--\eqref{eq:ssmse:positive_maps}. Prediction, initialization, posterior update, and the memory shift are fixed linear maps; their action on a joint second moment is a congruence $\mathbf Z\mapsto\mathbf U\mathbf Z\mathbf U^\top$. Independent process-noise injection adds a fixed positive-semidefinite covariance. Thus every stage preserves positive-semidefinite order.

If $\rho_2\ge\rho_1\ge\rho_0$, \eqref{eq:ssmse:positive_maps}--\eqref{eq:ssmse:positive_additive} show that each slot map at $\rho_2$ is no larger than the same map at $\rho_1$, on every positive-semidefinite input. Starting from the same zero initial joint state, induction over the slots and control cycles orders the full second moments. Extracting the error block and taking its trace preserves this order. Taking the steady-state limits established in \ref{app:mse:stationary} proves $\mathcal M_\star(\alpha,\rho_2;N_r)\le\mathcal M_\star(\alpha,\rho_1;N_r)$. This argument uses the full joint moment, rather than trying to order the generally signed cross term $\mathcal C_{\rho,N_r}$ separately, and completes the monotonicity assertions in part (i).

\subsection{Uniform Bounds Independent of Noise Variance and SNR}\label{app:mse:bounds}
The mean-square bounds established in \ref{app:mse:radio} make the following constants finite:
\begin{equation}\label{eq:ssmse:uniformconstants}
 \begin{aligned}
 K_w&=\left\lVert \mathbf L_w\right\rVert_2^2,\\
 K_a&=\sup_{\rho\ge\rho_0}\mathbb{E}\left\lVert \mathbf U_{\mathrm a}(\rho)\right\rVert_2^2,\\
 K_x&=\sup_{\rho\ge\rho_0}\mathbb{E}\left\lVert \mathbf U_x(\rho)\right\rVert_2^2.
 \end{aligned}
\end{equation}
For example, $\left\lVert\mathbf L_w\boldsymbol\Omega_0\mathbf L_w^\top\right\rVert_{\mathrm F}\le\left\lVert\mathbf L_w\right\rVert_2^2\left\lVert \boldsymbol\Omega_0\right\rVert_{\mathrm F}$. Applying this inequality to the process term, and using Cauchy--Schwarz for the mixed term, gives
\begin{equation}\label{eq:ssmse:Gwbound}
 \begin{aligned}
 \left\lVert \mathbf G_w(\rho)\right\rVert_{\mathrm F}&\le K_w\left\lVert \boldsymbol\Omega_0\right\rVert_{\mathrm F}+K_x\left\lVert \mathbf X_0\right\rVert_{\mathrm F}\\
 &\quad+2\sqrt{K_aK_x}\left\lVert \mathbf Y_0\right\rVert_{\mathrm F}=:C_G,\\
 \left\lVert \mathbf G_\eta(\rho)\right\rVert_{\mathrm F}&\le\left\lVert \mathbf J_\eta\right\rVert_2^2 C_T.
 \end{aligned}
\end{equation}
The norm of a vectorized matrix equals its Frobenius norm. Use \eqref{eq:ssmse:inversebound}, \eqref{eq:ssmse:Gwbound}, and \eqref{eq:ssmse:coefficients} to choose
\begin{equation}\label{eq:ssmse:finalconstants}
 \begin{aligned}
 C_w&=2\left\lVert \boldsymbol\ell_e\right\rVert_2\left\lVert \mathbf R_{\Xi,\infty}\right\rVert_2 C_G,\\
 C_n&=2\left\lVert \boldsymbol\ell_e\right\rVert_2
 \left\lVert \mathbf R_{\Xi,\infty}\right\rVert_2\\
 &\phantom{=}\times\left\lVert \mathbf J_\eta\right\rVert_2^2 C_T.
 \end{aligned}
\end{equation}
Then $0\le a(\rho,N_r)\le C_w$ and $0\le b(\rho,N_r)\le C_n/\rho$. This proves \eqref{eq:ssmse:bound}, with constants independent of both $\alpha$ and $\rho$.

\subsection{High-SNR Coefficients and the Joint Limit}\label{app:mse:limit}
By \eqref{eq:ssmse:radio_limits} and \eqref{eq:ssmse:operator_limits},
\begin{equation}\label{eq:ssmse:Glimits}
 \begin{aligned}
 \mathbf G_w(\rho)&\longrightarrow\mathbf L_w\boldsymbol\Omega_0\mathbf L_w^\top,\\
 \mathbf G_\eta(\rho)&\longrightarrow \mathbf J_\eta \mathbf R_{\eta,T,\infty}\mathbf J_\eta^\top.
 \end{aligned}
\end{equation}
Combining \eqref{eq:ssmse:Glimits} with $\mathbf R_\Xi(\rho)\to \mathbf R_{\Xi,\infty}$ in \eqref{eq:ssmse:coefficients} proves \eqref{eq:ssmse:limits}. In particular, their explicit expressions are
\begin{equation}
\begin{aligned}
a_\infty(N_r)
={}&\boldsymbol\ell_e^\top\mathbf R_{\Xi,\infty}\\
&\times\operatorname{vec}\mathbb E[\mathbf L_w\boldsymbol\Omega_0
\mathbf L_w^\top],\\
c_\infty(N_r)
={}&\boldsymbol\ell_e^\top\mathbf R_{\Xi,\infty}\\
&\times\operatorname{vec}(\mathbf J_\eta
\mathbf R_{\eta,T,\infty}\mathbf J_\eta^\top).
\end{aligned}
\label{eq:ssmse:limitingconstants}
\end{equation}
Both limits are nonnegative because $a(\rho,N_r)$ and $\rho b(\rho,N_r)$ are nonnegative. With $\alpha=0$, the stationary plant moment and error--plant cross moment in \eqref{eq:ssmse:xy_limits} are zero, even for nonzero initial states. The stationary error equation then reduces exactly to
\begin{equation}
\begin{aligned}
\operatorname{vec}\boldsymbol\Sigma_{\mathrm a}^\star
={}&\rho^{-1}\mathbf R_\Xi(\rho)\\
&\times\operatorname{vec}[
\mathbf J_\eta\mathbf R_{\eta,T}(\rho)\mathbf J_\eta^\top].
\end{aligned}
\label{eq:ssmse:noiseless_exact}
\end{equation}
This proves \eqref{eq:ssmse:noiseless}. Persistent fading at a finite $N_r$ affects $\mathbf R_{\Xi,\infty}$, but it does not produce an additive stationary forcing when both the process noise and the high-SNR receiver noise are absent.

Finally, \eqref{eq:ssmse:bound} gives, along any sequences in part (iii),
\begin{equation}\label{eq:ssmse:squeeze}
 0\le\mathcal M_\star(\alpha_j,\rho_j;N_r)
          \le C_w\alpha_j+C_n/\rho_j\longrightarrow0.
\end{equation}
At a fixed $\alpha$, \eqref{eq:ssmse:affine} and \eqref{eq:ssmse:limits} give $\lim_{\rho\to\infty}\mathcal M_\star(\alpha,\rho;N_r)=\alpha a_\infty(N_r)$. At a fixed stable $\rho$, $\lim_{\alpha\to0}\mathcal M_\star(\alpha,\rho;N_r)=b(\rho,N_r)$, which vanishes as $\rho\to\infty$. Thus both iterated limits also equal zero. All conclusions concern the steady-state limit taken first in time; \ref{app:mse:stationary} ensures that this limit is independent of any admissible finite initial second moments.
\hfill$\blacksquare$

\section{Proof of Corollary~\ref{cor:finite_joint_ideal_mse_convergence}}
\label{app:proof_ideal_mse_convergence}
The explicit $\rho$-dependent factors in
\eqref{eq:aggregated_csi}, \eqref{eq:regularized_ota_term}, and
\eqref{eq:finite_pilot_noise_perturbation}, together with the
receive-array bounds in
\eqref{eq:finite_receive_array_terminal_scaling}, imply the uniform
bounds
\begin{equation}
\begin{aligned}
\sup_{\rho\geq1}\mathbb E\!\left[
\|\mathbf R_T(k)\|_{\mathrm F}^2
+\|\mathbf Q_T(k)\|_{\mathrm F}^2
\right]
&=\mathcal O(N_r^{-1}),\\
\sup_{\rho\geq1}\mathbb E\!\left[
\|\mathbf U_{\mathrm a}(k)\|_{\mathrm F}^2
+\|\mathbf U_x(k)\|_{\mathrm F}^2
\right]
&=\mathcal O(N_r^{-1}),\\
\sup_{\rho\geq1}\frac{1}{\rho}
\mathbb E[\|\tilde{\mathbf n}_T(k)\|^2]
&=\mathcal O(N_r^{-1}).
\end{aligned}
\label{eq:finite_joint_uniform_terminal_scaling}
\end{equation}
The constants hidden in
\eqref{eq:finite_joint_uniform_terminal_scaling} are independent of
$\rho$ and $N_r$: for $\rho\geq1$, every inverse power of $\rho$ is
bounded by one, while the fading-only terms are independent of $\rho$.
Hence, along any joint sequence with $\rho\to\infty$ and
$N_r\to\infty$,
$\mathbf U_{\mathrm a}(k)\to\mathbf0$,
$\mathbf U_x(k)\to\mathbf0$, and
$\boldsymbol\nu^{\mathrm a}(k)\to\boldsymbol\nu_{\mathrm w}^{\mathrm a}(k)$ in mean square.
By \eqref{eq:finite_second_moment_operators}, the mean-square bounds in
\eqref{eq:finite_joint_uniform_terminal_scaling} yield
\begin{equation}
\begin{aligned}
\lim_{\substack{\rho\to\infty\\N_r\to\infty}}
\mathcal H_{\rho,N_r}(\mathbf Z)
&=\mathbf F_{\mathrm a}\mathbf Z\mathbf F_{\mathrm a}^\top,\\
\lim_{\substack{\rho\to\infty\\N_r\to\infty}}
\mathcal P_{\rho,N_r}(\mathbf Z)&=\mathbf0,\\
\lim_{\substack{\rho\to\infty\\N_r\to\infty}}
\mathcal C_{\rho,N_r}(\mathbf Z)&=\mathbf0.
\end{aligned}
\label{eq:finite_joint_ideal_operator_limits}
\end{equation}
The limiting second-moment recursion retains only the time-invariant
forcing $\boldsymbol\Sigma_{\nu_{\mathrm w}}$ in
\eqref{eq:finite_process_only_augmented_disturbance} and therefore
satisfies
\begin{equation}
\boldsymbol\Sigma_{\mathrm a}(k+1)
=\mathbf F_{\mathrm a}\boldsymbol\Sigma_{\mathrm a}(k)\mathbf F_{\mathrm a}^\top
+\boldsymbol\Sigma_{\nu_{\mathrm w}}.
\label{eq:finite_joint_ideal_covariance_recursion}
\end{equation}
By
Theorem~\ref{thm:pcsi_mean_convergence} and \ref{app:proof_nominal_augmented_stability},
$r(\mathbf F_{\mathrm a})<d<1$, so
\eqref{eq:finite_joint_ideal_covariance_recursion} has a unique finite
stationary second moment and is mean-square stable. Subtracting the
associated stationary equation from
\eqref{eq:finite_joint_ideal_covariance_recursion} eliminates the
process-noise forcing and gives
\begin{equation}
\mathbf D_{\mathrm a}(k+1)
=\mathbf F_{\mathrm a}\mathbf D_{\mathrm a}(k)\mathbf F_{\mathrm a}^\top.
\label{eq:finite_joint_ideal_centered_covariance_recursion}
\end{equation}
Vectorizing and iterating \eqref{eq:finite_joint_ideal_centered_covariance_recursion}
from $k=1$ yields
\begin{equation}
\begin{aligned}
\operatorname{vec}(\mathbf D_{\mathrm a}(k))
={}&(\mathbf F_{\mathrm a}\otimes\mathbf F_{\mathrm a})^{k-1}\times\operatorname{vec}(\mathbf D_{\mathrm a}(1)).
\end{aligned}
\label{eq:finite_joint_ideal_centered_covariance_solution}
\end{equation}
Choose any $q$ such that $r(\mathbf F_{\mathrm a})<q<d$. The standard
finite-dimensional matrix-power bound and
$r(\mathbf F_{\mathrm a}\otimes\mathbf F_{\mathrm a})=r(\mathbf F_{\mathrm a})^2$ give a constant
$K_q>0$ such that
\begin{equation}
\|\mathbf D_{\mathrm a}(k)\|_{\mathrm F}
\leq K_q q^{2(k-1)}
\|\mathbf D_{\mathrm a}(1)\|_{\mathrm F}.
\label{eq:finite_joint_ideal_covariance_decay_bound}
\end{equation}
Leading-block extraction followed by the trace is a bounded linear
functional. Therefore, \eqref{eq:finite_tracking_mse},
\eqref{eq:finite_joint_ideal_covariance_decay_bound}, and $q<d$ imply
\eqref{eq:finite_joint_ideal_mse_decay}. Equivalently, the centered MSE
transient is $o(d^{2k})$.

If $\boldsymbol\Sigma_{\tilde w}=\mathbf0$, then
\eqref{eq:finite_process_only_augmented_disturbance} gives
$\boldsymbol\Sigma_{\nu_{\mathrm w}}=\mathbf0$. Equation
\eqref{eq:finite_receive_array_stationary_limit} then gives
$\boldsymbol\Sigma_{\mathrm a}^\star=\mathbf0$, so the steady-state MSE
is zero.
\hfill$\blacksquare$


\bibliographystyle{elsarticle-harv}
\bibliography{bibtex/bib/ref}

\end{document}